\documentclass[aps,prb,twocolumn,footinbib,superscriptaddress,longbibliography]{revtex4-2}
\usepackage{graphicx}
\usepackage{rotating}
\usepackage{color}
\usepackage{bm}
\usepackage{braket}
\usepackage{amsmath,amssymb}
\usepackage{calc}
\usepackage[shortcuts]{extdash}
\usepackage{bbold}
\usepackage[cal=boondoxo]{mathalfa}
\allowdisplaybreaks
\usepackage[bookmarksnumbered,bookmarksopen]{hyperref}
\hypersetup{
   colorlinks=true,
   citecolor=blue,
}

\usepackage{url}
\usepackage[normalem]{ulem}

\usepackage{array}
\newcolumntype {L}{>{$}l<{$}}
\newcolumntype {C}{>{$}c<{$}}
\newcolumntype {R}{>{$}r<{$}}
\newcolumntype {s}[1]{@{\hspace*{#1}}}
\newcolumntype {S}[1]{@{\extracolsep{#1}}}
\newcolumntype {e}{@{\extracolsep {0pt}}}
\newcolumntype {E}{@{\extracolsep \fill}}
\newcolumntype {n}{@{}}

\newcommand{\marr}[2][@{}l@{}]{\mbox{\renewcommand{\arraystretch}{0.8}%
  \tabcolsep 0pt\begin{tabular}{#1} #2 \end{tabular}}}

\newcommand{\tbox}[3][0pt]{\raisebox{#1}{{\renewcommand{\arraystretch}{1.0}%
\begin{tabular}[t]{#2}#3\end{tabular}}}}

\usepackage{accents}
\newcommand{\dbtilde}[1]{\accentset{\approx}{#1}}

\newcommand* {\vek}[1]{{\bm{\mathrm{#1}}}}
\newcommand* {\vekc}[1]{{\bm{\mathcal{#1}}}}
\newcommand* {\kk}{\vek{k}}
\newcommand* {\gr}{\mathcal{g}}
\newcommand* {\gv}{\vekc{g}}
\newcommand* {\rr}{\vek{r}}
\newcommand* {\xx}{\vek{x}}
\newcommand* {\frack}[2]{{\Ts\frac{#1}{#2}}}

\newcommand* {\Ts}{\textstyle}

\renewcommand* {\cp}{\mbox{cp}}

\newcommand* {\gtis}{C_\theta}
\newcommand* {\gsis}{C_i}
\newcommand* {\gstis}{C_{i\theta}}
\newcommand* {\gsistis}{C_{i \times \theta}}
\newcommand* {\ggen}{C_\gamma}
\newcommand* {\ggend}{C_{\gamma_\mathrm{d}}}

\newcommand* {\Rit}{R_{i \times \theta}}

\newcommand* {\gin}{\zeta}
\newcommand* {\any}{\mathrm{any}}
\newcommand* {\inv}[2]{{(\mathrm{#1},#2)}}
\newcommand* {\gprop}{G_\mathrm{p}}
\newcommand* {\gtprop}{\tilde{G}_\mathrm{p}}
\newcommand* {\gttprop}{\dbtilde{G}_\mathrm{p}}
\newcommand* {\gmin}[1]{\langle#1\rangle}

\newcommand* {\dprime}{{\prime\prime}}

\DeclareFontFamily{U}{MnSymbolC}{}
\DeclareFontShape{U}{MnSymbolC}{m}{n}{
    <-6> MnSymbolC5
   <6-7> MnSymbolC6
   <7-8> MnSymbolC7
   <8-9> MnSymbolC8
  <9-10> MnSymbolC9
 <10-12> MnSymbolC10
 <12->   MnSymbolC12}{}
\DeclareFontShape{U}{MnSymbolC}{b}{n}{
    <-6> MnSymbolC-Bold5
   <6-7> MnSymbolC-Bold6
   <7-8> MnSymbolC-Bold7
   <8-9> MnSymbolC-Bold8
  <9-10> MnSymbolC-Bold9
 <10-12> MnSymbolC-Bold10
 <12->   MnSymbolC-Bold12}{}
\DeclareSymbolFont{MnSyC}      {U}{MnSymbolC}{m}{}
\SetSymbolFont{MnSyC}    {bold}{U}{MnSymbolC}{b}{}
\DeclareMathSymbol{\mcirc}{\mathbin}{MnSyC}{88}
\DeclareMathSymbol{\mbullet}{\mathbin}{MnSyC}{89}
\DeclareMathSymbol{\medcirc}{\mathbin}{MnSyC}{90}
\DeclareMathSymbol{\msquare}{\mathbin}{MnSyC}{104}
\DeclareMathSymbol{\mfilledsquare}{\mathbin}{MnSyC}{105}
\DeclareMathSymbol{\mdiamond}{\mathbin}{MnSyC}{108}
\DeclareMathSymbol{\mfilleddiamond}{\mathbin}{MnSyC}{109}
\DeclareMathSymbol{\mstar}{\mathbin}{MnSyC}{128}
\DeclareMathSymbol{\mfilledstar}{\mathbin}{MnSyC}{129}
\DeclareMathSymbol{\mast}{\mathbin}{MnSyC}{135}

\newcommand*{\mysymb}[1]{\ifcase#1 \rule{1.0ex}{0pt}\or
  \mcirc\or    \mbullet\or
  \mstar\or    \mfilledstar\or
  \mdiamond\or \mfilleddiamond\or
  \msquare\or  \mfilledsquare\or
  \fi}
\newcommand*{\mygroup}[4]{(%
  \raisebox{0.2ex}[0pt][0pt]{\renewcommand{\arraystretch}{1.0}%
  \extrarowheight 0pt
  $\begin{array}
  {@{}>{\scriptstyle}cs{0.08em}>{\scriptstyle}c@{}}
  \mysymb{#1} & \mysymb{#2} \\[-1.6ex] \mysymb{#3} & \mysymb{#4}
  \end{array}$})}

\makeatletter
\newcommand{\customlabel}[2]{%
   \protected@write \@auxout {}{\string \newlabel {#1}{{#2}{\thepage}{#2}{#1}{}}}%
   \@ifundefined{hypertarget}{}{\hypertarget{#1}{#2}}}

\newenvironment{subtables}{%
  \refstepcounter{table}%
  \mathchardef\c@parenttable\c@table
  \protected@edef\theparenttable{\thetable}%
  \protected@edef\theHparenttable{%
    \@ifundefined{theHtable}\thetable\theHtable
  }%
  \let\thetable\thesubtable
  \global\c@table\z@
  \def\theHtable{\theHparenttable\alph{table}}%
  \ignorespaces
  }{%
  \global\c@table\c@parenttable
  \global\@ignoretrue
  }
\newcommand{\thesubtable}{\theparenttable(\alph{table})}
\makeatother

\newcommand* {\wlra}{\;\leftrightarrow\;}

\graphicspath{{./FIG/}}

\begin{document}

\title{Standard Model of Electromagnetism and Chirality in Crystals}

\author{R. Winkler}
\affiliation{Department of Physics, Northern Illinois University,
DeKalb, Illinois 60115, USA}
\affiliation{Materials Science Division, Argonne National Laboratory,
Lemont, Illinois 60439, USA}

\author{U. Z\"ulicke}
\affiliation{MacDiarmid Institute, School of Chemical and Physical Sciences,
Victoria University of Wellington, PO Box 600, Wellington 6140, New
Zealand}

\date{\today}

\begin{abstract}
We present a general, systematic, and quantitative theory of
electromagnetism and chirality in crystalline solids.
Symmetry is its basic guiding principle, enabling us to consider
macroscopic multipole densities without reference to any specific
microscopic configurations.  We use a formal analogy between space
inversion $i$ and time inversion $\theta$ to identify two complementary,
comprehensive classifications of crystals, based on five categories of
electric and magnetic multipole order---called polarizations---and five
categories of chirality.
The five categories of polarizations (parapolar, electropolar,
magnetopolar, antimagnetopolar, and multipolar) embody the variety
of ways in which electromagnetic multipole order can be realized in
solids, thus expanding the familiar notion of electric
dipolarization in ferroelectrics and magnetization in ferromagnets
to higher-rank multipole densities characterizing, e.g.,
altermagnets and alterelectrics.  Group theory also permits the
absolute, gauge-invariant quantification of these multipole
densities.
The five categories of chirality (parachiral, electrochiral,
magnetochiral, antimagnetochiral, and multichiral) extend the notion
of enantiomorphism---conventionally associated with the lack of
spatial mirror symmetries---to include all possibilities for creating
non-superposable images by applying the
inversions $i$, $\theta$, and $i\theta$.  In particular, multichiral
systems lack all inversion symmetries and therefore have four different
enantiomorphs.  Each category of chirality is shown to arise from
particular superpositions of electric and magnetic multipole densities.
Our theory has great predictive power by relating electric and
magnetic multipolar order with, e.g., distinctive features in the
crystal structure, the electronic band structure, property tensors
and response tensors of quantum materials.  These findings are
critical for a large community working on fundamental and applied
aspects of novel quantum materials.
\end{abstract}

\maketitle

\section{Introduction}
\label{sec:intro}

Recent work has identified a multitude of new forms of electric and
magnetic order in quantum materials that transcends the traditional
understanding of crystalline solids.  Beyond the textbook examples of
ferroelectrics and ferromagnets, these new forms include, for
example, multiferroics \cite{fie16}, unconventional hidden-order
phases \cite{aep20}, altermagnets \cite{sme22, sme22a, fer24}, and
chirality in novel domains \cite{yan25, jur25}.  The richness of
materials and phenomena is reminiscent of particle physics when, mid-20th
century, more and more ``elementary'' particles were
discovered until the standard model of particle physics was worked
out that could organize and interpret these findings systematically.
Then, as now, symmetry plays a crucial role for obtaining a deeper
understanding of how seemingly disparate phenomena are
connected~\cite{wil06}.  Here, we use symmetry as the guiding concept
to develop a standard model of electromagnetism and chirality in
crystalline solids that provides a physically transparent, unified
predictive framework for the fascinating world of quantum materials.

Symmetry is the most fundamental organizing principle in solid-state
physics, linking the periodic crystal structure of solids with
physical properties.
Translational symmetry is usually the starting point for a
discussion of crystal symmetries; it yields the well-known 14
Bravais lattices \cite{ash76}.  These lattices are complemented by
point-group symmetries, including space inversion $i$, to obtain the
230 crystallographic space groups \cite{bra72}.  In a last step, to
account for magnetic phenomena, one may incorporate time inversion
$\theta$ to construct the 1651 magnetic space groups \cite{lud96}.

Important progress in our understanding of the physical properties
of crystalline solids was made early on based on Neumann's
principle~\cite{voi10, nye85} stating that macroscopic properties of
a crystal structure do not require a knowledge of the space group
characterizing a crystal structure.  Instead, these properties are
already determined by the simpler crystallographic point groups
(also called crystal classes \cite{bir74}) that include proper and
improper rotations as their group elements, but no
translations~\cite{wey52, kos63, bra72, bir74, lud96}.

In the present work, we follow Neumann's principle by not
considering translational symmetries, but we go one step further by
taking space inversion symmetry (SIS) and time inversion symmetry
(TIS) as the starting point.  SIS and TIS are among the most
fundamental symmetries describing the Universe.  Both SIS and TIS
are so-called black-white symmetries; they are represented by
two-element groups $\gsis = \{e, i\}$ and $\gtis = \{e, \theta\}$
that are obviously isomorphic \cite{misc:doubleG2}.  (Here $e$
denotes the unit element.)  We interpret the isomorphism as
\emph{SIS-TIS duality}.

\begin{figure}
  \centering
  \includegraphics[width=0.9\linewidth]{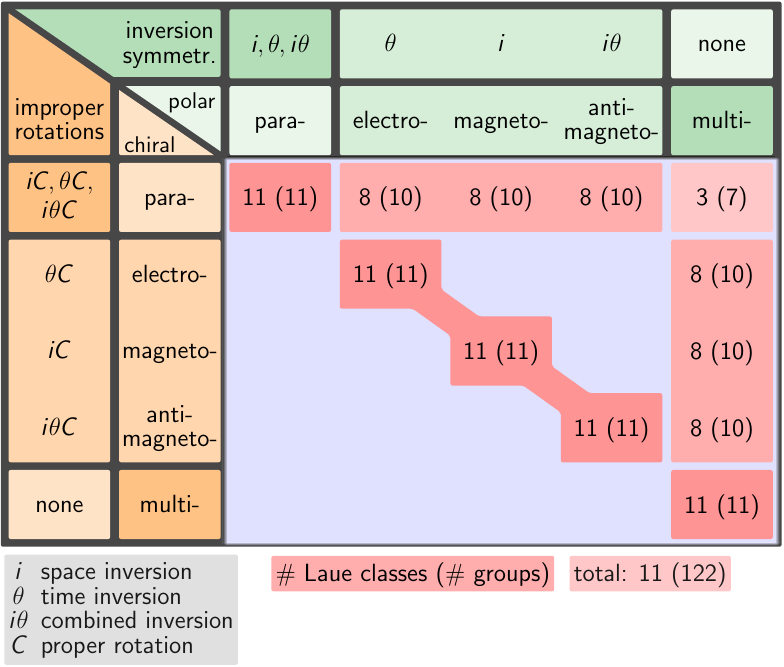}
  \caption[]{\label{fig:standardmodel} Standard model of broken
  crystal symmetries.  Crystal structures of materials fall into
  twelve types classified by the allowed polarizations and
  chirality.  The presence of inversion symmetries $i$, $\theta$,
  and $i\theta$ defines five categories of polarizations, while
  symmetry under improper rotations $iC$, $\theta C$, and $i\theta
  C$ defines five categories of chirality.  Triadic relations among
  the categories and types are highlighted by conjoint tiles.  For
  each type, the number of Laue classes and magnetic point groups
  (crystal classes) realizing it are indicated.}
\end{figure}

Space inversion $i$ and time inversion $\theta$ can be combined in
the full inversion group $\gsistis = \{e, i, \theta, i\theta \}$
that treats the three elements $i$ (SIS), $\theta$ (TIS), and
$i\theta$ (combined inversion symmetry, CIS) on an equal footing.
This allows us to define two complementary, comprehensive
classifications of crystals, using five categories of polarizations
and five categories of chirality.  In each scheme, the five
categories are related to the five subgroups of the group
$\gsistis$, see Table~\ref{tab:inversion-group}.  Jointly, the five
categories of polarizations and the five categories of chirality
yield the succinct and comprehensive description illustrated in
Fig.~\ref{fig:standardmodel} that we call the \emph{standard model}
of electromagnetism and chirality in crystals.

\begin{table}
  \caption{\label{tab:inversion-group} Inversion groups that can be
  formed from space inversion $i$, time inversion $\theta$, and the
  combined inversion $i\theta$.  The quantity $e$ denotes the unit
  element.}
  \renewcommand{\arraystretch}{1.2}%
  \begin{tabular*}{\linewidth}{nCLEln}
    \hline \hline \rule{0pt}{2.8ex}
    C_1 & = \{ e \} & trivial inversion group \\
    \gsis & = \{ e, i \} & space inversion group \\
    \gtis & = \{ e, \theta \} & time inversion group \\
    \gstis & = \{ e, i\theta \} & combined inversion group \\
    \gsistis & = \{ e, i, \theta, i\theta \} & full inversion group \\
    \hline \hline
  \end{tabular*}
\end{table}

\begin{table}
  \caption{\label{tab:tensor-signat} Signature $ss'$ of multipoles
  of order $\ell$.  Respective columns define the electric,
  magnetic, and toroidal multipoles represented via irreducible
  spherical tensors $T_\ell^{ss'}$ of the full rotation
  group~$\Rit$.}
  \renewcommand{\arraystretch}{1.2}%
  \begin{tabular*}{\linewidth}{cE*{4}{C}}
    \hline \hline \rule{0pt}{3.5ex}
    & \mbox{electric} & \mbox{magnetic}
    & \mbox{\marr{electro- \\ toroidal}}
    & \mbox{\marr{magneto- \\ toroidal}} \\ \hline
    $\ell$ even & ++ & -- & -+ & +- \\
    $\ell$ odd  & -+ & +- & ++ & -- \\ \hline \hline
  \end{tabular*}
\end{table}

In the following, we briefly summarize the essence of the standard
model and the obtained results.  We are particularly interested in
electric and magnetic multipole order as an organizing principle for
characterizing different phases of crystalline matter.  Such phases
are associated with spontaneously formed rank\=/$\ell$ multipole
densities, which we call \emph{polarizations}.  Following
Ref.~\cite{win23}, we identify four types of
po\-lar\-iza\-tions---electric, magnetic, electrotoroidal, and
magnetotoroidal, see Table~\ref{tab:tensor-signat}.  The
\emph{signature} $ss'$ indicates how a polarization behaves under
space inversion (even/odd if $s=+/-$) and time inversion (even/odd
if $s'=+/-$).

Examples of polarizations include the electric dipolarization in
ferroelectrics ($\ell = 1$ and signature $-+$) and the magnetization
in ferromagnets ($\ell = 1$, $+-$).  Early studies identified the
crystal classes permitting a bulk electric
dipolarization~\cite{voi10, nye85} or a bulk
magnetization~\cite{tav56, tav58, cor58a}.  More recent work focused
on higher-rank multipolar order~\cite{kur08, suz17, wat18, yat21,
kus23, win23}.  Some work also studied electrotoroidal~\cite{dub86,
hli16} and magnetotoroidal~\cite{gor94a} multipoles, though their
physical significance has remained unclear \cite{fer17}.  We clarify
how toroidal multipole densities emerge in crystals and manifest in
physical properties.

The treatment of multipole densities as strictly macroscopic
quantities distinguishes our theory from other work~\cite{suz17,
wat18, yat21, kus23}.  Group theory expanding on Neumann's principle
provides a precise formalism for discussing electric and magnetic
multipole order in solids and avoids all difficulties that have been
identified with previous extrapolations of classical electromagnetism
from finite systems to crystalline solids~\cite{mar74, kin93, res94,
res10, hir97}.  We have recently applied this method to obtain a
systematic, universally applicable absolute, gauge-invariant
\emph{quantification} of electric and magnetic multipole densities
in crystals~\cite{win25a}.
These multipole densities represent natural order parameters for
electric and magnetic phase transitions in crystals \cite{rei98, lan5e}.
Our approach is also complementary to numerical procedures proposed
to extract bulk polarizations from first-principles calculations of
atomic-site multipoles~\cite{spa13, suz18}.
Previous work assumed that multipole order beyond dipolar order
$\ell=1$ is an exotic, rare phenomenon \cite{hit14, hit16, suz18}.
We show that multipolar order $\ell > 1$ is actually common among
crystalline materials.

In the standard model, both the five categories of polarizations and
the five categories of chirality classify the electric and magnetic
order in crystal structures based on the presence or absence of the
inversion symmetries $i$, $\theta$ and $i\theta$, but from rather
different perspectives.

The five categories of polarizations reflect the presence or absence
of the inversion symmetries $i$, $\theta$ and $i\theta$ as
\emph{independent} elements in the symmetry group $G$ of a physical
system \cite{win23}.  See the first two rows in
Fig.~\ref{fig:standardmodel}.  In the \emph{parapolar} category, all
three inversion symmetries $i$, $\theta$ and $i\theta$ represent
good symmetries, so only even\=/$\ell$ electric multipole
densities called \emph{parapolarizations} are allowed, see
Table~\ref{tab:tensor-signat}.  These even\=/$\ell$ multipole
densities are also allowed in all other categories.  For the three
\emph{unipolar} categories, only one among the inversions $i$,
$\theta$ or $i\theta$ represents a good symmetry.  In the
\emph{electropolar} category, TIS is a good symmetry that permits
odd\=/$\ell$ electric multipole densities called
\emph{electropolarizations}.  In the \emph{magnetopolar} category,
SIS is a good symmetry that permits odd\=/$\ell$ magnetic multipole
densities called \emph{magnetopolarizations}.  In the
\emph{antimagnetopolar} category, CIS is a good symmetry that
permits even\=/$\ell$ magnetic multipole densities called
\emph{antimagnetopolarizations}.  Finally, in the \emph{multipolar}
category, all three symmetries $i$, $\theta$ and $i\theta$ are
broken, so all four types of polarizations may coexist.

\begin{figure*}[t]
  \centering
  \includegraphics[width=1.0\linewidth]{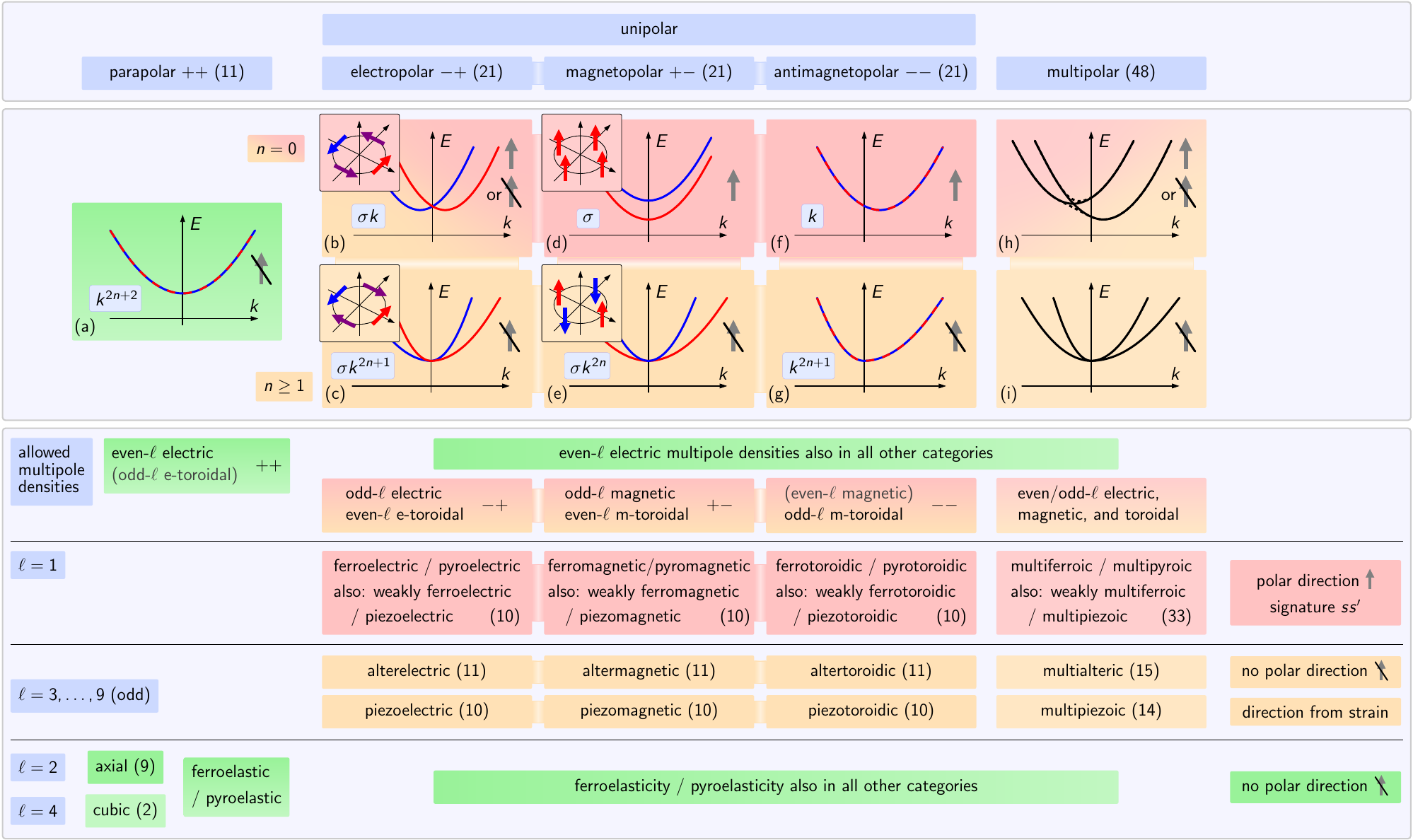}
  \caption[]{\label{fig:polar} Categories of polarizations.  The
  upper part with panels (a)-(i) shows characteristic electron-band
  dispersions $E_\sigma(\kk)$.  The symbol $k$ represents a generic
  wave-vector component, red and blue curves are associated with the
  two values of the spin quantum number $\sigma$, and $n$ is a
  non-negative integer.  Panels (b), (d), (f), and (h) visualize the case
  $n=0$ when a polar direction with signature $ss'$ (i.e., a
  dipolarization $T_1^{ss'}$) is symmetry allowed, as indicated by
  the gray arrows.  Panels (c), (e), (g), and (i) correspond to
  $n\ge 1$, when polar directions are forbidden (strikethrough gray
  arrows).  Realizations of the electropolar and multipolar cases
  with $n=0$ [panels (b) and (h)] without a symmetry-allowed polar
  direction are also possible.  In the parapolar case [panel
  (a)], polar directions are always forbidden (for any $n \ge 0$).
  The lower part summarizes how magnetic crystal classes belonging
  to different categories of polarizations permit different electric
  and magnetic multipole densities (rank $\ell$) that manifest
  themselves via characteristic physical properties.  The features
  listed in the lower part are directly linked with the
  characteristic electron-band dispersions shown in the upper part,
  see text.  Numbers in braces indicate the number of crystal
  classes that realize each case.  More details on all 122 crystal
  classes are provided in Table~\ref{tab:groups-multipoles}.}
\end{figure*}

Each category of polarizations gives rise to distinctive features in the
electronic band structure \cite{win23}, as illustrated in the upper part of
Fig.~\ref{fig:polar}.  The lower part in the figure summarizes materials
classes belonging to the five categories of polarizations.  The $\ell=1$
multipole density within each category corresponds to a ferroic order
parameter exhibiting a polar direction, while multipole densities with
rank $\ell>1$ have no such directionality.  All these cases can be clearly
distinguished in the band structure and they are characterized by
distinct observable (macroscopic) physical effects.

\begin{figure*}[t]
  \centering
  \includegraphics[width=1.0\linewidth]{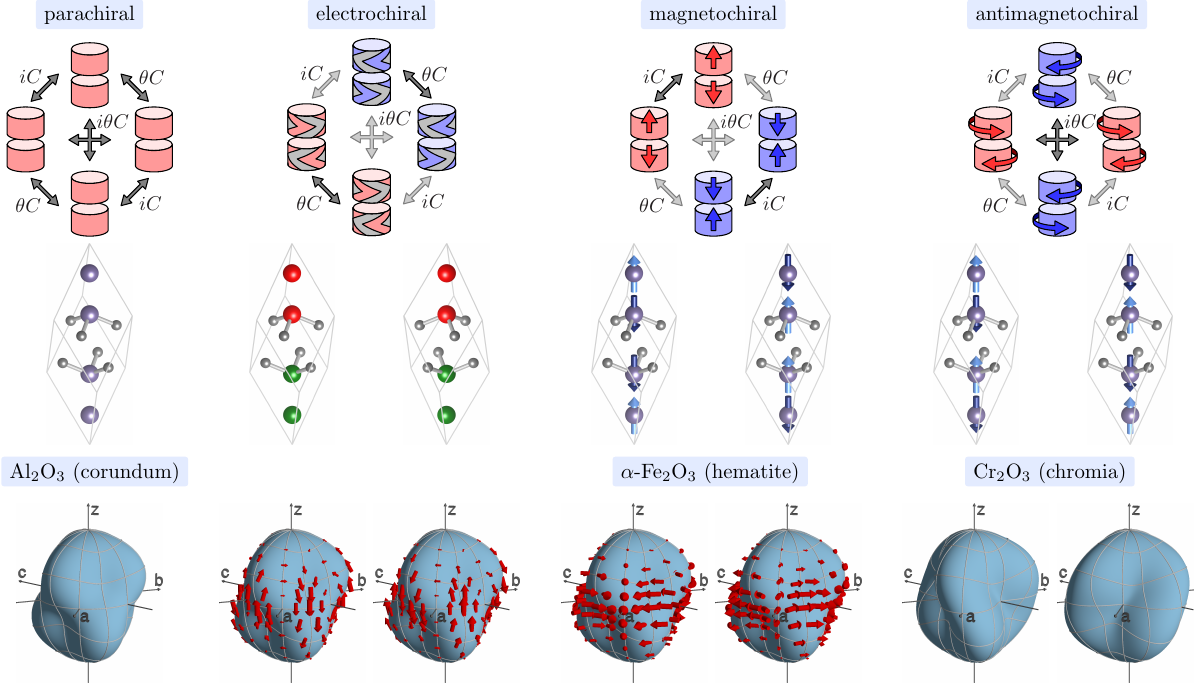}
  \caption[]{\label{fig:unichiral} Examples of parachiral and unichiral
  systems.  Straight (curved) colored arrows in the upper-row panels indicate
  translational (rotational) motion of the shown object.  The middle-row panels
  display the parachiral corundum structure and its unichiral derivatives.
  The bottom-row panels show energy surfaces with associated spin textures
  for crystal structures from the middle row.  The two enantiomorphs
  shown for each unichiral case are transformed into each other by two
  generalized improper rotations (indicated in the upper row by light-gray
  double arrows) but are invariant under the third (dark-gray double arrow).}
\end{figure*}

Very generally, one speaks of chirality when a system exists in, at
least, two versions called \emph{enantiomorphs} that cannot be
superposed upon each other by proper (pure) rotations
$C$~\cite{mis99, fla03, bar04, wag07, fec22, kis22, bou25, rom25}.
Ordinarily, this is tantamount to the case in which the enantiomorphs
are mapped onto each other by improper rotations $iC$.  By treating
the inversion symmetries $i$, $\theta$, and $i\theta$ on the same
footing, we identify five categories of chirality that reflect the
presence or absence in $G$ of any \emph{composite} symmetry elements
$\gamma C$ with $\gamma \in \{ i, \theta, i\theta\}$, i.e., the presence
or absence of $\gamma$\=/improper rotations.  See the first two
columns in Fig.~\ref{fig:standardmodel}.  Systems whose symmetries
include $i$\=/improper and $\theta$\=/improper (and thus also
$i\theta$\=/improper) rotations are \emph{parachiral} and show no
enantiomorphism.  See the left column in Fig.~\ref{fig:unichiral}
for an illustration.  In contrast, each of the three
\emph{unichiral} categories includes $\gamma$\=/improper rotations
for only one inversion symmetry $\gamma=i$, $\theta$ or $i\theta$.
Systems with only $\theta$\=/improper rotations among its symmetries
are \emph{electrochiral} and exist in two enantiomorphic versions
that are transformed into each other by $i$\=/improper and
$i\theta$\=/improper rotations.  In addition to this conventional
case, we identify the categories of \emph{magnetochiral} systems
(having only $i$\=/improper rotations) and \emph{antimagnetochiral}
systems (having only $i\theta$\=/improper rotations).  Each of these
categories permits two enantiomorphs.  The two enantiomorphs of
magnetochiral systems are superposed by $\theta$\=/improper and $i
\theta$\=/improper rotations, whereas $\theta$\=/improper and
$i$\=/improper rotations transform between the two enantiomorphs of
antimagnetochiral systems.  In addition, systems whose only
symmetries are proper rotations (none of them combined with $i$,
$\theta$ or $i\theta$) are \emph{multichiral} and exist in four
distinct enantiomorphs.

Examples of chiral media based on variations of the corundum structure
of Al$_2$O$_3$ are shown in Fig.~\ref{fig:unichiral}.  The new forms
of chirality identified here require magnetic order; they are
realized in magnetic crystals.  More specifically, we show that it
is the \emph{interplay} of different electric and magnetic multipole
densities that gives rise to the different categories of chirality.
Very generally, the enantiomorphism realized in unichiral and
multichiral media can be identified via observable effects such as
optical activity \cite{lan8e} and nonreciprocal directional
dichroism \cite{bro63} which have opposite signs for the different
enantiomorphs.  Multichiral systems require multiple such probes to
distinguish all four enantiomorphs.  The enantiomorphism is also
expressed in the electronic band structure, as demonstrated in the
bottom row of Fig.~\ref{fig:unichiral}.

Studying the morphology of the 122 magnetic point groups has been
essential for gaining a basic understanding of general materials
properties~\cite{tho65, sch73, ope74, kop76, kop79, kop06, gri91,
gri94, lit94, sch08}.  Our systematic approach reveals fundamental
patterns in the structure of magnetic point groups and the
composition of Laue classes assembled by them.  We identify 12
group types, where each type is characterized by a distinct
combination of a category of polarizations and a category of
chirality that act jointly like a unique identifier for each group
type, see Fig.~\ref{fig:standardmodel}.  These group types provide a
comprehensive, fine-grained classification of all 122 magnetic
crystallographic point groups and of the physical properties the
groups imply.  Figure~\ref{fig:standardmodel} also indicates how
many Laue classes contain groups of each type and how many
crystallographic magnetic point groups in total realize each type.
Quantification of multipole densities \cite{win25a} gives
practical measures of polarizations and chiralities that will be
studied in a future publication.

Group theory provides yet deeper insights.  For example, the group
morphology becomes remarkably similar for each of the three unipolar
categories (electropolar, magnetopolar, and antimagnetopolar) and
for each of the three unichiral categories (electrochiral,
magnetochiral, and antimagnetochiral) which reflects the
isomorphisms relating the two-element groups $\gsis$, $\gtis$, and
$\gstis$, see Table~\ref{tab:inversion-group}.  The groups in these
categories can thus be combined in \emph{triadic relations} that
imply striking correspondences between apparently dissimilar systems
and their physical properties.  We illustrate this point by
tabulating the form of material tensors for all magnetic point
groups.  Triadic sets of group types are highlighted in
Fig.~\ref{fig:standardmodel} by conjoint tiles.

The standard model simplifies and greatly extends earlier
studies~\cite{sch73, ope74, kop76, kop79, kop06, gri91, gri94,
lit94, sch08, sme22, sme22a}.  It has great predictive power by
relating electric and magnetic multipolar order with, e.g., distinctive
features in the crystal structure, the electronic band structure,
property tensors and response tensors of quantum materials.  It
elucidates hidden orders in solids~\cite{aep20}; the exhaustive
classification covers the full spectrum of conventional and
unconventional electric and magnetic order, including
altermagnetism~\cite{sme22, sme22a} and piezomagnetism~\cite{ikh22}
which are currently attracting great interest.  The standard model is
equally applicable to insulators and metals.  It is a strong base to
inspire the design of new experimental probes and
techniques~\cite{ye24}.

The remainder of this article is organized as follows.
Section~\ref{sec:multipoles} focuses on the description of electric and
magnetic multipole order in solids.  Here we explain the emergence of
the five categories of polarizations and their associated physical
properties.
Our theory of chirality in solids is developed in Sec.~\ref{sec:chirality}.
We discuss our results and their physical implications, as well
as future research directions building on them, in Sec.~\ref{sec:outlook}
alongside our conclusions.
Technical details are provided in Appendices that establish the
mathematical backbone for the standard-model classification
illustrated in Fig.~\ref{fig:standardmodel}.
Appendix~\ref{app:comp-groups} elucidates the general composition of
magnetic point groups.
The classification of continuous and discrete magnetic point groups
is discussed in Appendix~\ref{app:class-groups}.
Appendix~\ref{app:compound-multipoles} outlines the theoretical
basis for defining compound multipoles.
Appendix~\ref{app:spherTensTab} includes a tabulation of spherical
tensors and multipole order permitted by the 122 crystallographic
magnetic point groups.
The commonalities regarding the shape of material tensors implied by
symmetry is elucidated in Appendix~\ref{app:tensors}.
Finally, Appendix~\ref{app:chiral-materials} tabulates materials
candidates for the different categories of chirality introduced in
Sec.~\ref{sec:chirality}.

Throughout this work, we adhere to the Sch\"onflies notation to
denote symmetry elements and symmetry groups~\cite{lan3e}.  We use
the concept of \emph{Laue classes} \cite{bur13} as a basic
organizational principle for magnetic point groups: A Laue class $L
(\gprop)$ combines the point groups $G$ that represent the same set
of proper (pure) rotations $C$, when the inversion symmetries
$\gamma = i$, $\theta$, and $i\theta$ are stripped off from improper
rotations $\gamma C$ in $G$.  The group $\gprop$ formed by the
proper rotations $C$ is called the \emph{class root} of $L
(\gprop)$.  A thorough introduction to the structure of magnetic
point groups and their Laue classes is provided in
Appendix~\ref{app:comp-groups}.

\section{Multipole order in solids}
\label{sec:multipoles}

\emph{Tensors} denote quantities that transform in a well-defined
way under the symmetry group $G$ of a physical system.  Commonly,
discussions of tensors focus on the rotational symmetries of these
quantities \cite{edm60, tun85}.  In this work, we are
particularly interested in the effect of the inversion symmetries
$i$, $\theta$, and $i\theta$.  In Sec.~\ref{sec:spherTens}, we thus
extend the classification of tensors to also account for inversion
symmetries.

Very generally, physical quantities may be nonzero in a system with
symmetry group $G$ if these quantities are \emph{invariant} under
the symmetry elements in $G$.  According to Neumann's principle, the
pattern of invariant tensors characterizing the macroscopic
properties of a crystal structure is determined by the
crystallographic point group $G$ defining the crystal class of the
crystal structure \cite{voi10, nye85, bir74}.  An important example
of such tensors are electric and magnetic multipole densities that
represent electromagnetic order in solids
(Sec.~\ref{sec:categories-multipole}).  By focusing on how these
multipoles transform under the inversion symmetries $i$, $\theta$,
and $i\theta$, we identify five categories of multipole order, as
illustrated in Fig.~\ref{fig:standardmodel}.

To describe the full range of possible electric and magnetic orders
that can arise in crystals, the formation of compound tensors needs
to be discussed (Sec.~\ref{sec:compound}).  The construction of
irreducible tensors representing multipole densities is explained in
Sec.~\ref{sec:tens:mult-dens}.
In Sec.~\ref{sec:quant-mpol}, we briefly review the theory developed
in Ref.~\cite{win25a} for an absolute quantification of electric and
magnetic multipole densities in crystals.

In a microscopic analysis, the concept of invariant tensors also
allows one to classify the terms that may exist in the Hamiltonian
of a physical system and to discuss under what circumstances these
terms may be nonzero.  In the context of crystalline solids, this is
known as the theory of invariants \cite{bir74}.  It allows one to
derive, in a systematic way, the different terms that arise in a
Taylor expansion of the spin-dependent electronic band structure.
These terms can be classified according to the same categories in
Fig.~\ref{fig:standardmodel} so that we establish a one-to-one
correspondence between the categories of multipole order and
distinctive features of the band structure (Fig.~\ref{fig:polar}
and Sec.~\ref{sec:polar:bands}).

Insights gained from the standard model are used in
Sec.~\ref{sec:alter} to relate the band-structure features
elucidated in Sec.~\ref{sec:polar:bands} to specific materials
classes.  We introduce the term \emph{alteric} as a unifying concept
referring to multipole order with rank $\ell>1$.  For magnetopolar
systems, this term refers to the celebrated altermagnets~\cite{sme22a}, but
alterelectric and altertoroidic media also exist.  We discuss
weak ferroicity in Sec.~\ref{sec:weak-ferro}.  We briefly comment on
the general mechanisms for TIS breaking in magnetic systems and
associated physical phenomena in Sec.~\ref{sec:tis-soc}.

\subsection{Irreducible tensors}
\label{sec:spherTens}

An \emph{irreducible tensor} $T_\Gamma$ associated with a group $G$
represents a physical quantity that transforms according to the
irreducible representation (IR) $\Gamma$ of $G$.  If $G$ is a
crystallographic point group, we call $T_\Gamma$ an
\emph{irreducible crystallographic tensor}.  An \emph{irreducible
spherical tensor}
\begin{equation}
  \label{eq:tensor:def}
  T_\ell^{ss'} \equiv
  \{ T_{\ell m}^{ss'} : m = -\ell, -\ell + 1, \dots, \ell-1, \ell \}
\end{equation}
represents a physical quantity that transforms according to the IR
$D_\ell^{ss'}$ of the full rotation group $\Rit \equiv R \times
\gsistis$ with $R \equiv SO(3)$~\cite{edm60, tun85}.  (Detailed
information about $\Rit$ is provided in
Appendix~\ref{app:spher-groups}.)  Here $\ell = 0, 1, 2, \dots$
denotes the tensor's rank, and $m$ distinguishes its $2\ell + 1$
independent components.  The \emph{signature}
$ss'$~\cite{misc:sigtens} indicates how a tensor transforms under
SIS (even/odd if $s = +/-$) and TIS (even/odd if $s' = +/-$).

Spherical tensors $T_\ell^{ss'}$ transforming irreducibly under
$\Rit$ are generally reducible under crystallographic point groups
$G$, i.e., such tensors are not adapted to discuss crystal
properties.  Tensors $T_\ell^{ss'}$ can be decomposed into
crystallographic tensors $T_\Gamma$ transforming irreducibly under
subgroups $G$ of $\Rit$ by projecting the tensor $T_\ell^{ss'}$ onto
the IRs $\Gamma$ of $G$.  This decomposition follows the
compatibility relations tabulated in Ref.~\cite{kos63}, as discussed
in Ref.~\cite{win23}.  Equivalently, a crystallographic tensor
$T_\Gamma$ transforming according to an IR $\Gamma$ of a subgroup
$G$ of $\Rit$ can be projected onto the IRs $D_\ell^{ss'}$ of $\Rit$
\cite{lud96}.  The latter approach of associating a crystallographic
tensor $T_\Gamma$ with an IR $D_\ell^{ss'}$ of $\Rit$ can be viewed
as the inverse of using compatibility relations to decompose a
spherical tensor $T_\ell^{ss'}$ into components $T_\Gamma$ that
transform irreducibly under a crystallographic point group $G$.
While the compatibility relations connect infinitely many IRs
$D_\ell^{ss'}$ of $\Rit$ with a finite number of IRs $\Gamma$ of
each crystallographic point group, the correspondence between
spherical tensors $T_\ell^{ss'}$ and crystallographic tensors
$T_\Gamma$ can be made unique.  Crystallographic tensors $T_\Gamma$
transforming according to an IR $\Gamma$ of $G$ can be rearranged
into linear combinations $\tilde{T}_\Gamma$, each again transforming
according to the IR $\Gamma$ of $G$, such that under $\Rit$ each
$\tilde{T}_\Gamma$ transforms according to exactly one IR
$D_\ell^{ss'}$ of $\Rit$.  The preceding analysis clarifies the
relation between the spherical tensors frequently used in other work
to discuss crystal properties and symmetry-adapted crystallographic
tensors.  Examples will be discussed below.

A physical quantity $Q$ is allowed under a symmetry group $G$ if $Q$
is \emph{invariant} under $G$, i.e., $Q$ transforms according to the
identity representation $\Gamma_1$ of $G$.  Note that, according to
Neumann's principle~\cite{voi10, nye85}, the symmetry group relevant
for macroscopic properties such as material tensors is the
crystallographic point group $G$ defining the crystal class of a
crystal structure, i.e., space groups are not needed to discuss such
macroscopic properties.  We call the largest symmetry group that
leaves $Q$ invariant the \emph{maximal group} of $Q$.  According to
Neumann's principle, the maximal group of material tensors is always
a subgroup of $\Rit$.

More specifically, a physical quantity represented by a spherical
tensor $T_\ell^{ss'}$ becomes allowed if the symmetry is reduced
from $\Rit$ to a subgroup $G$ of $\Rit$ such that a linear
combination of components $T_{\ell m}^{ss'}$ of $T_\ell^{ss'}$
transforms according to the identity representation of $G$.  This is
equivalent to the condition that mapping the IR $D_\ell^{ss'}$ of
$\Rit$ onto the IRs of $G$ includes the identity representation
$\Gamma_1$ of $G$.  All continuous and discrete groups discussed in
this work are subgroups of $\Rit$.  Therefore, we use the
complete set of irreducible tensors $T_\ell^{ss'}$ of $\Rit$ as a
convenient reference to discuss functions transforming according to
IRs of subgroups of $\Rit$.  Note, however, that any linear
combination of tensors transforming according to the same IR
$\Gamma$ of $G$ yields a tensor that likewise transforms according to
the IR $\Gamma$ of $G$.  In that sense, it is generally not
possible, given a system with crystallographic point group $G$, to
associate distinct observable physics with crystallographic tensors
transforming according to the same IR of $G$ while transforming
according to different IRs of $\Rit$.

Irreducible rank\=/$0$ tensors $T_0^{ss'} = \{ T_{00}^{ss'} \}$ of
$\Rit$ are called \emph{scalars}.  Tensors $T_0^{++}$ are already
invariant under $\Rit$ and thus also invariant under all subgroups
of $\Rit$.  The maximal groups of the remaining tensors $T_0^{ss'}$
are the spherical subgroups $G_\mathrm{sp}$ with $R \subsetneq
G_\mathrm{sp} \subsetneq \Rit$ for which $T_0^{ss'}$ is odd under
those symmetry elements of $\Rit$ that are not contained in
$G_\mathrm{sp}$ (i.e., the tensors $T_0^{ss'}$ are odd under the
symmetry elements labeled ``$\circ$'' in Table~\ref{tab:spher:trafo}
of Appendix~\ref{app:spher-groups}).  Examples of
physical quantities transforming like tensors $T_0^{ss'}$ include
the four types of charges (electric, magnetic, electrotoroidal, and
magnetotoroidal).

Irreducible rank\=/$1$ tensors $T_1^{ss'}$ of $\Rit$ represent
\emph{spherical vectors}, they are equivalent to Cartesian vectors.
The component $m=0$ of rank\=/1 tensors $T_1^{ss'}$, and more
generally the component $m=0$ of any rank\=/$\ell$ tensor
$T_\ell^{ss'}$ with $\ell > 0$, becomes allowed under the continuous
cyclic groups $G_{c\infty}$ and the continuous dihedral groups
$G_{d\infty}$.  These groups, including their IRs and basis functions,
are discussed in more detail in Appendix~\ref{app:axial-groups}.
The maximal groups of tensor components $T_{\ell 0}^{ss'}$ are
generally the dihedral groups $G_{d\infty}$.  More specifically, the
maximal groups for the $m=0$ tensor components depend on the parity
of the index $\ell$ \cite{misc:parity}, i.e., we must distinguish
between $\ell$ even (``$g$'') and $\ell$ odd (``$u$'').  Beyond
that, the index $\ell$ is not relevant, i.e., for any even or odd
$\ell \ge 0$, tensor components $T_{\ell 0}^{ss'}$ transform the
same under the continuous axial groups, and for any $\ell > 0$ they
have the same maximal axial groups.  The cyclic groups $G_{c\infty}$
are subgroups of the dihedral groups $G_{d\infty}$.  Therefore, the
tensor components $T_{\ell 0}^{ss'}$ are likewise invariant under
the cyclic groups $G_{c\infty}$; but the distinction based on the
parity of $\ell$ is lost.

Interestingly, the maximal groups of tensor components
$T_{u0}^{-+}$, $T_{u0}^{+-}$, and $T_{u0}^{--}$ are the dihedral
groups $D_\infty [C_\infty] \times \gtis$, $D_\infty (C_\infty)
\times \gsis$, and $D_\infty (C_\infty) \times \gstis$,
respectively.  However, a tensor component $T_{u0}^{++}$ is not
invariant under any of the continuous dihedral groups.  It is only
invariant under continuous cyclic groups $G_{c\infty}$, including
$C_\infty \times \gsistis$.  This is shown explicitly in
Table~\ref{tab:c-dihedral:trafo} in Appendix~\ref{app:axial-groups}.

\begin{table*}
  \caption{\label{tab:spher:phys-example} Physical quantities
  transforming like irreducible tensor components $T_{\ell 0}^{ss'}$
  of $\Rit$.  The symbol $\vekc{E}$ ($\vekc{B}$) denotes a
  homogeneous electric (magnetic) field, and $\vek{v}$ is a velocity
  field.  An electric dipolarization $\vekc{P}$ (magnetization
  $\vekc{M}$) transforms in all respects like an electric field
  $\vekc{E}$ (magnetic field $\vekc{B}$).  The product $\star$ is a
  generic placeholder for mixed terms regarding the respective
  quantities.}
  \renewcommand{\arraystretch}{1.2}%
  \begin{tabular*}{\linewidth}{nCEl*{4}{C}Les{0.2em}Ln}
    \hline \hline \rule{0pt}{2.7ex}
    & &
    (\vekc{E}, \vekc{P}), (\vekc{B}, \vekc{M}) &
    (\vekc{E}, \vekc{P}) \star (\vekc{B}, \vekc{M}) &
    \vek{v} &
    \vek{v} \star (\vekc{E}, \vekc{B}, \vekc{P}, \vekc{M}) &
    \multicolumn{2}{c}{maximal group} \\ \hline \rule{0pt}{2.7ex}
    T_{00}^{++} & electric &
    \vekc{E} \cdot \vekc{P}, \; \vekc{B} \cdot \vekc{M} & & & &
    R \times \gsistis & = \Rit \\
    T_{00}^{-+} & electrotoroidal & & & &
    \vek{v} \cdot \vekc{B}, \: \vek{v} \cdot \vekc{M} &
    R \times \gtis & = R_\theta \\
    T_{00}^{+-} & magnetotoroidal & & & &
    \vek{v} \cdot \vekc{E}, \: \vek{v} \cdot \vekc{P} &
    R \times \gsis & = R_i \\
    T_{00}^{--} & magnetic & &
    \vekc{E} \cdot \vekc{B}, \; \vekc{P} \cdot \vekc{M} & & &
    R \times \gstis & = R_{i \theta} \\ \hline \rule{0pt}{2.7ex}
    T_{u0}^{++} & electrotoroidal & & & & &
    C_\infty \times \gsistis & = C_{\infty h} \times \gtis \\
    T_{u0}^{-+} & electric &
    \vekc{E}, \vekc{P} & & &
    \vek{v} \times \vekc{B}, \vek{v} \times \vekc{M} &
    D_\infty [C_{\infty}] \times \gtis & = C_{\infty v} \times \gtis \\
    T_{u0}^{+-} & magnetic &
    \vekc{B}, \vekc{M} & & &
    \vek{v} \times \vekc{E}, \vek{v} \times \vekc{P} &
    D_\infty (C_\infty) \times \gsis & = D_{\infty h} (C_{\infty h}) \\
    T_{u0}^{--} & magnetotoroidal & &
    \vekc{E} \times \vekc{B} &
    \vek{v} & &
    D_\infty (C_\infty) \times \gstis & = D_{\infty h} (C_{\infty v})
    \\ \hline \hline
  \end{tabular*}
\end{table*}

Examples of physical quantities transforming like tensor components
$T_{u0}^{ss'}$ are given in Table~\ref{tab:spher:phys-example}.  For
instance, $D_\infty [C_\infty] \times \gtis = C_{\infty v} \times
\gtis$ represents the maximal group of a physical system in the
presence of a homogeneous electric field $\vekc{E}$.  Similarly, the
dual group $D_\infty (C_\infty) \times \gsis = D_{\infty h}
(C_{\infty h})$ is the maximal group of a physical system in the
presence of a homogeneous magnetic field $\vekc{B}$.  Scalar
products of spherical vectors yield scalars \cite{edm60}; see
examples in Table~\ref{tab:spher:phys-example}.  Note that the
maximal group of a tensor component $T_{\ell m}^{ss'}$ may be larger
than the symmetry group permitting a physical realization of
$T_{\ell m}^{ss'}$.  For example, a scalar $\vekc{B}^2$ is clearly
even under TIS.  Nonetheless, it will arise only if a magnetic field
$\vekc{B}$ has broken TIS.

\subsection{Categories of multipole order in crystals}
\label{sec:categories-multipole}

Beyond Table~\ref{tab:spher:phys-example}, examples of spherical
tensors include the electric and magnetic multipoles~\cite{jac99,
raa05} as well as their toroidal complements~\cite{dub90, nan16},
where the tensor rank $\ell$ represents the multipole order.
Specifically, $\ell =0$ corresponds to the monopole, $\ell =1$ to
the dipole, $\ell =2$ to the quadrupole, etc.  Therefore, spherical
tensors are directly related to multipole order in solids.  The
distinct behavior of different multipoles under inversion symmetries
is summarized in Table~\ref{tab:tensor-signat} which gives the
signatures $ss'$ for electric, magnetic and toroidal multipoles.
Here, we follow the common convention that electric charges have the
signature $++$, which implies that magnetic charges have the
opposite signature~$--$~\cite{jac99}.  The transformational behavior
of electric, magnetic, and toroidal multipoles is given in
Table~\ref{tab:categories}.

\begin{table*}[t]
  \caption{\label{tab:categories} Categories of polarizations and
  chirality in crystalline solids.  Based on the decomposition
  (\ref{eq:decompose}) of magnetic point groups in terms of the
  five inversion groups $C_\gin$, categories of polarizations
  characterizing multipole order are defined in the upper part of
  the table.  Symmetry operations present (absent) in a given
  inversion group are indicated by ``$\bullet$'' (``$\circ$'').
  Electric, magnetic, and toroidal multipole densities with
  even/odd order $\ell$ and given signature $ss'$ that are allowed
  (forbidden) under an inversion group are also labeled by
  ``$\bullet$'' (``$\circ$'').
  The lower part of the table defines categories of chirality.  Here
  ``$\bullet$'' (``$\circ$'') indicates that symmetry elements
  $\gamma C$ are present (absent) in a point group belonging to a category.
  The number $n$ of distinct system realizations (enantiomorphs) for
  each chirality is given, and the scalars $T_0^{ss'}$ that are allowed
  (forbidden) under a chirality are indicated by ``$\bullet$''
  (``$\circ$'').}
  \newcommand{\mb}[1]{\makebox[1em][c]{#1}}%
  \newcommand{\mbm}[1]{\makebox[1em][c]{$#1$}}%
  \newcommand{\boxb}[1]{{\renewcommand{\arraystretch}{0.95}%
  \begin{tabular}[b]{ncn}#1\end{tabular}}}%
  \renewcommand{\arraystretch}{1.2}%
  \newcommand{\md}[1]{\multicolumn{4}{c}{\makebox[0pt]{#1}}}%
  \begin{tabular*}{\linewidth}{nlEcs{0.6em}*{4}{C}C*{4}{s{1.0em}Cs{0.4em}C}s{1em}n}
    \hline \hline \rule{0pt}{3.7ex}%
    & & & & & & &
    \multicolumn{2}{cs{1.2em}}{\mb{electric}} &
    \multicolumn{2}{cs{1.2em}}{\mb{magnetic}} &
    \multicolumn{2}{cs{1.2em}}{\mb{\tbox[1ex]{ncn}{electro-\\ toroidal}}} &
    \multicolumn{2}{cs{1.2em}}{\mb{\tbox[1ex]{ncn}{magneto-\\ toroidal}}} \\
    Polarization & &
    \mbm{e} & \mbm{i} & \mbm{\theta} & \mbm{i\theta} &
    C_\gin & \rule{0pt}{4.8ex}%
    \mb{\boxb{even\\ $++$}} & \mb{\boxb{odd\\ $-+$}} &
    \mb{\boxb{even\\ $--$}} & \mb{\boxb{odd\\ $+-$}} &
    \mb{\boxb{even\\ $-+$}} & \mb{\boxb{odd\\ $++$}} &
    \mb{\boxb{even\\ $+-$}} & \mb{\boxb{odd\\ $--$}} \\
    \hline \rule{0pt}{2.7ex}%
    parapolar & (PP) & \bullet & \bullet & \bullet & \bullet & \gsistis &
    \bullet & \circ & \circ & \circ &
    \circ & \bullet & \circ & \circ \\
    electropolar & (EP) & \bullet & \circ & \bullet & \circ & \gtis &
    \bullet & \bullet & \circ & \circ &
    \bullet & \bullet & \circ & \circ \\
    magnetopolar & (MP) & \bullet & \bullet & \circ & \circ & \gsis &
    \bullet & \circ & \circ & \bullet &
    \circ & \bullet & \bullet & \circ \\
    antimagnetopolar & (AMP) & \bullet & \circ & \circ & \bullet & \gstis &
    \bullet & \circ & \bullet & \circ &
    \circ & \bullet & \circ & \bullet \\
    multipolar & (MuP) & \bullet & \circ & \circ & \circ & C_1 &
    \bullet & \bullet & \bullet & \bullet &
    \bullet & \bullet & \bullet & \bullet \\
    \hline \hline \rule{0pt}{2.7ex}%
    Chirality & & \mbm{C} & \mbm{iC} & \mbm{\theta C} & \mbm{i\theta C}
     & n &
     T_0^{++} & & T_0^{--} & & T_0^{-+} & & T_0^{+-} &
    \\ \hline \rule{0pt}{2.7ex}%
    parachiral & (PC) & \bullet & \bullet & \bullet & \bullet & 1 &
     \bullet & & \circ & & \circ & & \circ & \\
    electrochiral & (EC) & \bullet & \circ & \bullet & \circ & 2 &
     \bullet & & \circ & & \bullet & & \circ & \\
    magnetochiral & (MC) & \bullet & \bullet & \circ & \circ & 2 &
     \bullet & & \circ & & \circ & & \bullet & \\
    antimagnetochiral & (AMC)& \bullet & \circ & \circ & \bullet & 2 &
     \bullet & & \bullet & & \circ & & \circ & \\
    multichiral & (MuC) & \bullet & \circ & \circ & \circ & 4 &
     \bullet & & \bullet & & \bullet & & \bullet & \\
    \hline \hline
  \end{tabular*}
\end{table*}

The columns of Table~\ref{tab:tensor-signat} define the electric,
magnetic, and toroidal IRs $D_\ell^{ss'}$ of the full rotation group
$\Rit$.  The IRs $D_\ell^{ss'}$ and associated irreducible spherical
tensors $T_\ell^{ss'}$ fall into eight fundamentally distinct
\emph{families} that are defined by three parities associated with
$s$, $s'$, and $\ell$ \cite{misc:parity}.

As discussed above, a physical quantity transforming according to an
IR $D_\ell^{ss'}$ of $\Rit$ becomes permitted by symmetry in a
system with symmetry group $G \subset \Rit$ if mapping
$D_\ell^{ss'}$ onto the IRs of $G$ includes the identity
representation $\Gamma_1$ of $G$.  Following Refs.~\cite{voi10,
nye85, win23} and exploiting the general correspondence discussed
above between irreducible spherical tensors $T_\ell^{ss'}$ under
$\Rit$ and crystallographic tensors $T_\Gamma$ under
crystallographic point groups $G$, we \emph{define} electric,
magnetic, and toroidal order via Table~\ref{tab:tensor-signat},
i.e., the presence of multipole order is characterized via the
condition that mapping $D_\ell^{ss'}$ onto the IRs of $G$ includes
the identity representation $\Gamma_1$ of $G$.  This generalized
definition of multipole order based on the transformation properties
of the rotation group \cite{ros54, edm60, win23} is independent of a
microscopic model for multipole order.  It embraces how electric and
magnetic multipoles are introduced by electrodynamics for finite
systems \cite{jac99, raa05}.  Yet it applies equally to finite and
infinite systems.  Also, it applies equally to insulators and metals.
The quantitative theory of multipole order developed in
Ref.~\cite{win25a} and reviewed in Sec.~\ref{sec:quant-mpol}
conforms with the generalized definition of multipole order.

We note that, in electrostatics, a finite charge distribution $\rho$
inside a sphere may be replaced by a suitable collection of point
electric multipoles located at the center of the sphere; these
multipoles yield the same potential distribution and thus observable
properties outside the sphere as $\rho$ itself \cite{mor53a, low97}.
Similarly, in magnetostatics, a finite equilibrium current
distribution may be replaced by a collection of point magnetic
multipoles.  The point electric and magnetic multipoles omit those
``degrees of freedom'' of the charge or current distribution which
do not give rise to any potential outside the sphere \cite{mor53a,
low97}.  This applies, in particular, to static toroidal moments
that do not contribute to measurable fields $\vekc{E}$ or $\vekc{B}$
outside the sphere (nor do these moments interact with external
fields $\vekc{E}$ or $\vekc{B}$) \cite{dub74, dub90}, which is the
reason why toroidal moments have been associated with ``hidden
order''.  In the following, we will not pursue such considerations
of multipole order based on electrostatics and magnetostatics.  We
find below that electric, magnetic, and toroidal moments arise
routinely, e.g., in the context of the electronic band structure.

In crystalline solids, multipole order is represented by macroscopic
multipole \emph{densities}.  Examples include the electric
dipolarization and the magnetization.  Whereas multipole densities
in crystals defy a simplistic definition using classical electromagnetism
(as explained in Refs.~\cite{mar74, kin93, res94, res10, hir97}, see
also Sec.~\ref{sec:quant-mpol}), group theory provides a precise
formalism for discussing electric and magnetic multipole order in
solids \cite{voi10, nye85, win23, win25a}.  Analogous to the
multipoles in finite systems, the multipole densities transform as
spherical tensors $T_\ell^{ss'}$.  According to Neumann's principle,
the relevant symmetry group governing the appearance of these
multipole densities is the crystallographic point group associated
with the space group of a crystal structure \cite{voi10, nye85}.

The basic classification of electromagnetic multipole densities
permitted by different magnetic point groups derives from the
possibility to decompose each group $G$ as
\begin{equation}
  \label{eq:decompose}
  G = \tilde{G} \times C_\gin \quad ,
\end{equation}
where $C_\gin$ is the inversion group that can be formed from
inversion symmetries $\gamma$ that are contained as group elements
in $G$ (see Table~\ref{tab:inversion-group}), and $\tilde{G}$ is the
subgroup of $G$ that contains none of the inversion symmetries as an
individual group element~\cite{misc:specinv}.  The five different
inversion groups $C_\gin$ that may appear in Eq.\
(\ref{eq:decompose}) define five \emph{categories} of multipole
order \cite{win23}, see the upper part of Table~\ref{tab:categories}
for the association between categories and their corresponding
inversion groups.  Groups $G$ whose decompositions
(\ref{eq:decompose}) contain the same inversion group $C_\gin$
belong to the same category of multipole order.  As shown in
Table~\ref{tab:categories}, the five categories are each
characterized by a certain pattern of allowed and forbidden
electric, magnetic, electrotoroidal and magnetotoroidal multipole
densities.
More specifically, Tables~\ref{tab:tensor-nonmag}
and~\ref{tab:tensor-mag} in Appendix~\ref{app:spherTensTab} list the
lowest order of multipole densities permitted by each
crystallographic magnetic point group~$G$.

The \emph{parapolar} category subsumes crystal structures that have
both SIS and TIS, which therefore permit only even\=/$\ell$ electric
multipole densities~\cite{kos63} (also called
\emph{parapolarizations}~\cite{win23}).  Even\=/$\ell$ electric
multipole densities are also allowed in all other categories
discussed below.  Media that are both paraelectric and paramagnetic
fall into this category.

\emph{Electropolar} crystals have only TIS; thus, odd\=/$\ell$
electric multipole densities
(\emph{electropolarizations}~\cite{win23}) are allowed, in addition.
Pyroelectric and ferroelectric media with their bulk electric
dipolarization ($\ell = 1$)~\cite{voi10, nye85} are familiar
examples from the electropolar category, as are the zincblende
materials with their bulk electric octupolarization ($\ell =
3$)~\cite{win23}.  A macroscopic electropolarization can also emerge
from microscopic magnetic order.  For example, odd-parity magnets
\cite{hel23, yu25} fall into the electropolar category, see
Sec.~\ref{sec:msg}.

Having only SIS, the \emph{magnetopolar} category
permits odd\=/$\ell$ magnetic multipole densities
(\emph{magnetopolarizations}~\cite{win23}) and therefore includes
both ferromagnets ($\ell = 1$)~\cite{tav56, tav58, cor58a} and
altermagnets ($\ell \ge 3$)~\cite{sme22, sme22a, bho24, win23}, see
Sec.~\ref{sec:alter}.  The \emph{antimagnetopolar} category is
characterized by SIS and TIS both being broken, but CIS still being
a good symmetry, which allows even\=/$\ell$ magnetic multipole
densities (\emph{antimagnetopolarizations}~\cite{win23}) to exist.
The $i\theta$\=/symmetric antiferromagnets~\cite{tav58a, wan20,
wat21, fei21} belong to the antimagnetopolar category.

Crystal structures without SIS, TIS and CIS are in the
\emph{multipolar} category.  The low symmetry of these systems
allows electropolarizations, magnetopolarizations, and
antimagnetopolarizations to coexist.  Figure~\ref{fig:polar}
summarizes the five categories of polarizations and the multipole
densities permitted for each category.

A more thorough discussion of the five categories will be given in
Secs.~\ref{sec:polar:bands} to ~\ref{sec:tis-soc}, when we elaborate
on how the categories of multipole order are linked to distinct
features in the electronic band structure and related observable
effects.

\subsection{Compound tensors}
\label{sec:compound}

In physical systems, quantities behaving like higher-rank
irreducible tensors often arise as products or polynomials of
quantities transforming like lower-rank irreducible tensors.  Before
discussing important examples, we briefly review how such
higher-rank tensors can be formed.

Two irreducible tensors $T_{\ell_1}^{s_1s_1'}$ and
$T_{\ell_2}^{s_2s_2'}$ can be combined to form \emph{compound
tensors} \cite{wyb74}.  Such compound tensors are generally
reducible, i.e., they can be decomposed into irreducible compound
tensors.  This conforms to the multiplication rules for the IRs of
$\Rit$.  The product $D_{\ell_1}^{s_1s_1'} \times
D_{\ell_2}^{s_2s_2'}$ of two IRs $D_{\ell_1}^{s_1s_1'}$ and
$D_{\ell_2}^{s_2s_2'}$ of $\Rit$ can be decomposed into IRs
$D_\ell^{(s_1s_2) (s_1's_2')}$.  Given the ranks $\ell_1$ and
$\ell_2$ of $D_{\ell_1}^{s_1s_1'}$ and $D_{\ell_2}^{s_2s_2'}$, the
range of ranks $\ell$ of $D_\ell^{(s_1s_2) (s_1's_2')}$ obeys the
relation \cite{edm60}
\begin{equation}
  \label{eq:l:ang-tot}
  |\ell_1 - \ell_2| \le \ell \le \ell_1 + \ell_2
\end{equation}
and the signature $(s_1s_2) (s_1's_2')$ of the IRs
$D_\ell^{(s_1s_2) (s_1's_2')}$ is the product of the signatures
$s_1s_1'$ and $s_2s_2'$ of the IRs it is derived from.
Similarly, the product of two irreducible spherical tensors
$T_{\ell_1}^{s_1s_1'}$ and $T_{\ell_2}^{s_2s_2'}$ can be decomposed
into irreducible spherical compound tensors
$T_\ell^{(s_1s_2) (s_1's_2')}$ using~\cite{edm60}
\begin{equation}
  \label{eq:tensor-prod}
  T_{\ell m}^{(s_1s_2) (s_1's_2')}
  = \sum_{m_1, m_2} T_{\ell_1 m_1}^{s_1s_1'} T_{\ell_2 m_2}^{s_2s_2'}
  (\ell_1 m_1 , \ell_2 m_2 | \ell_1 \ell_2, \ell m) \, ,
\end{equation}
where $(\ell_1 m_1 , \ell_2 m_2 | \ell_1 \ell_2, \ell m)$ are
Clebsch-Gordan coefficients.  Given the ranks $\ell_1$ and $\ell_2$
of the tensors $T_{\ell_1}^{s_1s_1'}$ and $T_{\ell_2}^{s_2s_2'}$,
the range of ranks $\ell$ is given by Eq.\ (\ref{eq:l:ang-tot}).
Previous works \cite{suz17, wat18, suz19} have considered a subset
of the compound tensors (\ref{eq:tensor-prod}) with $\ell = \ell_1 +
\ell_2$; these compound tensors are also known as \emph{polarized
harmonics}~\cite{ros54, edm60}.  Equation (\ref{eq:tensor-prod})
allows one to derive, in a systematic way, \emph{all} irreducible
compound tensors that may be relevant for a physical system.

In a similar way, using the multiplication rules and Clebsch-Gordan
coefficients for crystallographic point groups $G$ \cite{kos63},
crystallographic tensors $T_\Gamma$ transforming irreducibly
according to IRs $\Gamma$ of $G$ can be combined to form
crystallographic compound tensors transforming irreducibly
under~$G$.

\subsection{Irreducible tensors representing multipole densities}
\label{sec:tens:mult-dens}

In this section, we discuss in more detail irreducible tensors
representing multipole densities.  In \emph{finite} systems,
electric and magnetic multipoles can be formulated in powers of
Cartesian components of position $\rr = (x,y,z)$ \cite{edm60, jac99,
mor53a, low97}, leading to alternating even and odd transformation
behavior under SIS as $\ell$ is increased; see
Table~\ref{tab:tensor-signat}.  In contrast, the behavior under TIS
is the same for all multipoles of a given kind (electric, magnetic,
electrotoroidal or magnetotoroidal) and coincides with the
transformation property of the corresponding charge.  Electric
multipoles in finite systems thus can be written as polynomials in
Cartesian components of position (in the following collectively
denoted $r$), while magnetic multipoles can be expressed using
polynomials in $r$ and components of spin (collectively denoted
$\sigma$) representing magnetic dipole moments \cite{low97}.
Multipole densities in \emph{infinite}, periodic systems can be
expressed as a function of $r, \sigma$ or, equivalently, as a
function of components of reciprocal lattice vectors (collectively
denoted $\gr$) and $\sigma$.  Furthermore, the electronic band structure
can be written as polynomials in components of the wave vector
(collectively denoted $k$) and~$\sigma$ (see
Table~\ref{tab:tensor-pow-spher} and Sec.~\ref{sec:polar:bands}).
Note that position $\rr$ has the signature $-+$, whereas the wave
vector $\kk$ (representing crystal momentum $\hbar\kk$) and
reciprocal lattice vectors $\gv$ have the signature $--$, and (spin)
angular momentum has the signature $+-$.

\begin{table*}
  \caption{\label{tab:tensor-pow-spher} Irreducible spherical
  tensors.  In a simplified notation, we give the powers of
  Cartesian components of position (collectively denoted $r$),
  components of the wave vector (collectively denoted $k$) and
  components of spin (collectively denoted $\sigma$) required for a
  polynomial representation of irreducible spherical tensors
  $T_\ell^{ss'}$ transforming according to the IRs $D_\ell^{ss'}$ of
  $\Rit$.
  The symbol $p$ denotes a non-negative integer.  For the
  even\=/$\ell$ magnetotoroidal IRs $D_\ell^{+-}$, polynomials in
  $r,\sigma$ or $k,\sigma$ can only be formed for $\ell > 0$, but no
  such polynomials exist that transform according to $D_0^{+-}$.  A
  dash indicates that no such polynomials can be constructed for any
  $\ell$.  Terms $2p$ in the exponents reflect the fact that we can
  always multiply a tensor $T_\ell^{ss'}$ with a scalar $r^{2p}$ or
  $k^{2p}$ without changing its transformational properties.}
  \renewcommand{\arraystretch}{1.2}%
  \newcommand{\nothing}{\hspace{0.8em}-}%
  \newcommand{\ptp}{+2p}%
  \let\mc\multicolumn
  \tabcolsep 0pt
  \begin{tabular*}{\linewidth}{lE*{4}{Ce@{\;:\;\;}LE}}
    \hline \hline \rule{0pt}{2.8ex}
    & \mc{2}{c}{electric} & \mc{2}{c}{magnetic}
    & \mc{2}{c}{electrotoroidal}
    & \mc{2}{c}{magnetotoroidal} \\ \hline \rule{0pt}{2.8ex}%
    $\ell$ even
    & ++ & r^{\ell \ptp} & -- & r^{\ell \pm 1 \ptp} \sigma
    & -+ & \nothing & +- & r^{\ell \ptp} \sigma
    \quad\mbox{(only $\ell > 0$)} \\
    $\ell$ odd
    & -+ & r^{\ell \ptp} & +- & r^{\ell \pm 1 \ptp} \sigma
    & ++ & \nothing & -- & r^{\ell \ptp} \sigma \\ \hline \rule{0pt}{2.8ex}%
    $\ell$ even
    & ++ & k^{\ell \ptp} & -- & \nothing
    & -+ & k^{\ell \pm 1 \ptp} \sigma & +- & k^{\ell \ptp} \sigma
    \quad\mbox{(only $\ell > 0$)} \\
    $\ell$ odd
    & -+ & k^{\ell \ptp} \sigma & +- & k^{\ell \pm 1 \ptp} \sigma
    & ++ & \nothing & -- & k^{\ell \ptp}
    \\ \hline \hline
  \end{tabular*}
\end{table*}

\begin{table}
  \caption{\label{tab:tensor-pow-cryst} Irreducible crystallographic
  tensors.  In a simplified notation, we give the powers of
  Cartesian components of position (collectively denoted $r$),
  components of the wave vector (collectively denoted $k$) and
  components of spin (collectively denoted $\sigma$) required for a
  polynomial representation of irreducible crystallographic tensors
  with signature $ss'$.
  The symbol $n$ denotes a non-negative integer.  The first column
  indicates in square brackets the parity of the index $\ell$ of the
  spherical electric and magnetic tensors (as defined in
  Table~\ref{tab:tensor-signat}) that the respective
  crystallographic tensors correspond~to.}
  \renewcommand{\arraystretch}{1.2}%
  \let\mc\multicolumn \tabcolsep 0pt
  \begin{tabular*}{1.0\linewidth}{lE*{2}{Ce@{\;:\;\;}LE}}
    \hline \hline \rule{0pt}{2.8ex}
    & \mc{2}{c}{electric} & \mc{2}{c}{magnetic} \\ \hline \rule{0pt}{2.8ex}%
    {[$\ell$ even]} & ++ & r^{2n+2} & -- & r^{2n+1} \sigma \\
    {[$\ell$ odd]}  & -+ & r^{2n+1} & +- & r^{2n} \sigma \\ \hline \rule{0pt}{2.8ex}%
    {[$\ell$ even]} & ++ & k^{2n+2} & -- & k^{2n+1} \\
    {[$\ell$ odd]}  & -+ & k^{2n+1} \sigma & +- & k^{2n} \sigma \\ \hline \hline
  \end{tabular*}
\end{table}

Equation (\ref{eq:tensor-prod}) implies that electric
and magnetic multipoles can be combined to form \emph{compound
multipoles} that are likewise associated with tensors in
Table~\ref{tab:tensor-pow-spher}.  These compound tensors can act
as mediators for the observability of electric and magnetic multipoles.
When constructing compound multipoles from electric and magnetic
multipoles, we can view even\=/$\ell$ electric multipoles as
\emph{trivial} building blocks because such multipoles exist in any
crystal structure (see Sec.~\ref{sec:categories-multipole}).  On the
other hand, odd\=/$\ell$ electric and all magnetic multipoles are
\emph{nontrivial} as they arise only for certain categories of
polarized matter.  As shown in Appendix~\ref{app:compound-multipoles},
all toroidal tensors
(except $T_1^{++}$ \cite{misc:T1++}) can be obtained as products
(\ref{eq:tensor-prod}) combining one nontrivial and one or two
trivial electromagnetic tensors such that the compound tensor
inherits the signature of the nontrivial tensor.  Of course, similar
to electric and magnetic multipole densities, whether or not
specific compound multipole densities are realized in a specific
crystal structure also depends on the point group $G$ of that
structure.  See Tables \ref{tab:tensor-nonmag} and
\ref{tab:tensor-mag} in Appendix~\ref{app:spherTensTab}.

Explicit expressions for the irreducible spherical tensors in
Table~\ref{tab:tensor-pow-spher} can be derived from the rank\=/1
tensors
\begin{subequations}
  \label{eq:base-dipole}
  \begin{align}
    T_{[r]1}^{-+} & = \bigl[-\frack{1}{\sqrt{2}} (x+iy), z,
    \frack{1}{\sqrt{2}} (x-iy) \bigr] \\
    T_{[k]1}^{--} & = \bigl[-\frack{1}{\sqrt{2}} (k_x+ik_y), k_z,
    \frack{1}{\sqrt{2}} (k_x-ik_y) \bigr] \\
    T_{[\sigma]1}^{+-} & = \bigl[-\frack{1}{\sqrt{2}} (\sigma_x+i\sigma_y),
    \sigma_z, \frack{1}{\sqrt{2}} (\sigma_x-i\sigma_y) \bigr]
    \label{eq:spin-dipole}
  \end{align}
\end{subequations}
using Eq.\ (\ref{eq:tensor-prod}).  However, it turns out that,
using the rank\=/1 tensors $T_{[r]1}^{-+}, T_{[\sigma]1}^{+-}$
(upper part of Table~\ref{tab:tensor-pow-spher}) and $T_{[k]1}^{--},
T_{[\sigma]1}^{+-}$ (lower part of
Table~\ref{tab:tensor-pow-spher}), such tensors can be realized for
only six out of the eight multipole families, as indicated by dashes
in Table~\ref{tab:tensor-pow-spher}.  The missing two families
reflect the fact that the sets of building blocks ($T_{[r]1}^{-+},
T_{[\sigma]1}^{+-}$ and $T_{[k]1}^{--}, T_{[\sigma]1}^{+-}$) are
incomplete.  To fill these gaps in Table~\ref{tab:tensor-pow-spher},
we would need, e.g., (fictitious or engineered) electric dipole
moments (signature $-+$) that complement the magnetic dipole moments
$\sigma$ (signature $+-$) as building blocks \cite{dub86,
misc:T1++}.

Specifically, it is impossible to form irreducible tensors in $(k,
\sigma)$ transforming according to the magnetic IRs $D_\ell^{--}$
with $\ell$ even.  Tensors with signature $--$ can only be realized
via odd powers $k^{2n+1}$ representing magnetotoroidal tensors
transforming according to the IRs $D_\ell^{--}$ with $\ell$ odd.  In
the electronic band structure, the antimagnetopolar category with
signature $--$ thus can only be represented via magnetotoroidal
terms proportional to $k^{2n+1}$.  The remarkable consequences of
this fact are discussed in Sec.~\ref{sec:polar:bands} when we
examine the band structure for the different categories of
polarizations.

Similar to the construction of irreducible spherical tensors using
Eqs.\ (\ref{eq:tensor-prod}) and (\ref{eq:base-dipole}), one can
construct irreducible crystallographic tensors in $r, \sigma$ and
$k, \sigma$ using the Clebsch-Gordan coefficients for the
crystallographic point groups tabulated in Ref.~\cite{kos63}.  This
is known as the theory of invariants \cite{bir74}.  In the present
context, this was discussed in more detail in Ref.~\cite{win23}.
The essential difference between irreducible spherical tensors and
irreducible crystallographic tensors lies in the fact that the index
$\ell$ represents a good quantum number for the full rotation group
$\Rit$, but not for the crystallographic point groups $G$, i.e.,
under groups $G$ the number of parities characterizing the multipole
densities is reduced from three to two.  Therefore, in crystalline
systems, the four toroidal families each must be merged with the
respective electromagnetic family that has the same signature $ss'$.
In crystalline systems, we thus have only four distinct families of
tensors, see Table~\ref{tab:tensor-pow-cryst}.
This result reflects the fact that the electromagnetic and toroidal
tensors are fundamentally distinct only under the full rotation
group $\Rit$.  In a crystalline environment characterized by a
finite crystallographic point group $G$, electromagnetic and
toroidal tensors cannot be differentiated anymore, as only the
signature $ss'$ remains a distinguishing feature of tensors
\cite{win23}.  The generic exponents listed in
Table~\ref{tab:tensor-pow-cryst} for the crystallographic tensors
summarize the different exponents of the corresponding
electromagnetic and toroidal spherical tensors in
Table~\ref{tab:tensor-pow-spher} with the same signature.

In physical terms, toroidal tensors as in
Tables~\ref{tab:tensor-pow-spher} and \ref{tab:tensor-pow-cryst}
arise as compound tensors (\ref{eq:tensor-prod}) representing
products of electric and magnetic tensors; they describe the
combined effect of electric and magnetic multipoles as commonly
studied in quantum-mechanical perturbation theory.  Interestingly,
this contrasts with the equations of classical electromagnetism that
are linear \cite{jac99} so that, within the framework of classical
electromagnetism, a superposition of electric and magnetic
multipoles cannot give rise to such qualitatively new terms.

\subsection{Absolute quantification of electric and magnetic
multipole densities}
\label{sec:quant-mpol}

For crystals, the definition of electric dipolarization as the
dipole moment per unit cell is unsatisfactory, it depends on the
arbitrary choice of the unit cell \cite{mar74, kin93, res94, res10}.
Similar difficulties arise, for example, with orbital magnetic
dipoles \cite{hir97} and higher multipole moments associated with
clusters of atoms \cite{suz17}.  Important progress towards
addressing this issue was made by the modern theories of electric
dipolarization and magnetization that provide gauge-invariant
expressions for the electric dipolarization and magnetization based
on properties of Bloch functions, i.e., independent of individual
atoms or ions constituting a crystal structure~\cite{kin93, res94,
res07, res10, van18}.  However, the modern theories still have a
number of limitations.  Their quantification of the electric
dipolarization works only for insulators~\cite{res02} and only
pertains to polarization \emph{changes} that must be evaluated
relative to a reference state, though thermodynamics requires that a
state variable such as the polarization in a ferroelectric crystal
must have an absolute meaning independent of how a state has been
reached \cite{rei98}.  Most importantly, the extension of the modern
theories to higher-rank multipole densities beyond dipolar order
turns out to be difficult~\cite{ono19, dai20}.

Group theory provides not only a formal definition of multipole
densities in crystals.  But it can also provide an absolute
quantification of electric and magnetic multipole densities in
crystals, as shown in Ref.~\cite{win25a}.  In crystalline solids,
lattice-periodic quantities $q$ can be formulated in real space as
continuous functions $q(\rr)$ or, equivalently, in terms of their
discrete Fourier components $q(\gv)$ in reciprocal space.  Here $q$
may represent, e.g., scalar quantities such as the charge density
$\rho$ or tensorial quantities such as a magnetization density
$\vek{m}$.  In the following, we denote such quantities collectively
as $q(\xx)$, where $\xx$ may denote position $\rr$ or reciprocal
lattice vectors $\gv$.

If $G$ is the symmetry group of the crystal, observable quantities
$q(\xx)$ must transform according to the identity IR $\Gamma_1$ of
$G$.  However, if $G$ is a supergroup of the symmetry group of the
crystal (see below), parts of $q(\xx)$ may transform according to
different IRs $\Gamma_\alpha$ of $G$.  To identify these parts, we
can project $q(\xx)$ onto the IRs $\Gamma_\alpha$ of
$G$~\cite{lud96}
\begin{subequations}
  \label{eq:project:f}
  \begin{equation}
    \label{eq:project:op}
    q_\alpha (\xx) = \Pi_\alpha\, q(\xx)
    \equiv \frac{n_\alpha}{h} \sum_{g \in G} \chi_\alpha(g)^\ast\,
    P_g \, q (\xx) \; ,
  \end{equation}
  such that $q_\alpha (\xx)$ transforms according to
  $\Gamma_\alpha$.  Here $\chi_\alpha (g)$ are the characters of the
  IR $\Gamma_\alpha$ and $P_g$ are the symmetry operators
  corresponding to the group elements $g \in G$, $n_\alpha$ is the
  dimension of $\Gamma_\alpha$, and $h$ is the order of $G$.  The
  projectors $\Pi_\alpha$ can be formulated and evaluated
  independent of a particular choice of origin or choice of a unit
  cell \cite{str71}.  The projection yields the decomposition
  \begin{equation}
    \label{eq:project:decomp}
    q(\xx) = \sum_\alpha q_\alpha (\xx) \; ,
  \end{equation}
\end{subequations}
because the projection operators $\Pi_\alpha$ are orthogonal and complete
\cite{lud96}.  Previously, the projection operators $\Pi_\alpha$
have been used, e.g., in electronic band structure calculations for
the method of symmetrized plane waves \cite{str71}.

The symmetry group $G$ used in Eqs.\ (\ref{eq:project:f}) need not
be the symmetry group of the system.  For example, if $G$ is the
parent point group above the critical temperature $T_c$ of a
ferroelastic or ferroelectric phase transition and $q(\xx)$
represents the charge density $\rho(\xx)$, the projection
(\ref{eq:project:f}) becomes a crystallographic multipole expansion
as introduced in Sec.~\ref{sec:categories-multipole} that allows one
to evaluate quantitatively the crystallographic multipole densities
$\rho_\alpha (\xx)$ arising in such ferroic phase transitions.
Above $T_c$, the entire charge density $\rho(\xx) = \rho_1 (\xx)$
transforms according to the trivial IR $\Gamma_1$ of the parent
point group $G$, whereas below $T_c$ (when the new symmetry group
$U$ is a subgroup of $G$ \cite{lan5e}), some fractions $\rho_\alpha
(\xx)$ of the total density transform according to nontrivial IRs
$\Gamma_\alpha$ of $G$ (say, an electric dipole density in
ferroelectric transitions, and an electric quadrupole density in
ferroelastic transitions).
The quantities $q_\alpha (\xx)$ provide a complete characterization
of the respective multipole density they represent.  We can estimate
the magnitude of $q_\alpha (\xx)$ by evaluating, for example,
\begin{equation}
  \label{eq:mpol-mag}
  Q_\alpha = \sum_{\xx} \; |q_\alpha (\xx)| \; .
\end{equation}
Numerical examples for electric and magnetic multipole densities in
crystals based on Eqs.\ (\ref{eq:project:f}) have been given in
Ref.~\cite{win25a}.

The decomposition (\ref{eq:project:f}) provides a unified,
quantitative theory of electric and magnetic multipole densities in
crystalline media, treating all these quantities on the same
footing.  It is in line with established phenomenological
(group-theory based) studies of electric dipolarization and
magnetization in crystalline media \cite{voi10, nye85, new05} and
also with the modern theory of polarization~\cite{kin93, res94,
res07, res10, van18}.
The decomposition (\ref{eq:project:f}) is valid for insulators,
semimetals, and metals.
The projected multipole densities $q_\alpha (\xx)$ and $Q_\alpha$
are gauge-invariant quantities with absolute values independent of a
reference state.  Therefore, these quantities are well-suited as
intensive state variables for thermodynamic theories \cite{rei98}
including the order parameters in Landau's theory of phase
transitions \cite{lan5e}.
The projection technique lends itself to be incorporated into the
suite of computational materials-science tools, thus complementing
current approaches that focus on the modern theories \cite{res10}
and local multipole configurations~\cite{spa13, suz18, pic24}.

\subsection{Multipole densities and band structure}
\label{sec:polar:bands}

The irreducible crystallographic tensors listed in the bottom part
of Table~\ref{tab:tensor-pow-cryst} arise prominently in a Taylor
expansion of the spin-dependent electronic band structure $E_\sigma
(\kk)$ (obtained systematically via the theory of invariants
\cite{bir74}).  Table~\ref{tab:tensor-pow-cryst} implies that the
spin-dependent electronic band structure $E_\sigma (\kk)$ in crystal
structures is qualitatively different when some or all inversion
symmetries are broken, as compared with the case that these
symmetries are preserved \cite{her37, her37a, bir74}.

More specifically, the spin-dependent band structure $E_\sigma
(\kk)$ directly reflects the presence of multipolar order in
solids~\cite{wat18, hay18c, win23}.  This is illustrated for the
five categories of polarization in Fig.~\ref{fig:polar} that
juxtaposes typical band dispersions $E_\sigma (\kk)$
\cite{misc:theoinv} with the multipole order (tensors
$T_\ell^{ss'}$) allowed in these materials.

Among the tensors $T_\ell^{ss'}$ in a family $ss'$, the tensors
$\ell=1$ stand out; they define for crystal structures a unique
\emph{polar direction} with signature $ss'$ \cite{nye85}.  Familiar
examples include the vector of electric dipolarization in
ferroelectric and pyroelectric media and the magnetization vector in
ferromagnetic and pyromagnetic media.  We illustrate in
Fig.~\ref{fig:polar}(a)-(i) using gray arrows and strikethrough
arrows whether systems possess a polar direction or whether a polar
direction is forbidden.  Polar directions $ss'$ are defined more
accurately in Appendix~\ref{app:axial-groups} via the IRs of the
full axial group $D_\infty \times \gsistis$.

Panels Fig.~\ref{fig:polar}(b), (d), (f), and (h) consider the case
$n=0$, where the integer $n$ is defined in
Table~\ref{tab:tensor-pow-cryst}, whereas panels (c), (e), (g), and
(i) consider $n \ge 1$.  No such distinction is needed for
panel~(a).  We note that rows and columns of Fig.~\ref{fig:polar}
each represent specific distinct subsets of the 122 magnetic crystal
classes such that we obtain a complete classification of the 122
magnetic crystal classes and important physical properties they
entail.  The detailed classification of all 122 crystal classes is
listed in Table~\ref{tab:groups-multipoles}.  See also
Tables~\ref{tab:tensor-nonmag} and~\ref{tab:tensor-mag}.

In the following, we discuss in more detail the electronic band
structure associated with the different categories of polarizations
(upper part of Fig.~\ref{fig:polar}).

(i) In parapolar materials, both SIS and TIS are good symmetries.
According to Table~\ref{tab:tensor-pow-spher}, we have only
even\=/$\ell$ electric tensors $k^{2n+2}$ that can be formed with
signature $++$, but no odd\=/$\ell$ electrotoroidal tensors.
Therefore, parapolar materials have a spin-degenerate dispersion
illustrated in Fig.~\ref{fig:polar}(a) that is symmetric about $\kk
=0$, i.e., we have $E(\kk) = E (-\kk)$.  A polar direction associated
with a tensor $T_1^{++}$ \cite{hli16} cannot arise in the band
structure of parapolar materials.

(ii) The electropolar category is characterized by a spin-split
dispersion $E_\sigma (\kk)$ that does not break TIS.  An example of
a $k$-linear spin splitting depicted in Fig.~\ref{fig:polar}(b)
is the Rashba term proportional to $\kk \times \vek{\sigma}$ that reflects
the electric dipolarization $T_1^{-+}$ in ferroelectric or
pyroelectric media such as the hexagonal wurtzite
structure~\cite{ras59a}.  A $k$-linear spin splitting also arises in
materials that permit an electrotoroidal scalar $T_0^{-+}$ or an
electrotoroidal quadrupole density $T_2^{-+}$, see
Sec.~\ref{sec:chirality} and Appendix~\ref{app:examples-compound}.
The electropolar case with $n=0$ can thus be realized with and
without a polar direction $-+$.

A dispersion of the form sketched in Fig.~\ref{fig:polar}(c) with
$n>0$ occurs in materials permitting odd\=/$\ell$
electropolarizations $T_\ell^{-+}$ with $\ell > 1$ such as the cubic
zincblende structure, where the Dresselhaus term~\cite{dre55}
represents an electric octupolarization $T_3^{-+}$.  Further
examples are discussed in Appendix~\ref{app:examples-compound}.

(iii) Magnetopolar materials break TIS.  They can exhibit a finite
spin-dependent splitting proportional to $\sigma$ around $\kk= \mathbf{0}$
as illustrated in Fig.~\ref{fig:polar}(d).  Note that the spin
matrices $\sigma_j$ jointly transform like a vector $T_1^{+-}$, see
Eq.\ (\ref{eq:base-dipole}).  Therefore, the spin-dependent
splitting of the band structure near $\kk= \mathbf{0}$ due to terms
$\propto\sigma$ reflects not just the signature $+-$ of any
magnetopolar system but also the existence of a polar
direction $+-$ that coincides with the macroscopic magnetization
$T_1^{+-}$; i.e., these systems must be ferromagnetic.  In contrast,
magnetopolar systems that show a $\kk$\=/dependent spin splitting as
in Fig.~\ref{fig:polar}(e), that vanishes for $\kk= \mathbf{0}$, do
not permit a magnetization $T_1^{+-}$.  These systems are
altermagnetic, they permit magnetic multipolarizations $T_\ell^{+-}$
with odd~$\ell$ and $\ell >1$ such as a magnetic octupolarization,
see Sec.~\ref{sec:alter}.  Similar to the electropolar category, the
electronic band structure of magnetopolar systems may also include
terms that transform like magnetotoroidal tensors $T_\ell^{+-}$ with
even~$\ell$.

(iv) Antimagnetopolar materials have a spin-degenerate $\kk$\=/asymmetric
dispersion that reflects the fact that these materials break both
SIS and TIS, but CIS remains a good symmetry.  As discussed in
Sec.~\ref{sec:tens:mult-dens}, the signature $--$ characteristic of
antimagnetopolar order can only be realized via magnetotoroidal
terms proportional to $k^{2n+1}$.  If the system permits a magnetotoroidal
dipolarization $T_1^{--}$, this is reflected in the band structure
by a $k$\=/linear term that dominates the dispersion near $k=0$,
leading to a shifted parabola as depicted in
Fig.~\ref{fig:polar}(f) \cite{gor94a, yan14, hay14, win20,
win23}.  The magnetotoroidal octupolarization $T_3^{--}$ present in a
bulk diamond antiferromagnet is indicated by terms proportional to $k^3$
\cite{win20, win23}, so the dispersion is an example of the one
shown in Fig.~\ref{fig:polar}(g).

The fact that the electronic band structure in antimagnetopolar
crystal structures only permits magnetotoroidal terms $\propto
k^{2n+1}$ has remarkable consequences.  The
antimagnetopolar category is characterized by the presence of
even\=/$\ell$ magnetic multipoles of the form $r^{2n+1}\sigma$, see
Table~\ref{tab:tensor-pow-cryst}.  Such even\=/$\ell$ magnetic
multipoles $T_\ell^{--}$ do not possess a polar direction $--$ for
any even~$\ell$.  In the band structure, magnetotoroidal terms linear
in $k$ represent a polar direction for the antimagnetopolar category
(with signature $--$) in the same way an electric dipole density
defines a polar direction for the electropolar category ($-+$) and a
magnetic dipole density defines a polar direction for the
magnetopolar category ($+-$); i.e., each unipolar category permits a
characteristic polar direction $ss'$.  Among the antimagnetopolar
crystal classes, we have no simple relation between the lowest even
$\ell>0$ permitting $T_\ell^{--} \ne 0$ and the lowest $n$ of
allowed magnetotoroidal terms $k^{2n+1}$.  In
Table~\ref{tab:groups-multipoles} all 122 crystal classes are
ordered according to the lowest-rank multipole densities they
permit.  See also Tables~\ref{tab:tensor-nonmag} and
\ref{tab:tensor-mag}.  As discussed in more detail in
Appendix~\ref{app:compound-multipoles}, magnetotoroidal vectors
arise as compound tensors.

(v) Multipolar materials break all inversion symmetries $i$,
$\theta$, and $i\theta$.  Therefore, the band dispersions do not
possess any symmetries, as shown in Figs.~\ref{fig:polar}(h)
and~\ref{fig:polar}(i).  These materials permit all families of
multipoles.  Similar to the electropolar case with $n=0$
[Fig.~\ref{fig:polar}(b)], the multipolar case with $n=0$ can be
realized with or without a polar direction.  The multipolar case
with $n\ge 1$ [Fig.~\ref{fig:polar}(i)] always excludes the
possibility of having a polar direction~$ss'$.

An electric dipole density $T_1^{-+}$ (ferroelectricity or
pyroelectricity) may exist in 10 electropolar crystal classes and 21
multipolar crystal classes permitting a polar direction with
signature $-+$.  Similarly, a magnetic dipole density $T_1^{+-}$
(ferromagnetism or pyromagnetism) may exist in 10 magnetopolar
crystal classes and 21 multipolar crystal classes.  A
magnetotoroidal dipole density $T_1^{--}$ (ferrotoroidicity or
pyrotoroidicity) may exist in 10 antimagnetopolar crystal classes
and 21 multipolar crystal classes.  The electronic band structure of these
ferroic systems generally corresponds to the case $n=0$ shown in the
upper row [panels (b), (d), (f) and (h)] of Fig.~\ref{fig:polar}.
Note that the higher-order terms $n>0$ depicted in the lower row
[panels (c), (e), (g) and (i)] of Fig.~\ref{fig:polar} require a
higher crystal symmetry than the terms with $n=0$.  Therefore, the
higher-order terms $n>0$ are generally also present in the band
structure of systems realizing the scenarios illustrated in the
upper row.  Weakly ferroic systems (see Sec.~\ref{sec:weak-ferro})
are a subset of the systems permitting, by symmetry, the terms $n=0$
shown in the upper row, though weak ferroicity implies that the band
structure will be dominated by the higher-order terms $n>0$ shown in
the lower row.  For example, the spin-dependent band structure of
weak ferromagnets is dominated by the altermagnetic terms $\sigma
k^{2n}$ with $n \ge 1$ \cite{yin19}.

\subsection{Altermagnetism and piezomagnetism}
\label{sec:alter}

The spin-dependent band dispersion shown in Fig.~\ref{fig:polar}(e)
that does not give rise to a macroscopic magnetization $T_1^{+-}$
has recently been identified as a characteristic feature of
altermagnets~\cite{sme22}.  The presence of magnetopolarizations
$T_\ell^{+-}$ with odd $\ell > 1$, while a magnetic dipolarization
$T_1^{+-}$ is forbidden by symmetry (no polar direction $+-$), can
be considered a defining feature for altermagnets~\cite{win23,
bho24}.  According to this definition, eleven magnetopolar and
twelve multipolar crystal classes are altermagnetic.

Unlike recent theories \cite{sme22, sme22a}, the above definition of
altermagnetic systems does not depend on local configurations of
magnetic and nonmagnetic atoms in a crystal structure.  The
altermagnetic band structure displayed in Fig.~\ref{fig:polar}(e)
can be realized in crystal structures made solely of magnetic atoms,
as illustrated in Ref.~\cite{win23}.  The above definition applies
both to collinear and to noncollinear magnetic structures.

Similarly, eleven electropolar and twelve multipolar crystal classes
are alterelectric.  These classes are characterized by the condition
that they permit electropolarizations $T_\ell^{-+}$ with odd $\ell >
1$, while an electric dipolarization $T_1^{-+}$ with polar direction
$-+$ is forbidden.  Zincblende is an example of a cubic
alterelectric crystal structure that permits an electric
octupolarization $T_3^{-+}$ \cite{win23}.  The spin-dependent band
dispersion of electropolar alterelectric media is shown in
Fig.~\ref{fig:polar}(c).  If alterelectric order arises in an
order-disorder phase transition similar to the phase transition in
ferroelectric media, it can be identified, e.g., by a
specific-heat anomaly.

In addition, eleven antimagnetopolar and twelve multipolar crystal classes
are altertoroidic.  These classes are characterized by the condition
that they permit magnetotoroidizations $T_\ell^{--}$ with odd $\ell
> 1$, while a dipolarization $T_1^{--}$ with polar direction $--$ is
forbidden.  The antiferromagnetic diamond structure that was studied
in Refs.~\cite{win20, win23} is an example of an altertoroidic
structure.  The spin-degenerate band dispersion of antimagnetopolar
altertoroidic media is shown in Fig.~\ref{fig:polar}(g).  Finally,
we have fifteen multialteric crystal classes.

All magnetopolar altermagnetic crystal classes except for the
octahedral class $O_h = O \times \gsis$ are also piezomagnetic.  In
piezomagnets, external stress (which does not change the signature
$ss'$) defines the polar direction that permits ferromagnetism
and/or pyromagnetism.  (Ferromagnets/pyromagnets are commonly
likewise counted as piezomagnets \cite{new05}.)  In strained
altermagnetic crystals, an external magnetic field can thus be used
to manipulate the magnetic order.  Similarly, all electropolar
alterelectric crystal classes except for the octahedral class $O
\times \gtis$ are also piezoelectric, all antimagnetopolar
altertoroidic classes except for the octahedral class $O_h (O) = O
\times \gstis$ are also piezotoroidic, and all multialteric classes
except for the octahedral class $O$ are also multipiezoic.

\subsection{Weak ferroicity and the anomalous Hall effect}
\label{sec:weak-ferro}

The ferroic crystal classes summarized in Fig.~\ref{fig:polar} also
include the possibility of weak ferroicity such as weak
ferromagnetism \cite{dzy58, misc:weak}.  In weakly ferroic systems,
the symmetry of the magnetic crystal class \emph{permits} ferroicity
(i.e., symmetry permits a dipolarization $T_1^{ss'}$ representing,
e.g., a magnetization); nonetheless the ferroicity in a specific
material may be very small or even zero.  Generally, more crystal
classes permit higher-rank spherical tensors than lower-rank
tensors, see Table~\ref{tab:groups-multipoles}.  Thus when a
lower-rank tensor like $T_1^{ss'}$ becomes nonzero, this is
generally accompanied by a symmetry reduction from some group $G$ to
a subgroup $U \subsetneq G$.  However, in weak ferroics the crystal
class that permits a dipole density $T_1^{ss'}$ does not change when
reorienting local moments on the atoms such that $T_1^{ss'}
\rightarrow 0$ \cite{dzy58}, while higher-rank odd\=/$\ell$
multipole densities $T_\ell^{ss'}$ remain nonzero.  Therefore, even
in the limit of a vanishing dipolarization $T_1^{ss'}$, these
systems possess a polar direction that is realized, e.g., by an
octupole density $T_3^{ss'}$, i.e., under the crystallographic point
group $G$ of the respective crystal class some component of the
octupole density $T_3^{ss'}$ transforms according to the same IR
$\Gamma_\alpha$ of $G$ as a dipole density $T_1^{ss'}$ \cite{kos63}.
For example, let $\vek{m} (\xx)$ denote the magnetization density
due to the local distribution of magnetic moments in the crystal,
and let $\Gamma_\alpha$ denote the IR under $G$ of both a magnetic
dipole density $T_1^{+-}$ and (part of) an octupole density
$T_3^{+-}$.  In a weak ferromagnet, the projection of $\vek{m}
(\xx)$ onto the IR $\Gamma_\alpha$ can be nonzero [compare Eq.\
(\ref{eq:project:f})], yielding a projected density $\vek{m}_\alpha
(\xx)$ that represents a magnetic octupole density $T_3^{+-}$
\cite{suz17}, while the net magnetization density $T_1^{+-}$
vanishes in the crystal, $\sum_\xx \vek{m}_\alpha (\xx) = 0$.  This
result reflects the fact discussed above that multiple IRs
$D_\ell^{ss'}$ of the full rotation group $\Rit$ correspond to the
same IR $\Gamma_\alpha$ of a crystallographic point
group~$G$~\cite{kos63}.

The proper identification of the material's crystal class is
relevant in the context of related effects such as the anomalous
Hall effect (AHE).  While the AHE has commonly been associated with
ferromagnetic materials \cite{nag10}, recent studies have identified
large anomalous Hall conductivities in antiferromagnetic materials
\cite{che14, bet23, kue14, nak15}.  It was shown early on in
Ref.~\cite{sht65} that the AHE may exist only in materials belonging
to one of the 31 ferromagnetic crystal classes, i.e., classes that
\emph{permit} by symmetry a macroscopic magnetization density
$T_1^{+-}$, see also Ref.~\cite{gri93}.  Indeed, consistent with
Ref.~\cite{sht65} the antiferromagnetic materials showing nonzero
anomalous Hall conductivities turn out to be weak ferromagnets,
where the Hall vector is aligned with the polar direction permitted
by the respective crystal class.  For example, intermetallic
ferromagnetic RGa (R${}={}$rare earth Pr, Nd, Sm, Gd, Tb, Dy, Ho,
Er) \cite{fuj71, bar72}, the collinear antiferromagnet MnTe
\cite{bet23}, and the noncollinear antiferromagnets Mn$_3$Sn and
Mn$_3$Ge \cite{kue14, nak15} all belong to the ferromagnetic crystal
class $D_{2h} (C_{2h}) = m'm'm$.  Therefore, these materials share
properties such as an AHE, even when the net magnetization in some
of these materials may be zero, and the Hall vector must align
parallel to the polar direction of the crystal class $D_{2h}
(C_{2h})$.  Furthermore, natural Faraday rotation (i.e., Faraday
rotation in the absence of an external magnetic field) may exist in
crystal structures for which ferromagnetism is permitted by symmetry
\cite{lan8e}, which includes weak ferromagnets.  Weak
ferroelectricity has previously been considered in
Ref.~\cite{sch73}.

\subsection{Odd-parity magnets: space-group symmetries versus point-group symmetries}
\label{sec:msg}

The present work focuses on point-group symmetries of crystal
structures that are responsible for macroscopic crystal properties,
consistent with Neumann's principle.  A more fine-grained
classification of crystal symmetries is provided by the 1651
magnetic space groups (MSGs), where generally multiple MSGs belong
to the same crystal class; i.e., the respective MSGs have the same
magnetic point group \cite{cam24}.  For example, all MSGs with
black-white Bravais lattices (sometimes called type\=/IV MSGs)
describe magnetically ordered structures.  However, all type\=/IV
MSGs belong to type\=/II (see Appendix~\ref{app:BWform}) crystal
classes that also describe nonmagnetic crystal structures.
Therefore, magnetic crystals with type\=/IV MSGs only allow
macroscopic tensor properties that would also be permitted in
nonmagnetic crystal structures belonging to the same type\=/II
crystal class \cite{cor58a, cam24}.  In the standard model, crystals
with type\=/IV MSGs thus belong to the parapolar and electropolar
categories.  Parapolar magnetic media include, e.g., NiO and MnO
(point group $C_{2h} \times \gtis = C_2 \times \gsistis$
\cite{pom24}).  Every type\=/II crystal class has at least one
realization in terms of type\=/IV MSGs.

The type\=/II crystal class of a magnetic type\=/IV crystal generally
has a lower symmetry than the type\=/II crystal class of the
same material in the absence of magnetic order (i.e., above the
critical temperature for magnetic ordering).  Therefore, the
magnetic phase transition of type\=/IV materials can manifest itself
via phenomena usually associated with phase transitions in
nonmagnetic media.  Recently, crystal structures with type\=/IV MSGs
called odd-parity magnets have sparked renewed interest because
their magnetic order can give rise to properties such as a Rashba
effect \cite{hel23, yu25}, i.e., odd-parity magnets fall into the
electropolar category of the standard model.  Previously, the Rashba
effect had been discussed predominantly in the context of
nonmagnetic materials.  A more systematic discussion of how space-group symmetries relate to crystal properties within the standard
model will be published elsewhere.

\subsection{TIS breaking and spin-orbit coupling in magnetically
ordered systems}
\label{sec:tis-soc}

The essence of the standard model developed in this work is a
coarse-grained, yet comprehensive classification of crystalline
matter based on the three fundamental inversion symmetries $i$,
$\theta$, and $i\theta$.  For example, broken TIS represents the
fundamental cause for all physical phenomena that distinguish
magnetic systems from nonmagnetic systems \cite{lan8e}.

We want to compare our theory with recent complementary efforts
aiming for a more fine-grained classification of crystalline matter
beyond the 122 magnetic point groups \cite{sme22, sme22a}.  In the
theory of magnetic groups, symmetry operations act jointly on
orbital and spin (magnetic) degrees of freedom.  The theory of spin
groups \cite{lit74} considers separate operations acting on orbital
and spin degrees of freedom.  Recent interest in this approach
\cite{sme22, sme22a} is motivated by the fact that magnetic order is
generally dominated by exchange coupling \cite{whi07}.  In the
nonrelativistic Schr\"odinger theory, the electron spins are then
completely decoupled from the electrons' orbital motion
\cite{bih07}.  For collinear spins, this yields decoupled
Schr\"odinger equations for spin up and down that preserve TIS, as
noted previously in the context of electronic-structure calculations
for magnetic systems in, e.g., Refs.~\cite{gos15, yua20, sme22}.
Very generally, this is due to the fact that the respective
Hamiltonians for spin up and down are real \cite{win25}.  Therefore,
the Schr\"odinger-Pauli theory is fundamentally incomplete in its
account of broken TIS in magnetically ordered systems.  Spin-orbit
coupling is needed in magnetically ordered systems to break TIS by
making the Hamiltonian complex.  For this reason, spin-orbit
coupling has been an indispensable ingredient in microscopic
theories of the anomalous Hall effect \cite{nag10} and the
magnetoelectric effect \cite{don19}.

It is interesting to compare the Schr\"odinger-Pauli theory with the
Dirac theory.  The Dirac theory provides a relativistically
invariant formulation of quantum mechanics, while the
Schr\"odinger-Pauli theory is generally believed to give a faithful
representation of the nonrelativistic and weakly relativistic limit
of the Dirac theory.  Indeed, unlike the Schr\"odinger-Pauli theory,
the Dirac theory properly accounts for TIS breaking in magnetic
systems even in the nonrelativistic limit \cite{win25}.  This is
closely related to the fact that the well-known electron magnetic
moment $e\hbar/(2m)$ appears explicitly only in the
Schr\"odinger-Pauli theory.  However, in the Dirac theory, this
moment is realized via orbital equilibrium currents that break TIS
even in the nonrelativistic limit \cite{cha08}.

In consequence, a proper account of magnetic phenomena is naturally
achieved in electronic-structure calculations based on the Dirac
theory.  Such fully relativistic calculations of magnetic systems,
though less common than calculations based on the
Schr\"odinger-Pauli theory, have been reported, e.g., in
Refs.~\cite{low10, wag19}.  However, such a description does not
lend itself to a decoupling of real-space order and magnetic order
that is assumed in applications of spin-group theories \cite{sme22,
sme22a, lit74}.

\section{Chirality in solids}
\label{sec:chirality}

Chirality literally means ``handedness'', referring to one of the
most fundamental and ubiquitous types of asymmetry in
nature~\cite{misc:chiral}.  On a conceptional level, chirality is
associated with the existence of at least two inequivalent
versions of a physical entity---called \emph{enantiomorphs}---that
cannot be superposed by proper rotations $C$ and/or
translations~\cite{mis99, fla03, bar04, wag07, fec22, kis22, bou25,
rom25}.  Instead, interconversion of enantiomorphs requires
$i$\=/improper rotations $iC$, i.e., transformations that involve
space inversion $i$ such as a mirror reflection (which is a $\pi$
rotation about an axis perpendicular to the reflection plane
followed by space inversion $i$).  Equivalently, an object has been
called \emph{chiral} if its symmetry group does not contain any
$i$\=/improper rotation.  An \emph{achiral} object can be mapped
onto itself by an $i$\=/improper rotation; i.e., it shows no
enantiomorphism.

The space groups of chiral crystal structures allow two distinct
cases.  Either an enantiomorphic pair of crystal structures
transforms according to an enantiomorphic pair of distinct space
groups, or the pair of crystal structures transforms according to
the same space group.  In total, 65 nonmagnetic space groups
describe chiral structures.  These are the 65 Sohncke groups that
were identified early on in crystallography and that represent a
subset of the 230 nonmagnetic space groups \cite{bur13}.  However,
in agreement with Neumann's principle, the macroscopically
observable features of chirality only depend on the crystallographic
point group $G$ of the crystal class of a crystal structure
\cite{nye85}.  The 65 Sohncke groups are exactly the space groups
for which the respective point groups $G$ are one of the 11
(nonmagnetic) chiral point groups; i.e., the distinction between the
two cases described above is not relevant at the level of Neumann's
principle.  The Sohncke groups include symmorphic and nonsymmorphic
space groups.  Ignoring TIS, the 11 chiral point groups $G$ are
precisely the proper point groups $\gprop$ that form the roots
for the 11 Laue classes $L (\gprop)$ of crystallographic point
groups~\cite{misc:allBackground}.

Optical activity is often viewed as a unique hallmark of chirality.
While this is correct for liquids and molecules in solution, crystal
structures can be optically active though they are not chiral (point
groups $C_s$, $C_{2v}$, $S_4$ and $D_{2d}$) \cite{nye85, bar04,
hob67}.  This example shows how chirality in infinite crystal
structures is qualitatively different from chirality in finite
molecules.  The surge of recent interest in understanding chirality
in solids on a fundamental level~\cite{fec22, kis22, bou25, rom25}
is one of the motivations for the present work.

The importance of TIS for chirality has been noted early
on~\cite{zoc53, bar86}.  A distinction between two different types
of chirality was suggested: so-called \emph{true chirality} occurs
in situations where only $i$\=/improper rotations, but not
$\theta$\=/improper rotations, can interconvert between
enantiomorphs, whereas systems for which both $i$\=/improper and
$\theta$\=/improper rotations but not $i\theta$\=/improper rotations
interconvert the enantiomorphs have been called \emph{false
chiral}~\cite{bar04, bar86, bar20}.  Original discussions of false
chirality only considered TIS-breaking external magnetic fields
\cite{bar04, bar86, bar20}, but they did not consider magnetically
ordered systems breaking TIS inherently.  The utility of adopting a
broader definition of enantiomorphism has also been
emphasized~\cite{mis99}.

In the following, we develop the most general description of
enantiomorphism based on symmetry considerations.
We start by introducing the five categories of chirality in
Sec.~\ref{sec:chiralCat}.  Experimental signatures for unichiral and
multichiral systems are discussed in Sec.~\ref{sec:exp:chiral}.
Section~\ref{sec:chiral:bands} focuses on the intricate way
chirality manifests in the electronic band structure.  A brief
discussion of induced chirality is presented in
Sec.~\ref{sec:chiral:ind}.  Materials candidates for the different
categories of chirality are listed in
Appendix~\ref{app:chiral-materials}.

\subsection{Categories of chirality}
\label{sec:chiralCat}

In the present work, we treat SIS, TIS, and CIS on an equal footing
to obtain a more systematic understanding of chirality and to
acquire a unified description of associated physical phenomena.  In
the lower part of Table~\ref{tab:categories}, we identify five
categories of chirality that are distinguished by which
$\gamma$\=/improper rotations are present in the point group of a
system.  The \emph{parachiral} category refers to systems that have
all three kinds of $\gamma$\=/improper rotations among their
symmetries.  These systems have no enantiomorphism.  The
\emph{electrochiral}, \emph{magnetochiral} and
\emph{antimagnetochiral} systems have $\theta$\=/improper,
$i$\=/improper, and $i\theta$\=/improper rotations, respectively,
among their symmetries, so each of these categories has two
distinct enantiomorphs.  Finally, \emph{multichiral} systems have
only proper rotations among their symmetries, so these systems
have four distinct enantiomorphs.

Clearly, the defining features of the five categories of chirality
displayed in the lower part of Table~\ref{tab:categories} follow a
similar pattern as the defining features of the five categories of
polarizations displayed in the upper part of
Table~\ref{tab:categories}.  Therefore, each group type is
associated with one category of multipole order and one category of
chirality, which yields the symmetry classification of magnetic
point groups presented in Fig.~\ref{fig:standardmodel} (see also
Appendix~\ref{app:StructClass}, especially
Table~\ref{tab:group-types-def}).  The five categories of multipole
order (polarizations) and the five categories of chirality thus
represent two complementary classifications based on the presence or
absence of the inversion symmetries $i$, $\theta$, and $i\theta$.
Polarizations reflect the presence or absence of $i$, $\theta$, and
$i\theta$ as \emph{independent} elements in the symmetry group $G$
of a physical system.  Chiralities, on the other hand, reflect the
presence or absence in $G$ of any \emph{composite} symmetry elements
$iC$, $\theta C$, and $i\theta C$ (i.e., presence or absence of
$\gamma$\=/improper rotations $\gamma C$).  See
Fig.~\ref{fig:standardmodel} and Table~\ref{tab:categories}.

While the classifications of point groups with respect to the
categories of polarizations and those of chirality are
complementary, we nevertheless have an intricate connection between
multipole order and chirality that may exist in a system.
Specifically, the absence of $\gamma$\=/improper rotations in chiral
systems implies that the respective monopoles $T_0^{ss'}$ become
symmetry allowed, see Table~\ref{tab:categories}.  These monopoles
thus represent \emph{chiral charges}.  They are signatures of chiral
order, similar to how electric and magnetic multipoles are
signatures of polar order \cite{misc:noether}.  (Below we discuss
more generally how configurations of polarizations can give rise to
chirality.)  Our naming of the five categories of chirality reflects
the nature of the allowed scalars other than the universally allowed
$T_0^{++}$ (see Table~\ref{tab:categories}).  Compared with the
existing nomenclature~\cite{bar04} comprising (true) chirality,
false chirality and achirality, the five mutually exclusive
categories constitute a more comprehensive and systematic
classification of chirality in solids~\cite{misc:relateToBarron}.

\begin{figure*}[t]
  \centering
  \includegraphics[width=0.90\linewidth]{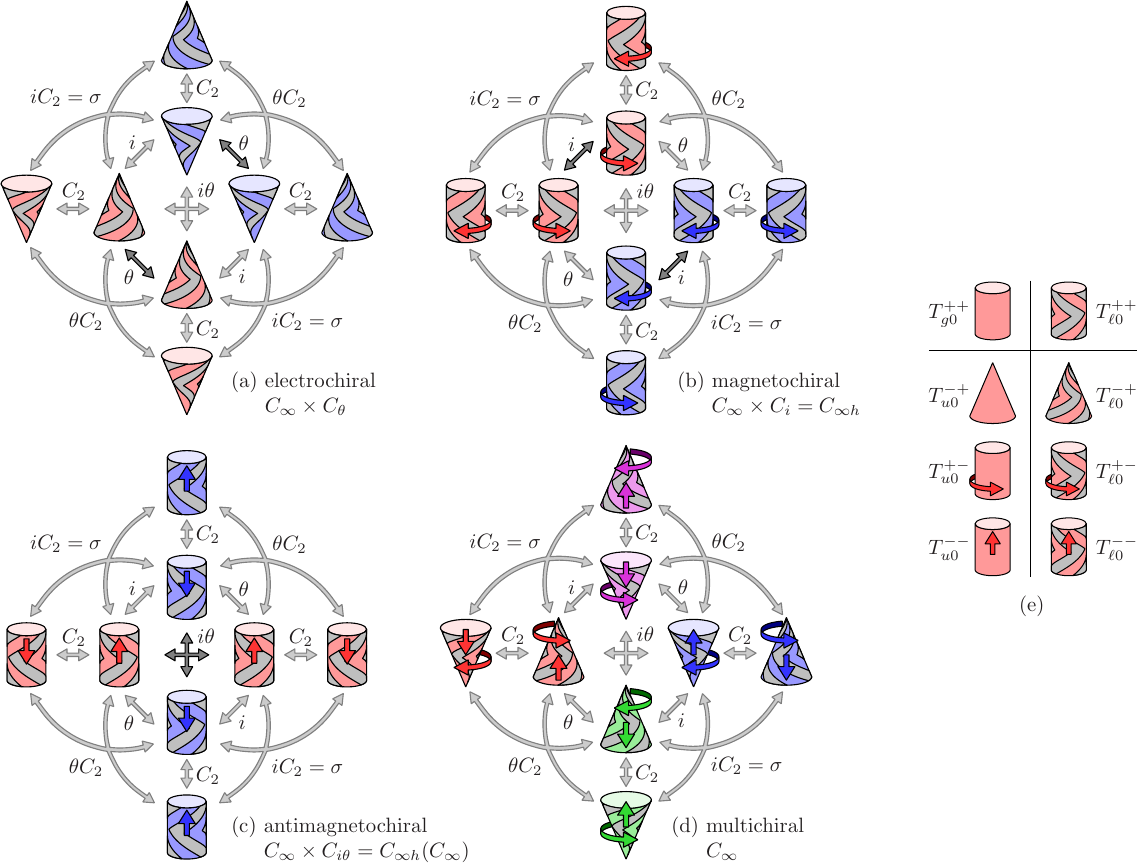}
  \caption{\label{fig:chiral:c-infty} Categories of chirality
  illustrated for axially symmetric objects representing the
  unipolar-unichiral and multipolar-multichiral groups in class
  $L (C_\infty)$~\cite{misc:allBackground}.  As a visual aid, different
  enantiomorphs are distinguished by color.  Straight (curved)
  colored arrows indicate translational (rotational) motion of an
  object.  Dark (light) gray double-arrows connect two objects that
  transform into each other under the specified symmetry operation,
  which is (is not) a good symmetry of the object.  $C_2$ represents
  a $\pi$ rotation about any axis perpendicular to the high-symmetry
  $z$ axis.  (a) In electrochiral systems, space inversion $i$
  interconverts the enantiomorphs, but time inversion $\theta$ does
  not.  (b) The situation is reversed for magnetochiral systems.
  (c) Both $i$ and $\theta$ separately interconvert enantiomorphs in
  antimagnetochiral systems, but their combination $i\theta$ leaves
  them invariant.  In contrast, $i\theta$ interconverts the two
  enantiomorphs present in electro- and magnetochiral systems.  (d)
  All three inversions $i$, $\theta$ and $i\theta$ interconvert the
  four enantiomorphs of a multichiral system.  The chiral objects in
  panels (a) to (c) transform like tensor components $T_{\ell
  0}^{ss'}$ as indicated in panel (e).  These tensor components
  $T_{\ell 0}^{ss'}$ can be derived either by reducing the symmetry
  of tensor components $T_{u 0}^{ss'}$ or by reducing the symmetry
  of tensor components $T_{\ell 0}^{++}$.}
\end{figure*}

\begin{figure*}[t]
  \centering
  \includegraphics[width=0.90\linewidth]{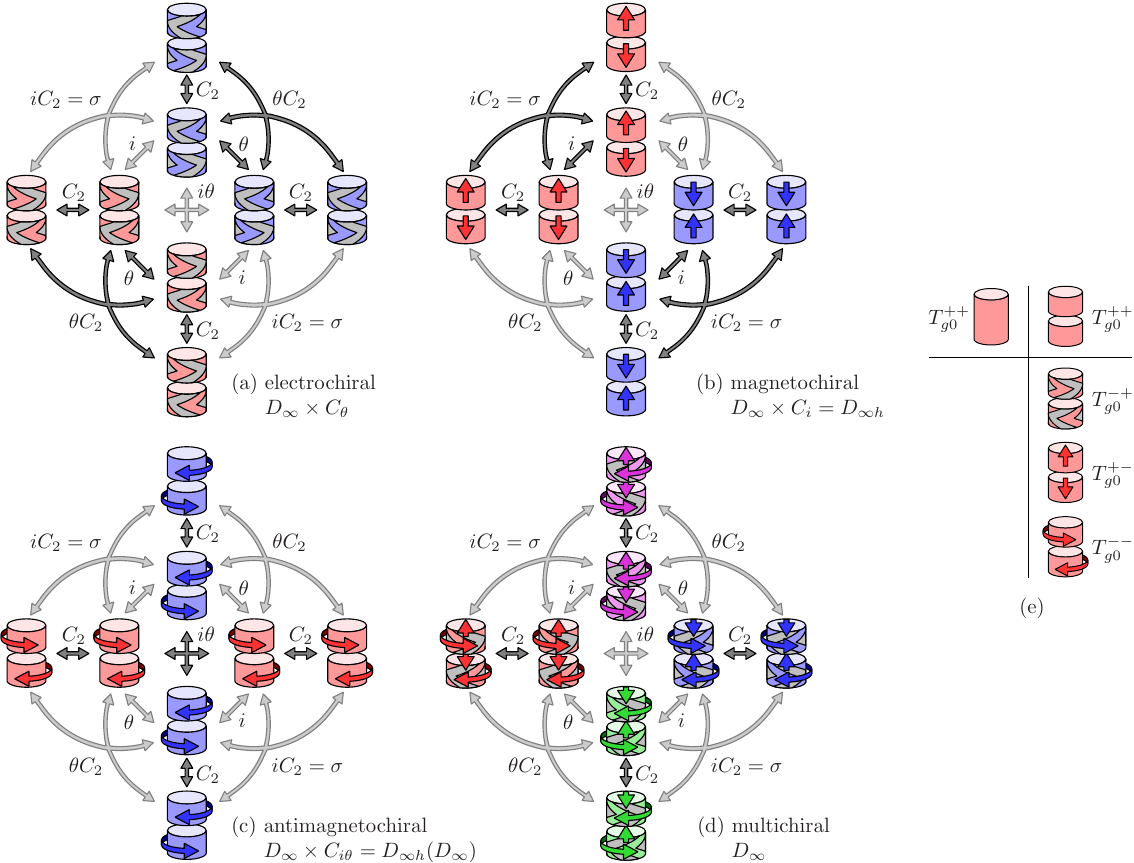}
  \caption{\label{fig:chiral:d-infty} Categories of chirality illustrated for
  axially symmetric objects representing the unipolar-unichiral and
  multipolar-multichiral groups in class
  $L (D_\infty)$~\cite{misc:allBackground}.
  Conventions are the same as in Fig.~\ref{fig:chiral:c-infty}.
  (e) The chiral objects in panels (a) to (c) transform like tensor
  components $T_{g 0}^{ss'}$ that can be derived by reducing the
  symmetry of tensor components $T_{g 0}^{++}$.}
\end{figure*}

\begin{figure*}[t]
  \centering
  \includegraphics[width=0.98\linewidth]{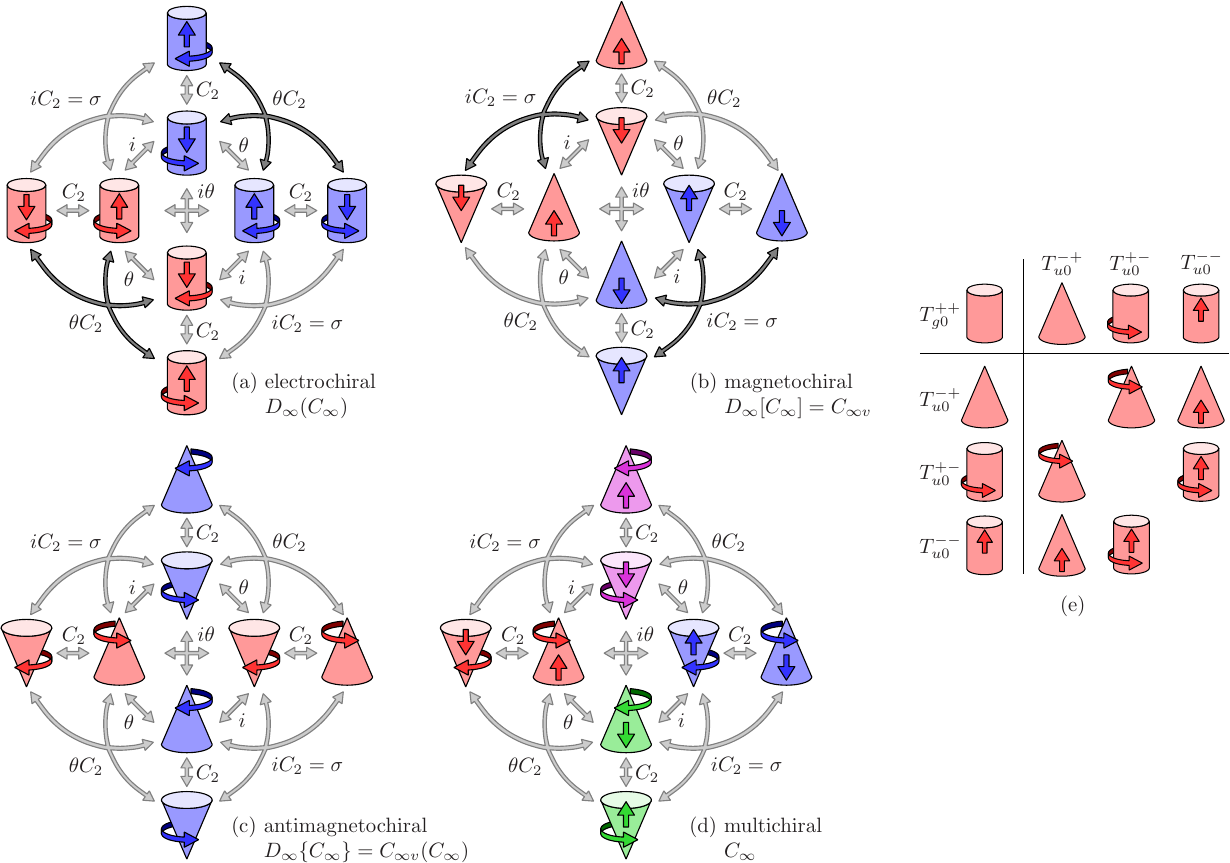}
  \caption{\label{fig:chiral:dc-infty} Categories of chirality illustrated for
   axially symmetric objects representing the multipolar-unichiral groups
   in class $L(D_\infty)$~\cite{misc:allBackground}.  The
   multipolar-multichiral group $C_\infty$ shown in panel (d) belongs to
   $L (C_\infty)$~\cite{misc:allBackground}.  Conventions are the same as in
   Fig.~\ref{fig:chiral:c-infty}.  The chiral objects in panels (a) to (d) can
   be viewed as superpositions of the parachiral objects in panel (e).}
\end{figure*}

We now provide a brief overview of each category.
Figures~\ref{fig:chiral:c-infty}, \ref{fig:chiral:d-infty}
and~\ref{fig:chiral:dc-infty} give illustrative examples for
these chiralities by exploring how different axially symmetric objects
are either invariant (i.e., mapped onto themselves), mapped onto a
rotated copy of themselves, or mapped onto distinct enantiomorphs.
Such simple shapes, typically incorporating angular and/or linear
momentum, provide generic representations for finite chiral systems
\cite{bar86, bar04, jur25}.  Formally, this approach depicts
categories of chirality for the continuous axial classes
$L(C_\infty)$ and $L (D_\infty)$ discussed in more detail in
Appendix~\ref{app:axial-groups}.  The basic symmetry properties of
the objects utilized in Fig.~\ref{fig:chiral:c-infty},
\ref{fig:chiral:d-infty} and~\ref{fig:chiral:dc-infty} are also
reviewed in Appendix~\ref{app:axial-groups}, see
Table~\ref{tab:c-infinity-objects}.

\subsubsection{Parachirality}
\label{sec:para-chiral}

Parachiral point groups contain $\gamma$\=/improper rotations
$\gamma C$ for all three inversions $\gamma = i$, $\theta$ and
$i\theta$.  Therefore, such systems do not allow enantiomorphism.
No scalars of $\Rit$ other than $T_0^{++}$ are permitted under
parachiral groups.  In particular, the scalar $T_0^{-+}$ is forbidden,
and parachiral systems are therefore achiral as per the definition in
Ref.~\cite{bar04}.

Examples of parachiral groups include the full axial rotation groups
$C_\infty \times \gsistis$ and $D_\infty \times \gsistis$ permitting
scalars $T_{\ell 0}^{++}$ and $T_{g0}^{++}$, respectively, as
illustrated in Figs.~\ref{fig:chiral:c-infty}(e), \ref{fig:chiral:d-infty}(e)
and~\ref{fig:chiral:dc-infty}(e).  See Fig.~\ref{fig:unichiral} for a
crystallographic example (corundum Al$_2$O$_3$, point group
$D_3\times\gsistis \equiv D_{3d}\times\gtis$).

\subsubsection{Electrochirality}
\label{sec:electro-chiral}

Electrochiral groups contain $\theta$\=/improper rotations $\theta
C$, but no $i$\=/improper or $i\theta$\=/improper rotations, i.e.,
an inequivalent enantiomorph is generated when a transformation $i
C$ or $i\theta C$ is applied to the crystal.  Electrochiral systems
permit the pseudoscalar $T_0^{-+}$.  See
Figs.~\ref{fig:chiral:c-infty}(a), \ref{fig:chiral:d-infty}(a),
and~\ref{fig:chiral:dc-infty}(a) for illustrative examples.

As seen from Fig.~\ref{fig:standardmodel}, electrochiral groups are
either electropolar or multipolar.  Electrochiral magnets~\cite{che22}
belong to the latter type.  Examples of nonmagnetic electrochiral
systems are shown in Figs.~\ref{fig:chiral:c-infty}(a) and
\ref{fig:chiral:d-infty}(a).  In these systems, chirality arises from the interplay
of a parapolarization with an electropolarization, i.e., between
even-$\ell$ and odd-$\ell$ electric-multipole densities.  Crystallographic
realizations of the scenarios from Figs.~\ref{fig:chiral:c-infty}(a) and
\ref{fig:chiral:d-infty}(a) have been encountered in Refs.~\cite{fav24}
and \cite{hay21} (with point groups $C_3\times\gtis$ and $D_4\times
\gtis$, respectively).  The situation depicted in
Fig.~\ref{fig:chiral:dc-infty}(a) is equivalent to an electrochiral
ferromagnet, where electrochirality emerges from the presence of
collinear magnetic and magnetotoroidal dipolarizations~\cite{che22},
which is the crystal analog of electrochirality arising when a magnetic
field is applied parallel to a beam of light~\cite{bar86}.  In such
multipolar electrochiral systems, the electrochiral enantiomorphs can
be viewed as magnetic domains that are related via $i$\=/improper and
$i\theta$\=/improper rotations, but are each invariant under improper
rotations of the form $\theta C$~\cite{che22}.  Such domains differ
fundamentally from ordinary magnetic domains that can be superposed
by proper rotations $C$.

\subsubsection{Magnetochirality}
\label{sec:magneto-chiral}

Magnetochiral groups contain $i$\=/improper rotations $i C$, but
$\theta$\=/improper and $i\theta$\=/improper rotations are absent.
Therefore, $\theta$\=/improper or $i\theta$\=/improper rotations
generate an inequivalent enantiomorph of a system described by
such a group.  See Figs.~\ref{fig:chiral:c-infty}(b),
\ref{fig:chiral:d-infty}(b), and~\ref{fig:chiral:dc-infty}(b) for
illustrative examples.

The nontrivial scalar allowed in magnetochiral groups is $T_0^{+-}$,
which has no indicator associated with it in the electronic band
structure, see Table~\ref{tab:tensor-pow-spher} and
Sec.~\ref{sec:chiral:bands}.  Some physical consequences arising
from the presence of $T_0^{+-}$ have recently been conjectured in
Ref.~\cite{hay23}.  As the scalar $T_0^{-+}$ is forbidden under
magnetochiral groups, magnetochiral systems would be considered
achiral according to Ref.~\cite{bar04}.  However, the
enantiomorphism arising under $\theta$\=/improper and
$i\theta$\=/improper rotations makes magnetochiral systems
fundamentally different from parachiral ones, as enantiomorphism is
completely absent in the latter.

Magnetochirality requires magnetic order.  Ordinary magnetic domains
can be superposed by proper rotations $C$.  Magnetochiral
enantiomorphs can be considered as magnetic domains that are related
by some improper rotations of the form $\theta C$ and $i\theta C$,
but these domains \emph{cannot} be superposed by any proper rotation
$C$, e.g., due to the presence of nonmagnetic atoms.  See
Fig.~\ref{fig:unichiral} for an example (hematite
$\alpha$\=/Fe$_2$O$_3$, point group $D_3\times\gsis\equiv D_{3d}$).
Thus, magnetic domains that are magnetochiral enantiomorphs
constitute fundamentally different, physically distinguishable
entities, unlike ordinary magnetic domains that differ solely by the
orientation of the magnetic order parameter.

Magnetochiral groups are either magnetopolar or multipolar (see
Fig.~\ref{fig:standardmodel}).  Examples of magnetopolar magnetochiral
systems are shown in Figs.~\ref{fig:chiral:c-infty}(b) and
\ref{fig:chiral:d-infty}(b).  Here enantiomorphism arises from the interplay
of a parapolarization with a magnetopolarization, i.e., even\=/$\ell$
electric and odd\=/$\ell$ magnetic multipole densities.  In the example
from Fig.~\ref{fig:chiral:c-infty}(b), the magnetopolarization is a magnetization,
i.e., this system is a ferromagnet.  Alternatively, as exemplified by the
system shown in Fig.~\ref{fig:chiral:d-infty}(b), magnetochiral systems can
also be altermagnets (having an odd\=/$\ell>1$ magnetic multipole density).
Hematite ($\alpha$\=/Fe$_2$O$_3$, point group $D_3\times\gsis\equiv D_{3d}$,
see Fig.~\ref{fig:unichiral}) is such an altermagnetic magnetochiral materials
candidate~\cite{hoy25}.  Its two enantiomorphs are the two different
antiferromagnetic domains that are interrelated via TIS and CIS.

The scenario in Fig.~\ref{fig:chiral:dc-infty}(b) is an example of
the multipolar magnetochiral type.  In this scenario,
magnetochirality arises from collinear electric and magnetotoroidal
dipolarizations, which is analogous to magnetochirality realized in
the situation where an electric field is applied parallel to a light
beam.

\subsubsection{Antimagnetochirality}
\label{sec:antimagneto-chiral}

Antimagnetochiral groups contain $i\theta$\=/improper rotations, but
$i$\=/improper and $\theta$\=/improper rotations are absent.  Thus,
$\gamma$\=/improper rotations $\gamma C$, where $\gamma = i$ or
$\theta$, generate an inequivalent enantiomorph of an
antimagnetochiral structure.  See the examples depicted in
Figs.~\ref{fig:chiral:c-infty}(c), \ref{fig:chiral:d-infty}(c),
and~\ref{fig:chiral:dc-infty}(c).

The antimagnetochiral category subsumes what has previously been
referred to as false chirality~\cite{bar86, bar20}: although such a
system exhibits enantiomorphism, it does not permit a scalar
$T_0^{-+}$ and is therefore not considered to be truly chiral.
Instead, a scalar $T_0^{--}$ is allowed in antimagnetochiral
systems, though this scalar has no indicator directly associated with
it in the electronic band structure, see Table~\ref{tab:tensor-pow-spher}
and Sec.~\ref{sec:chiral:bands}.

Antimagnetochiral groups are either antimagnetopolar or multipolar;
see Fig.~\ref{fig:standardmodel}.  Figures~\ref{fig:chiral:c-infty}(c) and
\ref{fig:chiral:d-infty}(c) show examples of antimagnetopolar
antimagnetochiral systems, where antimagnetochirality arises from the
combination of a parapolarization with an antimagnetopolarization.
Figure~\ref{fig:chiral:dc-infty}(c) shows a multipolar system,
representing the situation where antimagnetochirality arises
from the presence of an electric polarization collinear with a
magnetization, or, equivalently, from parallel electric and magnetic
fields~\cite{bar86}.

Antimagnetochiral enantiomorphs can be viewed as magnetic domains
that are related via $i$\=/improper and $\theta$\=/improper
rotations.  See Fig.~\ref{fig:unichiral} for a crystallographic
example [chromia Cr$_2$O$_3$, point group $D_3\times\gstis\equiv
D_{3d} (D_3)$].  Again, these should not be considered as ordinary
magnetic domains---antimagnetochiral enantiomorphs cannot be
superposed by proper rotations $C$ and, therefore, constitute
different physical entities~\cite{rad61, bou24}.

\subsubsection{Multichirality}
\label{sec:multi-chiral}

Multichiral groups do not contain $\gamma$\=/improper rotations
$\gamma C$ for any of the inversions $\gamma=i$, $\theta$ or
$i\theta$.  Hence, these groups coincide with the proper point groups.
See Figs.~\ref{fig:chiral:c-infty}(d), \ref{fig:chiral:d-infty}(d),
and~\ref{fig:chiral:dc-infty}(d) for illustrative examples.  As indicated
in these figures, multichiral systems have four distinct enantiomorphs,
as all three improper rotations $iC$, $\theta C$ and $i\theta C$ yield
distinct versions of the given object.

It is possible to group the four enantiomorphs of a multichiral system
into two pairs so that members of a pair are transformed into
each other by $i$\=/improper rotations.  Based on the existence of
enantiomorphs that interconvert via improper rotations $iC$, and the
fact that a scalar $T_0^{-+}$ is allowed, multichiral systems have been
conventionally classified as showing true chirality~\cite{bar04, che22}.
However, the enantiomorphs forming a pair related by $i$\=/improper
rotations are not invariant under $\theta$\=/improper rotations, as is the
case for electrochiral enantiomorphs.  In fact, both $\theta$\=/improper
rotations and $i\theta$\=/improper rotations transform one pair of
enantiomorphs related by $i$\=/improper rotations into the other pair
of such enantiomorphs.  Similarly, the four enantiomorphs could be
divided into two other pairs where members of a pair interconvert via
$\theta$\=/improper rotations.  While the same behavior is exhibited by
magnetochiral enantiomorphs, members of such a
multichiral-enantiomorph pair are again different because they are
not invariant under $i$\=/improper rotations.  Rather, the two pairs are
switched by both $i$\=/improper rotations and $i\theta$\=/improper
rotations.  Lastly, for the third possible grouping of the four multichiral
enantiomorphs into disjoint pairs, members of an individual pair can be
superposed by improper rotations $i\theta C$, and the two pairs are
interchanged by both $\theta$\=/improper rotations and $i$\=/improper
rotations.  Thus, the enantiomorphism of multichiral systems is
fundamentally different from the enantiomorphism of unichiral systems;
it is not a mixture or superposition of the unichiral enantiomorphisms.

The lower symmetry of multichiral systems permits all four scalars
$T_0^{ss'}$ to exist side by side.  More generally, they
simultaneously allow the same set of tensors $T_{lm}^{s s'}$ for all
signatures $s s'$.  Multichiral systems are therefore always
multipolar, see Fig.~\ref{fig:standardmodel}.  The interrelation of
different multipolar orders with (conventional) chirality has been
the focus of a recent experimental study~\cite{xu22} of a
multichiral material (BaCoSiO$_4$, point group $C_6$).  This system
is a realization of the scenario depicted in
Fig.~\ref{fig:chiral:c-infty}(d).  However, in such systems
(having a point group $C_n$), the simultaneous presence of electric,
magnetic and magnetotoroidal dipolarizations can mask physical
effects arising from multichirality, including optical
nonreciprocity.  More pristine realizations of multichiral systems
are those with a point group $D_{2n}$ [scenario depicted in
Fig.~\ref{fig:chiral:d-infty}(d)].

\begin{sidewaystable*}[tbp]
  \caption{\label{tab:chiral-invars} Lowest-degree invariant tensors
  realizing chiralities in the continuous classes $L(R)$ and
  $L(D_\infty)$, the cubic classes $L(O)$ and $L(T)$, and the axial
  classes $L(D_n)$ and $L(C_n)$ with $n = 2, 4, 3,
  6$~\cite{misc:allBackground}.  We have $R \supset O \supset T$
  and $D_\infty \supset D_n \supset C_n$, so that terms allowed under
  the respective supergroup are also allowed under the respective
  subgroup.  The symbol '$\cp$' denotes cyclic permutation of indices $x,y,z$ in the
  preceding term, and $(x \leftrightarrow y)$ denotes the preceding term
  with $x$ and $y$ interchanged.  For the crystallographic classes, we
  denote in square brackets the degree $\ell$ of multipolar order
  represented by the respective terms as well as the maximal group that
  leaves this term invariant.
  For electrochirality and magnetochirality, we only consider
  electric and magnetic tensors with odd $\ell$; we do not consider
  toroidal compound tensors with even $\ell$.  For
  antimagnetochirality, we consider the lowest-degree odd\=/$\ell$
  magnetotoroidal compound tensors representing an
  antimagnetopolarization.}
  \let\mc\multicolumn
  \renewcommand{\arraystretch}{1.2}%
  \begin{tabular*}{\linewidth}{nLlE*{3}{Les{0.8em}LE}n}
    \hline\hline
    \mc{2}{nl}{\rule{0pt}{2.7ex}class $L (\gprop)$}
    & \mc{2}{C}{L(R)} & \mc{2}{C}{L(O)} & \mc{2}{C}{L(T)} \\
    \hline \rule{0pt}{2.7ex} \gprop \times \gsistis & PC
    & &  k^2
    & [\ell = 4, O \times \gsistis] &  k_y^2 k_z^2 + \cp & & \\
    & & &
    & [\ell = 6, O \times \gsistis]
    & \left[k_x^4 (k_y^2 + k_z^2) + \cp\right] - 8 k_x^2 k_y^2 k_z^2
    & [\ell = 6, T \times \gsistis] &  (k_y^2-k_z^2) (k_z^2-k_x^2) (k_x^2-k_y^2) \\ \hline
    \rule{0pt}{2.7ex} \gprop \times \gtis &
    EC & &  \vek{\sigma} \cdot \kk
    & [\ell = 9, O \times \gtis] &
    \begin{tabular}[t]{nLn}
      \sigma_x k_x [k_y^2 k_z^2 (3 k_y^4 - 10 k_y^2 k_z^2 + 3 k_z^4) \\
    +  k_x^2 (k_x^2 - k_y^2 - k_z^2) (k_y^4 - 6 k_y^2 k_z^2 + k_z^4) ]
    + \cp
    \end{tabular}
    & [\ell = 3, O[T] \times \gtis] &
    \sigma_x k_x (k_y^2-k_z^2) + \cp \\
    \gprop \times \gsis & MC & &
    & [\ell = 9, O \times \gsis] &   \sigma_x k_y k_z (k_y^2-k_z^2) [5k_x^4
    -3k_x^2 (k_y^2 + k_z^2) + k_y^2 k_z^2] + \cp
    & [\ell = 3, O(T) \times \gsis] & \sigma_x k_y k_z + \cp \\
    \gprop \times \gstis & \makebox[0pt][l]{AMC} & &
    & [\ell = 9, O \times \gstis]
    & k_x k_y k_z (k_y^2-k_z^2) (k_z^2-k_x^2) (k_x^2-k_y^2)
    & [\ell = 3, O(T) \times \gstis] & k_x k_y k_z \\ \hline
    \mc{2}{nl}{\rule{0pt}{2.7ex}class $L (\gprop)$}
    & \mc{2}{C}{L(D_\infty)} & \mc{2}{C}{L(D_6)} & \mc{2}{C}{L(C_6)} \\
    \hline \rule{0pt}{2.7ex} \gprop \times \gsistis & PC
    & &  k_x^2 + k_y^2; k_z^2
    & [\ell = 6, D_6 \times \gsistis]
    & k_x^2 (k_x^2 - 3k_y^2)^2 ; \; k_y^2 (k_y^2 - 3k_x^2)^2
    & [\ell = 6, C_6 \times \gsistis]
    & k_x k_y (k_x^2 - 3k_y^2) (k_y^2 - 3k_x^2) \\ \hline
    \rule{0pt}{2.7ex} \gprop \times \gtis &
    EC & & \begin{tabular}[t]{nLn}
      \sigma_x k_x + \sigma_y k_y; \\ \sigma_z k_z
    \end{tabular}
    & [\ell = 7, D_{12} [D_6] \times \gtis] & \begin{tabular}[t]{nLn}
     \sigma_x k_x [k_y^2 (k_x^2-3k_y^2) (k_y^2-3k_x^2)
        + 3k_z^2 (k_x^4 - 10 k_x^2 k_y^2 + 5 k_y^4)] \\
      - (x \leftrightarrow y)
      - 3 \sigma_z k_z (k_x^2-k_y^2) (k_x^4 - 14 k_x^2 k_y^2 + k_y^4)
      \end{tabular}
    & [\ell = 1, D_\infty [C_\infty] \times \gtis] &
    \sigma_x k_y - \sigma_y k_x \\
    \gprop \times \gsis & MC & &
    & [\ell = 7, D_{12} (D_6) \times \gsis] &  \begin{tabular}[t]{nLn}
     3 \sigma_x k_y k_z (5k_x^4 - 10 k_x^2 k_y^2 + k_y^4)
      + (x \leftrightarrow y) \\
      - \sigma_z k_x k_y (k_x^2-3 k_y^2) (k_y^2-3 k_x^2)
      \end{tabular}
    & [\ell = 1, D_\infty (C_\infty) \times \gsis] & \sigma_z \\
    \gprop \times \gstis & \makebox[0pt][l]{AMC} & &
    & [\ell = 7, D_{12} (D_6) \times \gstis]
    & k_x k_y k_z (k_x^2 - 3k_y^2) (k_y^2 - 3k_x^2)
    & [\ell = 1, D_\infty (C_\infty) \times \gstis] & k_z \\ \hline
    \mc{2}{nl}{\rule{0pt}{2.7ex}class $L (\gprop)$}
    & \mc{2}{C}{} & \mc{2}{C}{L(D_3)} & \mc{2}{C}{L(C_3)} \\
    \hline \rule{0pt}{2.7ex} \gprop \times \gsistis & PC
    & &
    & [\ell = 4, D_3 \times \gsistis] & k_y k_z (k_y^2 - 3k_x^2)
    & [\ell = 4, C_3 \times \gsistis] & k_x k_z (k_x^2 - 3k_y^2) \\ \hline
    \rule{0pt}{2.7ex} \gprop \times \gtis &
    EC & &
    & [\ell = 3, D_6 [D_3] \times \gtis]
    & 2 \sigma_x k_z k_x k_y + \sigma_y k_z (k_x^2-k_y^2)
                   + \sigma_z k_y (k_y^2 - 3k_x^2)
    & [\ell = 1, D_\infty [C_\infty] \times \gtis]
    & \sigma_x k_y - \sigma_y k_x \\
    \gprop \times \gsis & MC & &
    & [\ell = 3, D_6 (D_3) \times \gsis]
    & \sigma_x (k_x^2-k_y^2) - 2 \sigma_y k_x k_y
    & [\ell = 1, D_\infty (C_\infty) \times \gsis] & \sigma_z \\
    \gprop \times \gstis & \makebox[0pt][l]{AMC} & &
    & [\ell = 3, D_6 (D_3) \times \gstis] & k_x (k_x^2 - 3k_y^2)
    & [\ell = 1, D_\infty (C_\infty) \times \gstis] & k_z \\ \hline
    \mc{2}{nl}{\rule{0pt}{2.7ex}class $L (\gprop)$}
    & \mc{2}{C}{} & \mc{2}{C}{L(D_4)} & \mc{2}{C}{L(C_4)} \\
    \hline \rule{0pt}{2.7ex} \gprop \times \gsistis & PC
    & &
    & [\ell = 4, D_4 \times \gsistis] & k_x^2 k_y^2
    & [\ell = 4, C_4 \times \gsistis] & k_x k_y (k_x^2 - k_y^2) \\ \hline
    \rule{0pt}{2.7ex} \gprop \times \gtis &
    EC & &
    & [\ell = 5, D_8 [D_4] \times \gtis] & \begin{tabular}[t]{nLn}
      \sigma_x k_x [k_x^2 (k_y^2 - k_z^2) - k_y^2 (k_y^2 - 3 k_z^2)]
      + (x \leftrightarrow y) \\
      + \sigma_z k_z (k_x^4 - 6 k_x^2 k_y^2 + k_y^4)
    \end{tabular}
    & [\ell = 1, D_\infty [C_\infty] \times \gtis]
    & \sigma_x k_y - \sigma_y k_x \\
    \gprop \times \gsis & MC & &
    & [\ell = 5, D_8 (D_4) \times \gsis]
    & \sigma_x k_y k_z (k_y^2 - 3k_x^2) - (x \leftrightarrow y)
    - \sigma_z k_x k_y (k_x^2 - k_y^2)
    & [\ell = 1, D_\infty (C_\infty) \times \gsis] & \sigma_z \\
    \gprop \times \gstis & \makebox[0pt][l]{AMC} & &
    & [\ell = 5, D_8 (D_4) \times \gstis] & k_x k_y k_z (k_x^2 - k_y^2)
    & [\ell = 1, D_\infty (C_\infty) \times \gstis] & k_z \\ \hline
    \mc{2}{nl}{\rule{0pt}{2.7ex}class $L (\gprop)$}
    & \mc{2}{C}{} & \mc{2}{C}{L(D_2)} & \mc{2}{C}{L(C_2)} \\
    \hline \rule{0pt}{2.7ex} \gprop \times \gsistis & PC
    & &
    & [\ell = 2, D_2 \times \gsistis]
    & k_x^2 - k_y^2; \; k_y^2 - k_z^2; \; k_z^2 - k_x^2
    & [\ell = 2, C_2 \times \gsistis] & k_x k_y \\ \hline
    \rule{0pt}{2.7ex} \gprop \times \gtis &
    EC & &
    & [\ell = 3, O [T] \times \gtis] & \sigma_x k_x (k_y^2-k_z^2) + \cp
    & [\ell = 1, D_\infty [C_\infty] \times \gtis]
    & \sigma_x k_y - \sigma_y k_x \\
    \gprop \times \gsis & MC & &
    & [\ell = 3, O (T) \times \gsis] & \sigma_x k_y k_z + \cp
    & [\ell = 1, D_\infty (C_\infty) \times \gsis] & \sigma_z \\
    \gprop \times \gstis & \makebox[0pt][l]{AMC} & &
    & [\ell = 3, O (T) \times \gstis] & k_x k_y k_z
    & [\ell = 1, D_\infty (C_\infty) \times \gstis] & k_z \\ \hline \hline
  \end{tabular*}
  \vspace*{-0.37\textheight}
\end{sidewaystable*}

\subsection{Experimental signatures of chirality}
\label{sec:exp:chiral}

Very generally, the enantiomorphism realized in unichiral and
multichiral media can be identified via observable effects such as
natural optical activity \cite{lan8e} and nonreciprocal directional
dichroism \cite{bro63} that have opposite signs for the different
enantiomorphs.  The occurrence of these effects is not limited to
chiral media, i.e., their observability represents a necessary but
not a sufficient criterion for enantiomorphism \cite{nye85, bar04,
hob67}.  However, in nonenantiomorphous media, these effects arise
such that symmetry-equivalent directions of the media show the same
effect with opposite signs \cite{nye85, hob67}.  If this case can be
excluded experimentally, i.e., opposite signs are observed only for
different enantiomorphs, a phenomenon like natural optical activity
becomes a necessary and sufficient criterion for enantiomorphism.

This aspect of enantiomorphs clearly distinguishes, e.g.,
(anti)magnetochiral domains from ordinary magnetic domains.  It also
allows one to differentiate between opposite enantiomorphs.
Multichiral systems require multiple such probes to distinguish all
four enantiomorphs.  In the following, we focus on optical
nonreciprocities.

\subsubsection{Optical nonreciprocity: General introduction}
\label{sec:exp:chiral-intro}

The reduced symmetry of chiral systems generally allows additional
contributions to the dielectric tensor that give rise to distinctive
optical effects~\cite{lan8e}.  To illustrate most clearly the dual
and triadic relationships between unichiral systems, in the
following, we focus on the phenomenon of \emph{nonreciprocal
directional dichroism} (NDD), i.e., the different rates of
absorption for unpolarized light beams propagating in opposite
directions.  Fundamentally, NDD occurs in materials allowing a
third-rank Cartesian tensor with signature $--$~\cite{bro63}.  A
subset of these materials are crystals permitting a magnetotoroidal
vector $T_1^{--}$ \cite{sza13}.
In the remainder of this section, we focus on the emergence of NDD
in chiral systems.

\subsubsection{Electrochirality}
\label{sec:exp:electro-chiral}

Electrochirality gives rise to \emph{magnetochiral dichroism}
\cite{wag82, bar84, rik97}, which is the version of NDD
characterized by the asymmetric absorption of unpolarized light
parallel and antiparallel to an applied magnetic field.  This
phenomenon is currently attracting great interest~\cite{atz20}.  It
arises because an electrochiral system subject to a magnetic field
or a magnetization has the reduced symmetries required for a nonzero
tensor $T_1^{--}$ \cite{sza13, xu24}.
This effect has equal magnitude but opposite sign for the two
enantiomorphs and can therefore be used to differentiate between
opposite enantiomorphs.

Electrochiral systems are also optically active~\cite{lan8e}, and
they show a linear current-induced magnetization~\cite{ivc78, fur21,
yan25}.  While these effects do not occur exclusively due to
electrochirality~\cite{nye85, bar04, win07a}, the magnitude of these
effects is again equal for the two enantiomorphs, but has the
opposite sign.

\subsubsection{Magnetochirality}
\label{sec:exp:magneto-chiral}

Directional dichroism also arises when a magnetochiral system is
placed into an electric field~\cite{sza13, che21, xu24, hay25}.
This \emph{electrochiral dichroism} is characterized by an asymmetry
in the absorption of light parallel and antiparallel to the applied
electric field (or electric polarization).  It can be considered as
the dual of magnetochiral dichroism discussed above.  Again, the
asymmetry exhibited by the two magnetochiral enantiomorphs is equal
in magnitude but has the opposite sign.

\subsubsection{Antimagnetochirality}
\label{sec:exp:antimagneto-chiral}

In antimagnetochiral materials that allow $T_1^{--}$, NDD occurs
with equal magnitude and opposite sign for the two enantiomorphs, thus
enabling their experimental differentiation~\cite{bro63, sat20}.  Placing an
antimagnetochiral material into an electric field generates the symmetry
conditions for both the Faraday effect and magnetic circular dichroism to
occur~\cite{smo75}, with the associated nonreciprocities again
having opposite sign for the two enantiomorphs~\cite{hay22a}.

\subsubsection{Multichirality}
\label{sec:exp:multi-chiral}

Multichiral enantiomorphs are more profoundly different than
unichiral enantiomorphs.  For physical quantities and effects, it is
generally the \emph{sign} that can be opposite for different chiral
domains.  Therefore, to distinguish between all four enantiomorphs of
a multichiral system, more than one physical quantity or effect must
be considered.
For example, multichiral systems show NDD when placed in a magnetic
or an electric field~\cite{xu24}, thus combining the experimental
signatures of electrochiral and magnetochiral systems.  Based on the
sign of the directional dichroisms exhibited in applied magnetic and
electric fields, the four multichiral enantiomorphs can be uniquely
identified.

\subsection{Chirality and band structure}
\label{sec:chiral:bands}

As discussed in Sec.~\ref{sec:multipoles}, the spin-dependent
electronic band structure $E_\sigma (\kk)$ directly reflects the
categories of polarizations in solids.  These categories manifest
themselves via characteristic terms in the band structure that
transform like irreducible tensors $T_\ell^{ss'}$.  In the
following, we argue that the band structure $E_\sigma (\kk)$
generally reflects the categories of chirality in solids via an
\emph{interplay} of parapolar and unipolar (i.e., electropolar,
magnetopolar, or antimagnetopolar) terms.

The categories of chirality can be defined based on which
monopoles $T_0^{ss'}$ are permitted in a system, see
Table~\ref{tab:categories}.  For example, in electrochiral crystal
structures, the electrotoroidal compound scalar $T_0^{-+}$ becomes
symmetry allowed.  In a band structure, such a scalar is realized by
a term $\vek{\sigma} \cdot \kk$.  The presence of this scalar in a
band structure has thus been identified as a key feature of
electrochiral systems \cite{kis22}.  It can be viewed as a minimal
model of an effective low-energy Hamiltonian for electrochirality in
solids.

In contrast, using the same building blocks $k, \sigma$, the charges
$T_0^{+-}$ and $T_0^{--}$ have no band-structure indicators
associated with them (see Table~\ref{tab:tensor-pow-spher}).  In
magnetochiral systems (fictitious or engineered), we can envision
compound scalars $T_0^{+-} \propto \vek{d} \cdot \kk$ and in
antimagnetochiral systems compound scalars $T_0^{--} \propto
\vek{\sigma} \cdot \vek{d}$, where $\vek{d}$ is an electric dipole
density (signature $-+$) similar to the magnetic dipole density
$\vek{\sigma}$ (signature $+-$) \cite{dub86}.

However, for a crystal structure to be chiral, it is not decisive
whether (and how) chiral charges $T_0^{ss'}$ are actually realized.
Electropolarizations, magnetopolarizations, and
antimagnetopolarizations $T_\ell^{ss'}$ with $\ell > 0$ combined
with parapolarizations $T_\ell^{++}$ with $\ell > 0$ can likewise
make a system chiral.  In the following, we show how tensors
$T_\ell^{ss'}$ with $\ell > 0$ can be used to define minimal models
of effective low-energy Hamiltonians for electrochirality,
magnetochirality and antimagnetochirality for all
magnetic-point-group classes $L (\gprop)$~\cite{misc:allBackground}.

Our theory is illustrated in Table~\ref{tab:chiral-invars}, which lists
the lowest-degree invariant tensors realizing chiralities in the
continuous classes $L(R)$ and $L(D_\infty)$, the cubic classes
$L(O)$ and $L(T)$, and the axial classes $L(D_n)$ and $L(C_n)$ with
$n = 2, 4, 3, 6$~\cite{misc:allBackground}.

\begin{figure*}
  \centering
  \includegraphics[width=\linewidth]{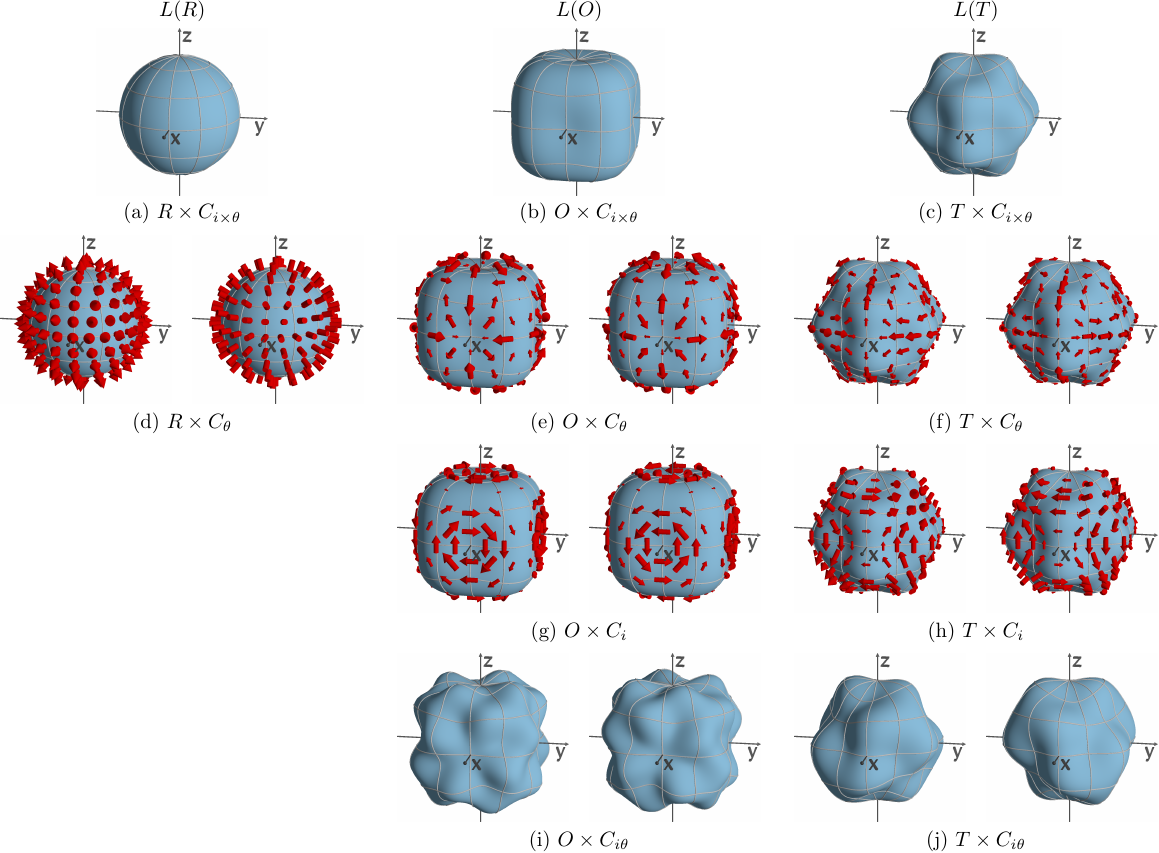}
  \caption{\label{fig:chiral-en-cub} Chirality exhibited in the band
  structure for unipolar-unichiral groups in classes $L(R)$
  (left columns), $L(O)$ (center columns), and $L(T)$ (right
  columns)~\cite{misc:allBackground}.  The first row shows energy
  surfaces of a parachiral system, whereas the second, third and
  fourth rows show energy surfaces for the two enantiomorphs allowed
  under the indicated electrochiral, magnetochiral and antimagnetochiral
  groups.}
\end{figure*}

\begin{figure}
  \centering
  \includegraphics[width=\linewidth]{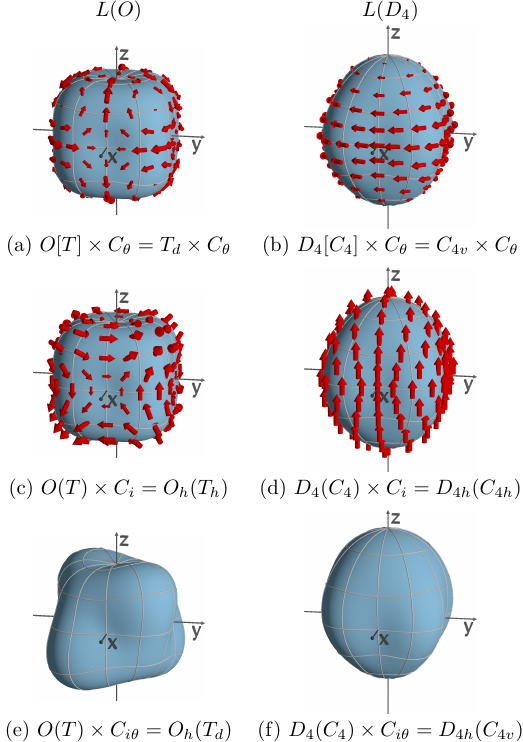}
  \caption{\label{fig:chiral-en-td} Unipolarizations giving rise to
  unichirality in classes $L(T)$ (Fig.~\ref{fig:chiral-en-cub})
  and $L(C_4)$ (Fig.~\ref{fig:chiral-en-d4}) may also exist in
  systems with point groups in classes $L(O)$ and
  $L(D_4)$~\cite{misc:allBackground}, respectively, where these
  terms, as shown here, \emph{do not} give rise to unichirality.}
\end{figure}

\begin{figure*}
  \centering
  \includegraphics[width=\linewidth]{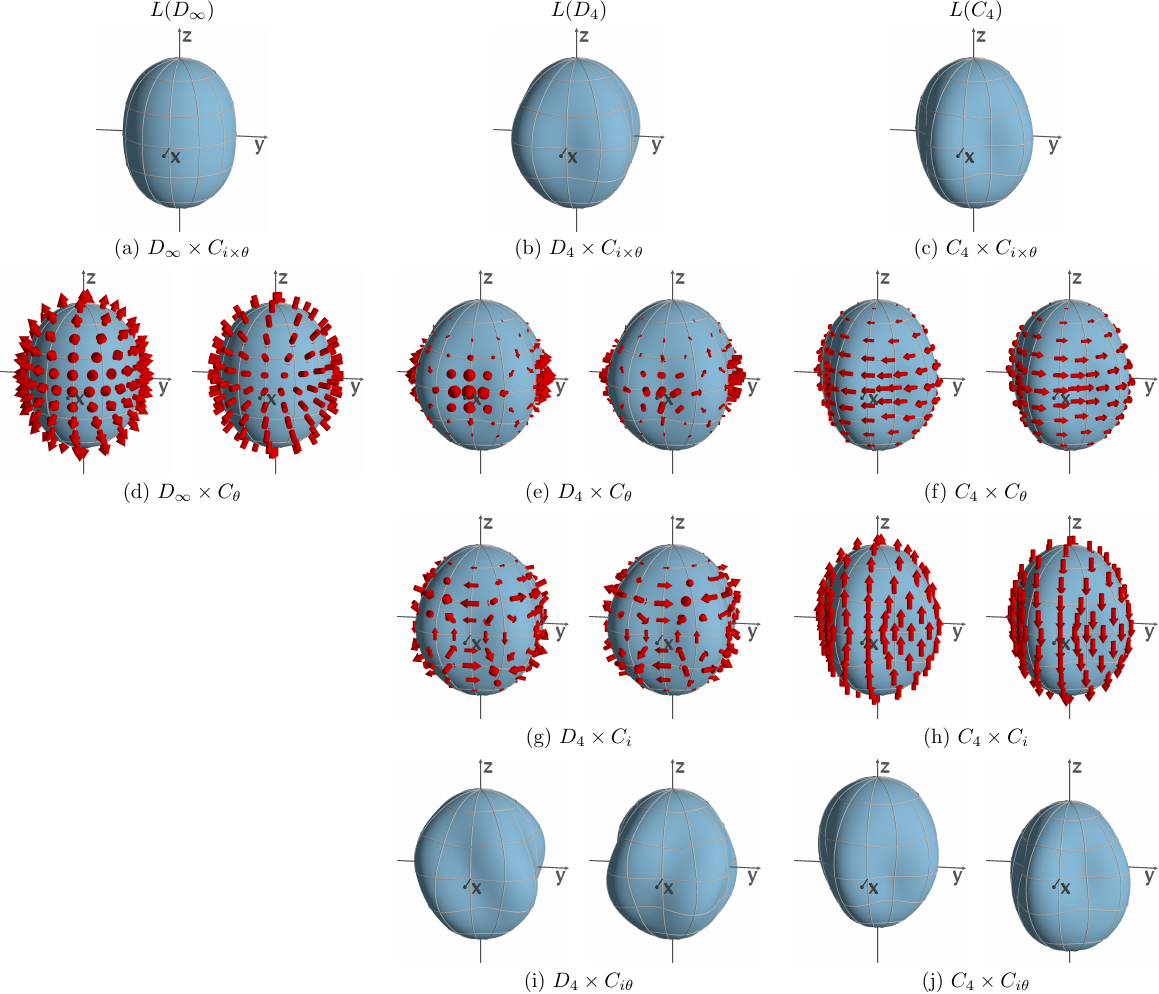}
  \caption{\label{fig:chiral-en-d4} Chirality exhibited in the band
  structure for unipolar-unichiral groups in classes
  $L(D_\infty)$ (left columns), $L(D_4)$ (center columns), and
  $L(C_4)$ (right columns)~\cite{misc:allBackground}.  The first row
  shows energy surfaces of a parachiral system, whereas the second,
  third and fourth rows show energy surfaces for the two enantiomorphs
  allowed under the indicated electrochiral, magnetochiral and
  antimagnetochiral groups.}
\end{figure*}

\subsubsection{Spherical class $L(R)$}
\label{sec:chiral:spher}

The full rotation group $\Rit$ is parachiral.  The only invariant
tensor permitted by this group is $T_0^{++}$ that is realized by the
scalar $k^2$ and functions thereof [Fig~\ref{fig:chiral-en-cub}(a)].
The electrochiral spherical group $R_\theta$ also permits a tensor
$T_0^{-+} \propto \vek{\sigma} \cdot \kk$ (and functions thereof).
The latter term gives rise to two distinct
enantiomorphs~\cite{cha18, win23, gos23}, see
Fig~\ref{fig:chiral-en-cub}(d).
As discussed before, the magnetochiral group $R_i$ and the
antimagnetochiral group $R_{i\theta}$ have no band-structure
indicators associated with them.

\subsubsection{Cubic crystal system [classes $L(O)$ and $L(T)$]}
\label{sec:chiral:cub}

Beyond the isotropic terms already permitted by the groups in the
spherical class $L(R)$, all groups in the cubic class $L(O)$ also
permit an electric hexadecapolarization ($\ell = 4$) that manifests
itself in lowest order via a term \cite{win23}
\begin{equation}
  \label{eq:class-O:e4:H}
  H^\inv{e}{4} \propto
  k_x^2 k_y^2 + k_y^2 k_z^2 + k_z^2 k_x^2 \, .
\end{equation}
This term represents a warping of the energy dispersion $E_\sigma
(\vek{k})$ of band electrons in cubic crystal structures
\cite{kan57, roe84}, see Fig~\ref{fig:chiral-en-cub}(b).

The group $O \times \gtis$ is electrochiral.  In lowest order
$\ell$, it permits an electropolarization with $\ell = 9$ that gives
rise to a term
\begin{align}
  H^\inv{e}{9} \propto {}
      & \sigma_x k_x [k_y^2 k_z^2 (3 k_y^4 - 10 k_y^2 k_z^2 + 3 k_z^4)
      \nonumber \\ &
    +  k_x^2 (k_x^2 - k_y^2 - k_z^2) (k_y^4 - 6 k_y^2 k_z^2 + k_z^4) ]
    + \cp \, ,
  \label{eq:class-O:e9:H}
\end{align}
see Fig.~\ref{fig:chiral-en-cub}(e).  The electrotoroidal scalar
$\vek{\sigma} \cdot \kk$ that is generally likewise present in such
systems can be interpreted as a compound moment density
(\ref{eq:tensor-prod}) due to the combined effect of the
electropolarization (\ref{eq:class-O:e9:H}) and parapolarizations
including the hexadecapolarization (\ref{eq:class-O:e4:H}).

By SIS-TIS duality, the group $O \times \gsis$ is magnetochiral.  In
lowest order $\ell$, it permits a magnetopolarization with $\ell =
9$ that gives rise to a term
\begin{align}
  H^\inv{m}{9} \propto {}
      & \sigma_x k_y k_z (k_y^2-k_z^2) [5k_x^4
    -3k_x^2 (k_y^2 + k_z^2) + k_y^2 k_z^2] \nonumber \\
    & \hspace{1em} + \cp \, ,
  \label{eq:class-O:m9:H}
\end{align}
see Fig.~\ref{fig:chiral-en-cub}(g).  Finally, the group $O \times
\gstis$ is antimagnetochiral.  It permits an antimagnetopolarization
($\ell = 0$) that manifests itself, in lowest order $\ell$, via a
magnetotoroidal density with $\ell = 9$ that gives rise to a term
\begin{align}
  H^\inv{am}{9} \propto {}
      &  k_x k_y k_z (k_y^2-k_z^2) (k_z^2-k_x^2) (k_x^2-k_y^2) \, ,
  \label{eq:class-O:am9:H}
\end{align}
see Fig.~\ref{fig:chiral-en-cub}(i).

Besides class $L(O)$, the cubic crystal system also includes class
$L(T)$.  As $T \subset O$, the polarizations permitted by the
point groups $G_O$ in $L(O)$ are also permitted by the respective
groups $G_T \subset G_O$ in $L(T)$.  Beyond that, the groups in
$L(T)$ permit polarizations not present in $L(O)$, as listed in
Table~\ref{tab:chiral-invars}.  These terms are of lower order than the
terms permitted by $L(O)$.

Specifically, the electrochiral group $T \times \gtis$ permits an
electric octupolarization $\ell = 3$ that manifests itself via a
Dresselhaus term
\begin{equation}
  \label{eq:class-T:e3:H}
  H^\inv{e}{3} \propto
  \sigma_x k_x (k_y^2-k_z^2) + \cp \, .
\end{equation}
As is well known \cite{dre55}, this term is already allowed for
systems with the parachiral point group $T_d \times \gtis = O[T]
\times \gtis$ [Fig.~\ref{fig:chiral-en-td}(a)] realized, e.g., by
the zincblende structure; i.e., by itself, this term is not
sufficient as an indicator for electrochirality in class $L(T)$.
The Dresselhaus term (\ref{eq:class-T:e3:H}) becomes an indicator
for electrochirality [Fig.~\ref{fig:chiral-en-cub}(f)] when combined
with the parapolarization [$\ell = 6$,
Fig.~\ref{fig:chiral-en-cub}(c)]
\begin{equation}
  \label{eq:class-T:e6:H}
  H^\inv{e}{6} \propto
  (k_y^2-k_z^2) (k_z^2-k_x^2) (k_x^2-k_y^2) \, .
\end{equation}
The parapolarization (\ref{eq:class-T:e6:H}) is forbidden by the
groups in class $L(O)$, but it is allowed in class $L(T)$.  A
parapolarization with $\ell = 6$ that is allowed already in class
$L(O)$ is listed in Table~\ref{tab:chiral-invars}.

Similarly, an altermagnetic magnetopolarization ($\ell = 3$)
\begin{equation}
  \label{eq:class-T:m3:H}
  H^\inv{m}{3} \propto \sigma_x k_y k_z + \cp
\end{equation}
may exist in systems with magnetochiral group $T \times \gsis$, but
also in parachiral systems with point group $O(T) \times \gsis$.  The
magnetopolarization (\ref{eq:class-T:m3:H}) becomes an indicator for
magnetochirality when combined with the parapolarization
(\ref{eq:class-T:e6:H}), see Fig.~\ref{fig:chiral-en-cub}(h).
Likewise, an antimagnetopolarization yields a magnetotoroidal
term with $\ell = 3$
\begin{equation}
  \label{eq:class-T:am3:H}
  H^\inv{am}{3} \propto k_x k_y k_z \, .
\end{equation}
This term may exist in systems with antimagnetochiral group $T
\times \gstis$, but also in parachiral systems with point group
$O(T) \times \gstis$.  The term (\ref{eq:class-T:am3:H}) becomes
an indicator for antimagnetochirality when combined with the
parapolarization (\ref{eq:class-T:e6:H}), see
Fig.~\ref{fig:chiral-en-cub}(j).

To summarize, in class $L(T)$, the unipolar terms
(\ref{eq:class-T:e3:H}), (\ref{eq:class-T:m3:H}), and
(\ref{eq:class-T:am3:H}) become indicators of electrochirality,
magnetochirality, and antimagnetochirality, respectively, only when
each of these terms is combined with the parapolarization
(\ref{eq:class-T:e6:H}); i.e., jointly, the terms define minimal
models for effective low-energy Hamiltonians realizing
electrochirality (point group $T \times \gtis$), magnetochirality
($T \times \gsis$), and antimagnetochirality ($T \times \gstis$),
respectively.  This is discussed further in
Sec.~\ref{sec:eff-model-chiral}.

\subsubsection{Axial systems [classes $L(D_n)$ and $L(C_n)$]}
\label{sec:chiral:ax}

Similar to the preceding discussion of the cubic classes $L(O)$ and
$L(T)$, Table~\ref{tab:chiral-invars} also lists the lowest-order
parapolar and unipolar tensors permitted by class $L(D_\infty)$,
as well as classes $L(D_n)$ and $L(C_n)$ with $n = 2, 4, 3, 6$.
Figure~\ref{fig:chiral-en-d4} illustrates the chiral band structure for
$L(D_\infty)$, $L(D_4)$, and $L(C_4)$.  The corresponding set of
plots for $L(D_3)$ is shown in Fig.~\ref{fig:unichiral}.  For the dihedral
classes $L(D_6)$, $L(D_3)$, $L(D_4)$, and $L(D_2)$ unichirality is
realized by combining suitable parapolar terms with unipolar terms
with $\ell = 7$, 3, 5, and 3, respectively.  Note that $D_2 \subset T$,
so class $L(D_2)$ largely reproduces the features of class $L(T)$.

For magnetochiral $\alpha$\=/Fe$_2$O$_3$ and antimagnetochiral
Cr$_2$O$_3$ depicted in Fig.~\ref{fig:unichiral}, the joint presence
of an electric hexadecapolarization and a magnetic (for
$\alpha$\=/Fe$_2$O$_3$) or a magnetotoroidal (for Cr$_2$O$_3$)
octupolarization generates chirality in the band structure.  A
hypothetical electrochiral corundum derivative (second column of
Fig.~\ref{fig:unichiral}) would similarly be characterized by having
both the electric hexadecapolarization and an electric
octupolarization present.

The electropolar cyclic groups $C_n \times \gtis$ permit a
spontaneous electric dipolarization ($\ell = 1$) that manifests
itself via a Rashba term
\begin{equation}
  \label{eq:class-C:e1:H}
  H^\inv{e}{1} \propto
  \sigma_x k_y - \sigma_y k_x \, .
\end{equation}
As is well known, this term is already allowed for systems with
parachiral point groups $C_{nv} \times \gtis = D_n [C_n] \times
\gtis$ realized, e.g., by the wurtzite structure ($C_{6v} \times
\gtis$) \cite{ras59a}.  See also Fig.~\ref{fig:chiral-en-td}(b).
When the Rashba term is combined with a suitable parapolar term
forbidden by class $L(D_n)$, but allowed by class $L(C_n)$
(see Table~\ref{tab:chiral-invars}), jointly these terms describe an
electrochiral system [Fig.~\ref{fig:chiral-en-d4}(f)].

Similarly, the cyclic groups $C_n \times \gsis$ are magnetopolar;
i.e., they permit a spontaneous magnetization ($\ell = 1$) that
manifests itself via an exchange (Zeeman) term
\begin{equation}
  \label{eq:class-C:m1:H}
  H^\inv{m}{1} \propto \sigma_z \, .
\end{equation}
Such a term is already allowed for systems with the parachiral point
groups $D_n (C_n) \times \gsis$, see Fig.~\ref{fig:chiral-en-td}(d).
This term becomes an indicator of magnetochirality when combined
with the respective parapolar terms listed for classes $L(C_n)$
in Table~\ref{tab:chiral-invars} [Fig.~\ref{fig:chiral-en-d4}(h)].

Finally, the cyclic groups $C_n \times \gstis$ are antimagnetopolar,
i.e., they permit a spontaneous antimagnetopolarization that
manifests itself via a magnetotoroidal term ($\ell = 1$)
\begin{equation}
  \label{eq:class-C:am1:H}
  H^\inv{am}{1} \propto k_z \, .
\end{equation}
Again, such a term is already allowed for systems with the
parachiral point groups $D_n (C_n) \times \gstis$, see
Fig.~\ref{fig:chiral-en-td}(f).  This term becomes an indicator of
antimagnetochirality when combined with the respective parapolar
terms listed for classes $L(C_n)$ in
Table~\ref{tab:chiral-invars} [Fig.~\ref{fig:chiral-en-d4}(j)].

\subsubsection{Minimal effective models for chirality}
\label{sec:eff-model-chiral}

For each tabulated tensor, Table~\ref{tab:chiral-invars} lists its
maximal group, i.e., the largest group that leaves the tensor
invariant.  For the odd\=/$\ell$ electropolar, magnetopolar and
antimagnetopolar tensors, the respective maximal group is always an
improper group beyond class $L (\gprop)$, so these tensors,
by themselves, cannot serve as minimal effective models for
chirality in class $L (\gprop)$, as illustrated in
Fig.~\ref{fig:chiral-en-td}.

Quite generally, a proper point group $\gprop$ is an invariant
subgroup of $i$\=/minor groups $\gtprop [\gprop]$
\begin{subequations}
  \begin{equation}
    \gprop \subset \gtprop [\gprop] \; ,
  \end{equation}
  see Appendix~\ref{app:BWform}.  Specifically, we have ($n = 1, 2, 3,
  \ldots$) \arraycolsep 0.2em
  \begin{align}
    T & \subset T_d = O [T] \\
    D_{2n} & \subset D_{(2n)d} = D_{4n} [D_{2n}] \\
    D_{2n+1} & \subset D_{(2n+1)h} = D_{4n+2} [D_{2n+1}] \\
    C_{2n} & \subset \left\{
      \begin{array}{rl}
        S_{4n} & = C_{4n} [C_{2n}] \\
        C_{(2n)v} & = D_{2n} [C_{2n}]
      \end{array} \right. \\
    C_{2n+1} & \subset \left\{
      \begin{array}{rl}
        C_{(2n+1)h} & = C_{4n+2} [C_{2n+1}] \\
        C_{(2n+1)v} & = D_{2n+1} [C_{2n+1}]
      \end{array} \right.
  \end{align}
\end{subequations}
and similarly for the supergroups $\gtprop (\gprop)$ and $\gtprop
\{\gprop\}$ of $\gprop$.  Of course, while a group like $D_4$ is a
crystallographic point group, the supergroup $D_{4d} = D_8 [D_4]
\supset D_4$ is not a crystallographic point group.  Nonetheless,
the improper point group $D_{4d}$ is, ignoring TIS, the maximal
group for the lowest-degree electropolar term permitted by the
crystallographic point group $D_4$, see
Table~\ref{tab:chiral-invars}.  So this electropolar term, by
itself, cannot serve as a minimal effective model for
electrochirality in class $L(D_4)$.

Among the finite point groups, the octahedral group $O$ is special
as it does not appear as a subgroup of a finite group outside class $L(O)$.  Therefore, the unipolar tensors with $\ell = 9$
listed in Table~\ref{tab:chiral-invars} for class $L(O)$
represent, by themselves, minimal effective models for unichirality
in class $L(O)$.

While the maximal groups for the unipolar tensors listed in
Table~\ref{tab:chiral-invars} generally go beyond the respective
class $L(\gprop)$, for each of the parapolar tensors in
Table~\ref{tab:chiral-invars} the maximal group is the full group
$\gprop \times \gsistis$ of the respective class.  Therefore,
\emph{jointly} the unipolar tensors combined with the respective
parapolar tensors represent minimal, effective models for
unichirality in class $L (\gprop)$.  Multichirality is realized
by combining at least two different unipolar tensors in a class $L
(\gprop)$.

The cooperation of a parapolarization and a unipolarization to
realize unichirality can also be recognized in
Fig.~\ref{fig:chiral:c-infty}.  In Fig.~\ref{fig:chiral:c-infty}(e)
we can envision that polar properties have been added to the
parapolar nonchiral object $T_{\ell 0}^{++}$.  Alternatively, we
could imagine that the unipolar nonchiral objects $T_{u 0}^{-+}$,
$T_{u 0}^{+-}$, and $T_{u 0}^{--}$ have acquired a second nonpolar
property, so jointly these properties realize chirality.  These
examples illustrate that we can generally have multiple pathways to
realize chirality.

The multipolar-unichiral case in Fig.~\ref{fig:chiral:dc-infty} is
different from the unipolar-unichiral case in
Figs.~\ref{fig:chiral:c-infty} and~\ref{fig:chiral:d-infty}.  The
objects shown in Fig.~\ref{fig:chiral:dc-infty}(e) visualize the
irreducible tensors $T_{g0}^{++}$ and $T_{u0}^{-+}$, $T_{u0}^{+-}$,
and $T_{u0}^{--}$ as indicated.  While individual objects
$T_{u0}^{ss'}$ in Fig.~\ref{fig:chiral:dc-infty}(e) are unipolar but
not chiral, it is the superposition of their polarizations that
yields unichirality [Figs.~\ref{fig:chiral:dc-infty}(a) to (c)] and
multichirality [Fig.~\ref{fig:chiral:dc-infty}(d)].

This rule applies also to minimal effective models for chirality in
multipolar-unichiral systems.  For example, the
multipolar-electrochiral group $D_4(C_4)$ permits the electropolar term
\begin{align}
  H^\inv{e}{5} \propto {}
      & \sigma_x k_x [k_x^2 (k_y^2 - k_z^2) - k_y^2 (k_y^2 - 3 k_z^2)]
      + (x \leftrightarrow y) \nonumber\\ &
      + \sigma_z k_z (k_x^4 - 6 k_x^2 k_y^2 + k_y^4) \, ,
  \label{eq:class-D4:e5:H}
\end{align}%
[because $D_4 (C_4) \subset D_4 \times \gtis \subset D_8 [D_4]
\times \gtis$, see Table~\ref{tab:chiral-invars} and
Appendix~\ref{app:StructClass}], the magnetopolar term
(\ref{eq:class-C:m1:H}) [because $D_4(C_4) \subset D_4(C_4) \times
\gsis$, see Table~\ref{tab:chiral-invars}], and the antimagnetopolar
term (\ref{eq:class-C:am1:H}) [because $D_4 (C_4) \subset D_4 (C_4)
\times \gstis$].  Any two of these terms are sufficient as a minimal
model for electrochirality under~$D_4(C_4)$ because the largest
subgroup common to any two of the three groups $D_4 \times \gtis$,
$D_4 (C_4) \times \gsis$, and $D_4 (C_4) \times \gstis$ is $D_4
(C_4)$.  Of course, $D_4(C_4)$ permits not only the electropolar
term (\ref{eq:class-D4:e5:H}), but also the parapolar term
\begin{equation}
  \label{eq:class-D4:e4:H}
  H^\inv{e}{4} \propto k_x^2 k_y^2
\end{equation}
because $D_4 (C_4) \subset D_4 \times \gsistis$, see
Table~\ref{tab:chiral-invars}.  Jointly, the terms
(\ref{eq:class-D4:e5:H}) and (\ref{eq:class-D4:e4:H}) can serve not
only as a minimal model for electrochirality under $D_4 \times
\gtis$, but also under $D_4 (C_4) \subset D_4 \times \gtis$.

Using Table~\ref{tab:chiral-invars}, minimal models for other
multipolar-unichiral systems can be constructed in a similar way.
Thus, we have a complete theory for minimal effective models
describing chirality in crystals \cite{misc:theoinv, misc:L-text}.

\subsection{Induced chirality}
\label{sec:chiral:ind}

From a symmetry point of view, an electropolar system with
spontaneous electric dipolarization (\ref{eq:class-C:e1:H}) is
equivalent to a parapolar system placed into an external electric
field.  In parapolar systems with crystallographic point group $C_n
\times \gsistis$, electrochirality can therefore be induced by
placing the system into an external electric field \cite{hli16}.
Similarly, magnetochirality can be induced by placing such a system
into an external magnetic field.  Finally, parallel electric and
magnetic fields make such systems multichiral; i.e., they not only induce
a Rashba term (\ref{eq:class-C:e1:H}) and a Zeeman term
(\ref{eq:class-C:m1:H}), but, jointly, the fields also induce a
magnetotoroidal term (\ref{eq:class-C:am1:H}).

\section{Conclusions and outlook}
\label{sec:outlook}

Our work introduces a standard model of electromagnetism and
chirality in crystals that provides a comprehensive classification
of broken symmetries and the physical properties they give rise to
in all 122 magnetic crystal classes.  The standard model is based on
five categories of electric and magnetic multipole order and five
categories of chirality that yield 12 types of crystal structures
(see Fig.~\ref{fig:standardmodel}), each with distinctive electric
and magnetic properties.  Multiple versions of chirality that were
previously overlooked, including multichirality with four distinct
enantiomorphs, are novel rubrics underlying our complete
classification.  Our work shows explicitly how chirality is linked
to electromagnetic order and band structure.  In particular, we show
how each category of chirality arises from distinct superpositions
of electric and magnetic multipole densities.  We also discuss
band-structure indicators of chirality.  For this purpose, we provide a
complete theory of minimal effective models characterizing the
different categories of chirality in different systems.  These
minimal models demonstrate explicitly how chirality in crystalline
solids requires the interplay of multiple polarizations, either the
interplay of unipolar and parapolar terms, or the interplay of
multiple unipolar terms.

\subsection{Macroscopic description of multipole order}
\label{sec:outlook-macro}

An overarching goal of our work is to provide a general theory of
electric and magnetic order in crystals that is freed from any
reference to microscopic gauge-dependent degrees of freedom.  To achieve this goal, we are generalizing the fundamental ideas developed earlier to
discuss electric~\cite{voi10, res10} and magnetic~\cite{tav58,
cor58a, res10} dipole order to higher-rank multipoles.  The five
categories of multipole order (i.e., polarizations; see columns in
Fig.~\ref{fig:standardmodel}) subsume phenomenology-based materials
classifications and directly reflect distinctive features in the
electronic band structure (Fig.~\ref{fig:polar}).

We present a careful and comprehensive discussion of how compound
multipoles arise in crystals, going beyond previous descriptions
that utilized polarized harmonics~\cite{suz17, wat18, suz19}.
These insights underpin the explanation for how
antimagnetopolarizations (i.e., even\=/$\ell$ magnetic-multipole
densities) are always represented by odd\=/$\ell$
magnetotoroidal terms in the spin-degenerate electronic band
structure.

\subsection{Hidden order parameters}
\label{sec:outlook-hidden}

The electric and magnetic dipole densities $\ell = 1$ (i.e.,
electric dipolarization and magnetization) represent the familiar
order parameters in the thermodynamic description of ferroelectrics
and ferromagnets, respectively~\cite{lan8e}.  Beyond these dipole
densities $\ell = 1$ in ferroelectric and ferromagnetic media,
multipole densities with rank $\ell>1$ can serve as physically
transparent order parameters of hidden orders in crystalline
matter, thus facilitating their experimental detection~\cite{lee18,
pat19, ye24}.  To this end, our work establishes a direct
association of multipole densities with the 122 magnetic crystal
classes, as illustrated in Fig.~\ref{fig:polar}.  Generally,
electric and magnetic multipole densities with odd $\ell >1$
represent alterelectrics and altermagnets, respectively, i.e.,
electric and magnetic orders without a polar direction, while the
even\=/$\ell$ antimagnetopolar (and odd\=/$\ell$ magnetotoroidal)
multipole densities describe antimagnetopolar media.  Application of
methods developed in this work enables the absolute and
gauge-invariant quantification of electric and magnetic multipole
densities of any rank~\cite{win25a}, which is a precondition for any
well-defined order parameter~\cite{rei98, lan5e}.

Using multipole densities as order parameters presents an attractive
alternative to utilizing staggered versions of electric or magnetic
dipoles represented, e.g., by a N\'eel vector~\cite{lan8e, tur62}.
For example, the fundamental physical differences between
magnetopolar antiferromagnets (i.e., altermagnets) and
antimagnetopolar antiferromagnets (e.g., magnetoelectrics) are
naturally captured by the different multipole densities representing
these orders, which also allows for clear association with
macroscopic physical properties (such as the distinctive
band-structure features shown in Fig.~\ref{fig:polar}).  Making such
connections based on N{\'eel-vector constructions is typically less
straightforward~\cite{mcc24, mos25}.

\subsection{Response tensors}
\label{sec:outlook-response}

In the present work, we have focused on \emph{irreducible} tensors,
i.e., we have focused on tensors that transform irreducibly under
point groups, including the spherical rotation groups and the
crystallographic point groups.  These irreducible tensors naturally
lend themselves to the description of crystal \emph{properties} such
as the multipole densities that have been a central topic of the
present work.  We note that Cartesian tensors that are commonly
reducible find widespread use as material tensors that describe
crystal \emph{responses} to external perturbations \cite{nye85, new05}.
Recent works have studied nonequilibrium effects such as
spin-dependent transport~\cite{bal18, dal24}, and light-tailored
multipole order~\cite{hay24, hay25a, yar25}.  A thorough analysis of
such effects needs to consider Onsager's reciprocal relations
accounting for the symmetry of fluctuations under time inversion
$\theta$ when studying irreversible processes \cite{lan5e}.  In the
context of magnetic systems breaking TIS, such an analysis is not
trivial.  For example, following early work \cite{bir64}, the
symmetry of the conductivity tensor describing the anomalous Hall
effect in magnetically ordered systems was discussed intensely till
a consensus was reached.  See Ref.~\cite{gri93}.  A proper
discussion of this important point is beyond the scope of the
present work.  It will be the subject of a future publication.

\subsection{More chirality from general improper rotations}
\label{sec:outlook-chirality}

The present work generalizes the concept of chirality, or
enantiomorphism, to include noninvariance under improper rotations
involving any one (or all three) among the inversions $i$, $\theta$
and $i\theta$, leading to the identification of five categories of
chirality (rows in Fig.~\ref{fig:standardmodel}, see also
Fig.~\ref{fig:unichiral}).  This categorization settles previous
arguments~\cite{mis99, bar04, bar20} about the place of TIS in
definitions of chirality.  We furthermore advance the search for
band-structure indicators of enantiomorphism, presenting minimal
models for each category of chirality realized in specific crystal
classes (see Sec.~\ref{sec:chiral:bands}, especially
Table~\ref{tab:chiral-invars}).  Our findings make it evident that
there is no universal indicator that is both necessary and
sufficient for identifying chirality in all crystal classes.
Rather, each unichiral or multichiral crystal class has its own
interplay between parapolar and unipolar multipole densities that
has its distinct fingerprints in the electronic band structure.  The
often-touted Dirac-like spin splitting term proportional to $\vek{\sigma}
\cdot \kk$ arises in electrochiral systems simply as a remnant of
the electrotoroidal scalar $T_0^{-+}$ having the Dirac term as an
indicator already under the electrochiral rotation group~$R_\theta$.

The newly proposed (anti\=/) magnetochiral and multichiral
categories will be of particular interest for further experimental
exploration.  In Sec.~\ref{sec:exp:magneto-chiral}, we describe the
unusual nonreciprocal effects associated with magnetochiral systems
(e.g., electrochiral dichroism), which have started to be
explored experimentally~\cite{hay25}.  Previous studies of
multichiral systems focused on materials that are also
multiferroic~\cite{xu22}, where effects due to enantiomorphism are
masked by multiferroic responses.  We encourage future experimental
investigations to focus on materials from the crystal classes $D_2$,
$D_4$, $D_3$, $D_6$, $T$ or $O$ [see Table~\ref{tab:materialsMulC}],
as the various nonreciprocities expected from multichirality should
be more pristinely exhibited in these classes that do not support a
polar direction.

\begin{acknowledgments}
  We thank the anonymous reviewers who helped us improve the
  presentation of this work.  We also acknowledge stimulating
  discussions with D.~F.\ Agterberg, S.~Bhowal, P.~M.~R.\ Brydon,
  Y.~Cao, R.~M.\ Fernandes, D.~M.\ Juraschek, Q.~Niu, M.~Norman,
  R.~Resta, C.~P.\ Romao, N.~A.\ Spaldin, and X.~H.\ Verbeek.  R.W.
  also benefited from discussions with B.~Assaf, S.~Bey, T.~Birol,
  L.~Buiarelli, R.~Cavanaugh, A.~Hoffmann, N.~Ghimire, L.~Lurio, and
  J.~Rau.  U.Z. would like to thank S.~Marsland for useful comments
  on an early version of the manuscript.  Work at Argonne was
  supported by Department of Energy, Office of Basic Energy Sciences
  under Contract No.\ DE-AC02-06CH11357.
\end{acknowledgments}

\section*{Data availability}

All data used for the research described in the article have been
included in the article.

\appendix

\section{Composition of magnetic point groups}
\label{app:comp-groups}

The symmetry of a crystal structure is fully characterized by its
(magnetic) space group $S$, which includes point symmetries,
translations, and combinations thereof \cite{bra72}.  However,
according to Neumann's principle~\cite{voi10, nye85, bir74},
macroscopic properties of a crystal structure such as material
tensors and electric and magnetic polarization densities that may be
realized in a structure depend only on the magnetic \emph{point
group} $G$ associated with the space group $S$.  We therefore focus
on point groups $G$ in this work.  For crystal structures
transforming according to a symmorphic space group $S_\mathrm{s}$,
the point group $G$ is the finite subgroup of $S_\mathrm{s}$
consisting of the elements of $S_\mathrm{s}$ that leave one point in
space fixed.  Nonsymmorphic space groups $S_\mathrm{n}$ also contain
group elements that combine point-group symmetries $g$ with
nonprimitive translations.  Here the elements $g$ are also elements
of the point group $G$, although these symmetry operations are not,
by themselves, elements of $S_\mathrm{n}$.  The latter case makes
crystallographic point groups defining crystal classes qualitatively
distinct from point groups of finite systems like molecules.

This appendix reviews the basic structure and properties of magnetic
point groups and introduces a new classification scheme where SIS
and TIS are systematically treated on the same footing.  The
relationships revealed through this classification underpin our
comprehensive discussion of multipole order
(Sec.~\ref{sec:multipoles}) and chirality (Sec.~\ref{sec:chirality})
in crystals.

\subsection{Black-white symmetries and inversion groups}
\label{app:BWsymm}

Rotation symmetries are represented by proper point groups
$\gprop$ that only contain proper rotations as symmetry elements.
In contrast, SIS and TIS are both examples of so-called
\emph{black-white}\ symmetries~\cite{zam57a, lud96, pad20}, which are
represented by two-element groups of the form
\begin{equation}
  \label{eq:BWgroup}
  \ggen = \{ e, \gamma \} \quad .
\end{equation}
Here $e$ is the identity, and $\gamma$ is the single nontrivial
group element that transforms between two qualities, represented
abstractly by ``black'' and ``white,'' satisfying $\gamma^2 = e$.
Space inversion $i$ is such an operation, as is time inversion
$\theta$ (for our systems of interest~\cite{misc:doubleG2}), and
their combination $i\theta$ (CIS).  We use the symbols $\gsis$,
$\gtis$, and $\gstis$ to refer to the inversion groups $\ggen$
defined in Eq.~(\ref{eq:BWgroup}) when $\gamma = i$, $\theta$, and
$i\theta$, respectively.  For completeness and notational
simplicity, we also define the full inversion group $\gsistis\equiv
\gsis \times \gtis = \{ e, i, \theta, i\theta \}$ and the trivial
inversion group $C_1 = \{ e \}$.  These groups are summarized in
Table~\ref{tab:inversion-group}.  In the present work, the symbol
$C_\gin$ stands for any of the five inversion groups $\gsistis$,
$\gtis$, $\gsis$, $\gstis$, or $C_1$.

\subsection{Combining a black-white symmetry with other symmetries}
\label{app:BWform}

There are three possible types of point groups $G$ that arise from
the combination of a black-white symmetry $\gamma$ with other
symmetries.  As we consider multiple black-white symmetries in this
work, we make their definitions specific:

\begin{enumerate}
  \renewcommand{\labelenumi}{(\alph{enumi})}
 \item The $\gamma$\=/\emph{lacking} groups $G^{(\mathrm{a})}$ do
  not contain $\gamma$ at all.

 \item The $\gamma$\=/\emph{major} groups are of the form
  \begin{equation}
    G^{(\mathrm{b})} = G^{(\mathrm{a})} \times \ggen \quad ,
  \end{equation}
  so $\gamma$ is, by itself, an element of $G^{(\mathrm{b})}$.

 \item Suppose a $\gamma$\=/lacking group $G^{(\mathrm{a})}$
  contains an invariant subgroup of index 2 denoted
  $\tilde{G}^{(\mathrm{a})}$, i.e.,
  \begin{subequations}
    \begin{equation}
      G^{(\mathrm{a})} = \tilde{G}^{(\mathrm{a})} + \bar{g} \,
      \tilde{G}^{(\mathrm{a})} \quad ,
    \end{equation}
    where $\bar{g} \in G^{(\mathrm{a})}$, but $\bar{g} \notin
    \tilde{G}^{(\mathrm{a})}$.  Then, the $\gamma$\=/\emph{minor}
    group is defined as
    \begin{equation}
      G^{(\mathrm{c})}
      \equiv G^{(\mathrm{a})} \gmin{\tilde{G}^{(\mathrm{a})}}_\gamma = \tilde{G}^{(\mathrm{a})}
      + \gamma \bar{g} \, \tilde{G}^{(\mathrm{a})}
      \quad .
    \end{equation}
  \end{subequations}
\end{enumerate}
By definition, the major group $G^{(\mathrm{a})} \times \ggen$ contains
the $\gamma$\=/lacking group $G^{(\mathrm{a})}$ and any
minor group $G^{(\mathrm{a})} \gmin{\tilde{G}^{(\mathrm{a})}}_\gamma$ as subgroups;
\begin{subequations}
  \label{eq:allInMaj}
  \begin{align}
    G^{(\mathrm{a})} & \subset G^{(\mathrm{a})} \times \ggen \,\, , \\
    \label{eq:minInMaj}
    G^{(\mathrm{a})} \gmin{\tilde{G}^{(\mathrm{a})}}_\gamma
    & \subset G^{(\mathrm{a})} \times \ggen \,\, .
  \end{align}
\end{subequations}

We adopt the convention established in the context of incorporating TIS,
whereby, for $\gamma = \theta$, the resulting groups $G^{(\mathrm{a})}$,
$G^{(\mathrm{b})}$ and $G^{(\mathrm{c})}$ have been referred to as type
I, type II and type III, respectively~\cite{zam57a, lan8e}, with the
particular type\=/III notation
\begin{subequations}
  \begin{equation}
    \label{eq:gcat:III}
    \gamma = \theta: \qquad
    G (\tilde{G}) \equiv G \gmin{\tilde{G}}_\theta \quad .
  \end{equation}
  Similarly, we label groups $G^{(\mathrm{a})}$, $G^{(\mathrm{b})}$
  and $G^{(\mathrm{c})}$ arising in the process of incorporating
  SIS, i.e., when $\gamma = i$, as type i, type ii and type iii,
  respectively, and denote type\=/iii groups by
  \begin{equation}
    \label{eq:gcat:iii}
    \gamma = i: \qquad
    G [\tilde{G}] \equiv G \gmin{\tilde{G}}_i \quad .
  \end{equation}
  Groups that contain both $i$ and $\theta$ can also be classified
  as major or minor groups with respect to $i\theta$.  We indicate
  $i\theta$\=/minor groups by
  \begin{equation}
    \gamma = i\theta: \qquad
    G \{\tilde{G}\} \equiv G \gmin{\tilde{G}}_{i\theta} \quad .
  \end{equation}
\end{subequations}
See Table~\ref{tab:group:types} for a summary of our nomenclature
and specifics of notation.

\begin{table}[t]
  \caption{\label{tab:group:types} Combining an inversion symmetry
  $\gamma = i$, $\theta$ and $i\theta$ with other symmetries.  The
  group $G$ does not contain $\gamma$ at all.}
  \centering
  \newcommand{\mc}[1]{\multicolumn{2}{c}{#1}}%
  \newcommand{\tms}{\!\times\!}%
  \renewcommand{\arraystretch}{1.2}%
  \begin{tabular*}{1.0\linewidth}{ncs{0.4em}ls{0.3em}C*{3}{Eces{0.4em}C}n}
    \hline\hline \rule{0pt}{2.5ex}
    & & & \mc{SIS: $i$} & \mc{TIS: $\theta$} & \mc{CIS: $i\theta$} \\
    \hline \rule{0pt}{2.5ex}%
    (a) & $\gamma$\=/lacking & G & i & G & I & G & J & G \\
    (b) & $\gamma$\=/major & G\tms\ggen &
    ii & G\tms\gsis & II & G\tms\gtis & JJ & G\tms\gstis \\
    (c) & $\gamma$\=/minor & G \gmin{\tilde{G}}_\gamma &
    iii & G (\tilde{G}) & III & G [\tilde{G}] & JJJ & G \{\tilde{G}\} \\
    \hline\hline
  \end{tabular*}
\end{table}

\subsection{Classes and types of magnetic point groups; duality}
\label{app:StructClass}

Proper point groups $\gprop$ are both $i$\=/lacking and
$\theta$\=/lacking, and thus also $i\theta$\=/lacking.  Combining
these with SIS using the formalism described in
Appendix~\ref{app:BWform} yields the ordinary improper point
groups~\cite{wey52, lud96} that can be either $i$\=/major or
$i$\=/minor.  These groups are still $\theta$\=/lacking and
$i\theta$\=/lacking.  Combining these groups with TIS in the same
way yields the magnetic point groups \cite{tav56}.  In this process
of extending the proper point groups by adding SIS and TIS, each
proper point group $\gprop$ spawns a \emph{class} $L (\gprop)$ of
magnetic point groups $G$~\cite{kop79, kop06, gri91, gri94} defined
via
\begin{equation}
  \label{eq:class-def}
  L (\gprop) = \{ G : G \mapsto \gprop
  \; \mbox{when} \; i, \theta \mapsto e \} \; .
\end{equation}
By definition, each magnetic point group $G$ belongs to one and only
one class $L (\gprop)$.  We call the proper point group $\gprop$ the
\emph{class root} of class $L (\gprop)$.

\begin{table*}[tbp]
  \caption{\label{tab:group-types-def} Classification of magnetic
  point groups based on the SIS-TIS duality
  relation~(\ref{eq:dual-def}).  We give the general form of each
  type and specify the number of groups of a given type among the
  122 crystallographic magnetic point groups.  Duality-related types
  are positioned underneath the self-dual type with which they form
  a triad.  In addition to the short type label, we provide a full
  label indicating the types with respect to SIS (i, ii or iii), TIS
  (I, II or III) and CIS (J, JJ or JJJ).  Small $2\times
  2$ matrices mark congruences between tensors with signature $ss'$
  as discussed in detail in Appendix~\ref{app:tensors}.  The
  categories of multipole order (i.e., polarization) and chirality
  are listed for each type.
  Notation: $\gprop$ denotes a proper point group, $\gtprop$ is an
  index\=/2 subgroup of $\gprop$, superscripts $(j)$ distinguish
  distinct index\=/2 subgroups of $\gprop$, unprimed and primed
  groups denote versions of the same index\=/2 subgroup of $\gprop$
  distinguished by different orientations of the coordinate system
  defining the group elements.  The inversion groups $C_\gin$ are
  defined in Appendix~\ref{app:BWsymm} and listed for convenience in
  Table~\ref{tab:inversion-group}.  See also
  Fig.~\ref{fig:hierarchy}.}
  \renewcommand{\arraystretch}{1.2}%
  \newcommand{\dual}[3][\widthof{electrochiral}]%
    {\makebox[#1][r]{#2}\hspace{0.4em}%
     $\leftrightarrow$\hspace{0.4em}\makebox[#1][l]{#3}}%
  \begin{tabular*}{\linewidth}{nlEs{-1.0em}cs{0.8em}cs{1.0em}cs{0.8em}}
    \hline \hline \rule{0pt}{2.8ex}%
    self-dual type
    & \customlabel{gcat:J}{J} $\equiv$ i-I-J
    & \customlabel{gcat:ii-II}{ii\=/II} $\equiv$ ii-II-JJ
    & \customlabel{gcat:JJJpp}{JJJ''} $\equiv$ iii'-III'-JJJ' \\
    general form
    & $\gprop$ & $\gprop \times \gsistis$
    & $\arraycolsep 0.12em \begin{array}{nrln}
     & \gprop [\gtprop] \,
          (\gtprop^\dprime [\gttprop]) \\
     = & \gprop (\gtprop^\prime) \,
     \{\gtprop (\gttprop)\} \\
     = & \gprop \{\gtprop^\dprime\} \,
      [\gtprop^\prime \{\gttprop\}]
      \footnotemark[1]
      \footnotetext{The multipolar parachiral types can be expressed in
      multiple equivalent ways.  We list only a few of these possibilities.}
    \end{array} $ \\
    allowed tensors &
    \mygroup{1}{1}{1}{1} &
    \mygroup{1}{0}{0}{0} &
    \mygroup{1}{3}{3}{3} \\
    polarization
    & multipolar & parapolar & multipolar \\
    chirality
    & multichiral & parachiral & parachiral \\
    number of groups
    & 11 & 11 & 1 \\
    \hline \hline \rule{0pt}{2.8ex}%
    self-dual type
    & \customlabel{gcat:JJ}{JJ} $\equiv$ i-I-JJ
    & \customlabel{gcat:JJp}{JJ'} $\equiv$ iii-III-JJ
    & \customlabel{gcat:JJJ}{JJJ} $\equiv$ i-I-JJJ \\
    general form
    & $\gprop \times \gstis$
    & $\gprop [\gtprop] \times \gstis
      = \gprop (\gtprop) \times \gstis$
    & $\gprop \{\gtprop\}$ \\
    allowed tensors &
    \mygroup{1}{0}{0}{1} &
    \mygroup{1}{0}{0}{3} &
    \mygroup{1}{3}{3}{1} \\
    polarization
    & antimagnetopolar & antimagnetopolar & multipolar \\
    chirality
    & antimagnetochiral & parachiral & antimagnetochiral \\
    number of groups
    & 11 & 10 & 10 \\
    \hline
    dual types
    & \dual{i-II-J $\equiv$ \customlabel{gcat:i-II}{i\=/II}}
           {\customlabel{gcat:ii-I}{ii\=/I} $\equiv$ ii-I-J}
    & \dual{iii-II-JJJ $\equiv$ \customlabel{gcat:iii-II}{iii\=/II}}
           {\customlabel{gcat:ii-III}{ii\=/III} $\equiv$ ii-III-JJJ}
    & \dual{i-III-J $\equiv$ \customlabel{gcat:i-III}{i\=/III}}
           {\customlabel{gcat:iii-I}{iii\=/I} $\equiv$ iii-I-J} \\
    general form
    & \dual{$\gprop \times \gtis$}
           {$\gprop \times \gsis$}
    & \dual{$\arraycolsep 0.16em
      \begin{array}{rln}
        & \gprop [\gtprop] \times \gtis \\
        = & \gprop \{\gtprop \} \times \gtis
      \end{array} \bigg \} \! $}
      {$\arraycolsep 0.16em \! \bigg\{
      \begin{array}{rl}
        & \gprop (\gtprop) \times \gsis \\
        = & \gprop \{\gtprop\} \times \gsis
      \end{array} $}
    & \dual{$\gprop (\gtprop)$}
           {$\gprop [\gtprop]$} \\
    allowed tensors
    & \dual{\mygroup{1}{0}{1}{0}}{\mygroup{1}{1}{0}{0}}
    & \dual{\mygroup{1}{0}{3}{0}}{\mygroup{1}{3}{0}{0}}
    & \dual{\mygroup{1}{3}{1}{3}}{\mygroup{1}{1}{3}{3}} \\
    polarization
    & \dual{electropolar}{magnetopolar}
    & \dual{electropolar}{magnetopolar}
    & \dual{multipolar}{multipolar} \\
    chirality
    & \dual{electrochiral}{magnetochiral}
    & \dual{parachiral}{parachiral}
    & \dual{electrochiral}{magnetochiral} \\
    number of groups &
    \dual{11}{11} & \dual{10}{10} & \dual{10}{10} \\
    \hline \hline
    \rule{0pt}{2.8ex}self-dual type
    & & \customlabel{gcat:JJJp}{JJJ'} $\equiv$ iii-III-JJJ' \\
    \raisebox{-0.3ex}{general form}
    & & \makebox[0pt]{$\gprop [\gtprop^{(1)}] \, (\gtprop^{(1')} [\gttprop])
     = \gprop (\gtprop^{(1')}) \, [\gtprop^{(1)} (\gttprop)]
     = \gprop \{\gtprop^{(1)}\} \, (\gtprop^{(2)} \{\gttprop\})
     $\footnotemark[1]
     \hspace{-0.50em}} \\
     allowed tensors &
     & \mygroup{1}{5}{5}{3} \\
     polarization &
     & multipolar \\
     chirality &
     & parachiral \\
     number of groups &
     & 2 \\
     \hline
     \rule{0pt}{2.8ex}dual types
     & & \dual{iii'-III-JJJ $\equiv$
                   \customlabel{gcat:iiip-III}{iii'\=/III}}
                  {\customlabel{gcat:iii-IIIp}{iii\=/III'} $\equiv$
                   iii-III'-JJJ} \\
     \raisebox{-0.3ex}{general form} &
     & \dual{$\arraycolsep 0.16em
       \left. \begin{array}{rl}
         \gprop \{\gtprop^{(1)}\} \, [\gtprop^{(1')} \{\gttprop\}] =
       & \gprop [\gtprop^{(1')}] \, \{\gtprop^{(1)} [\gttprop]\} \\
     = & \gprop [\gtprop^{(2)}] \, (\gtprop^{(1)} [\gttprop])
       \footnotemark[1]
       \end{array} \right\}$}
       {$\arraycolsep 0.16em  \left\{
       \begin{array}{cl}
         & \gprop (\gtprop^{(1)}) \, \{\gtprop^{(1')} (\gttprop)\}
         = \gprop \{\gtprop^{(1')}\} \, (\gtprop^{(1)} \{\gttprop\}) \\
         = & \gprop [\gtprop^{(1)}] \, (\gtprop^{(2)} [\gttprop])
         \footnotemark[1]
       \end{array} \right. $} \\
     allowed tensors &
     & \dual{\mygroup{1}{5}{3}{5}}{\mygroup{1}{3}{5}{5}} \\
     polarization &
     & \dual{multipolar}{multipolar} \\
     chirality &
     & \dual{parachiral}{parachiral} \\
     number of groups &
     & \dual{2}{2} \\
      \hline \hline
   \end{tabular*}
\end{table*}

\begin{figure}[t]
  \centering
  \includegraphics[width=1.0\linewidth]{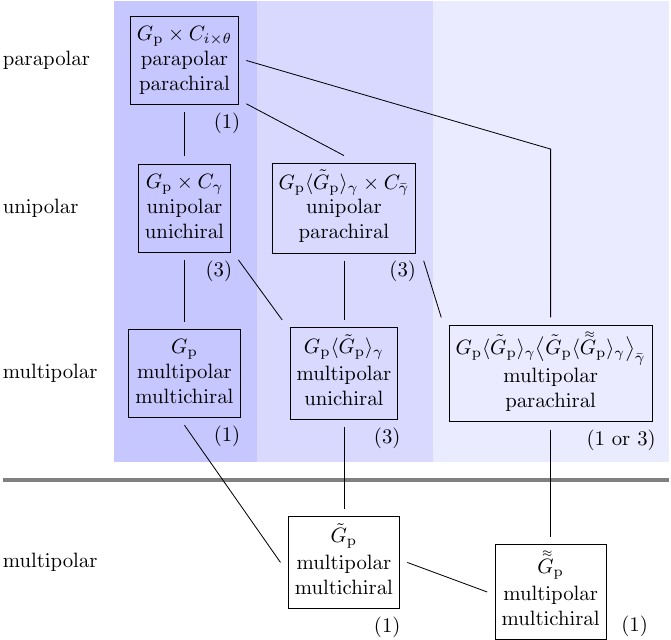}
  \caption[]{\label{fig:hierarchy} Subgroup relations and
  isomorphisms between groups in a given class $L
  (\gprop)$.  Subgroup relations are depicted by lines.  Groups
  placed within the same row are isomorphic.  The categories of
  polarization and chirality for each group are also indicated, with the
  prefix ``uni'' representing ``electro'', ``magneto'', or ``antimagneto''.
  Numbers $n$ at the lower right of a box specify how many groups
  represented by a box can be realized for inversion symmetries
  $\gamma, \bar{\gamma} = i$, $\theta$, and $i \theta$ with
  $\bar{\gamma}\ne\gamma$.  When $n=1$, these groups are
  self-triadic.  When $n=3$, these groups form a triad [Eq.\
  (\ref{eq:triadDef})].  When a line connects groups that depend on
  $\gamma, \bar{\gamma}$, the subgroup relation exists pairwise
  between the groups in the triads with the same $\gamma,
  \bar{\gamma}$.  Every class $L (\gprop)$ contains the groups
  shown in the left (dark-blue) column.  If $\gprop$ has an
  index\=/2 subgroup $\gtprop$, the class also includes
  the groups in the center (medium-blue) column.  If
  $\gtprop$ has an index\=/2 subgroup
  $\gttprop$, the class also includes the groups in
  the right (light-blue) column.  Below the dividing line at the
  bottom, subgroup relations between groups in class $L
  (\gprop)$ and the groups $\gtprop$ and
  $\gttprop$ are shown.  See also
  Table~\ref{tab:group-types-def} that presents the groups that may exist
  in a class $L (\gprop)$ in a more fine-grained way by
  indicating their types.}
\end{figure}

The separate consideration of the two black-white properties $i$ and
$\theta$ yields overlapping group types, i.e., the full set of
magnetic point groups can be viewed in terms of TIS as a collection
of types I, II and III, or seen equivalently~\cite{ope74, kop76} in
terms of SIS as a collection of types i, ii and iii.  Yet another
classification can be based on CIS.  We now develop a scheme that
treats SIS and TIS symmetrically and makes CIS explicit, thus
providing a systematic and complete classification of the magnetic
point groups.  This approach also reveals a general structure
underlying the composition of class $L (\gprop)$ of
magnetic point groups with class root $\gprop$.

It has previously been observed~\cite{ope74} that exchanging $i
\leftrightarrow \theta$ in all group elements of a magnetic point
group $G$ yields another magnetic point group $G_\mathrm{d}$.
Clearly, the groups $G$ and $G_\mathrm{d}$ are isomorphic.  We call
this relationship \emph{SIS-TIS duality} or $i \leftrightarrow
\theta$ \emph{duality}
\begin{equation}
  \label{eq:dual-def}
  G \wlra G_\mathrm{d}
\end{equation}
with \emph{dual partners} $G$ and $G_\mathrm{d}$.  It is easy to see
that dual partners $G$ and $G_\mathrm{d}$ belong to the same class
$L (\gprop)$, i.e., both $G$ and $G_\mathrm{d}$ satisfy Eq.\
(\ref{eq:class-def}) with the same class root $\gprop$.  The
concept of duality allows one to classify the groups in a given
class $L (\gprop)$ as follows.

\emph{Self-dual} groups $G = G_\mathrm{d} \in L (\gprop)$ are
invariant under exchange $i \leftrightarrow \theta$.  They belong to
one (and only one) of the seven types \ref{gcat:J}, \ref{gcat:ii-II},
\ref{gcat:JJ}, \ref{gcat:JJp}, \ref{gcat:JJJ}, \ref{gcat:JJJp}, or
\ref{gcat:JJJpp} defined in Table~\ref{tab:group-types-def}.
Besides the self-dual types, we have types associated with dual pairs
$G \wlra G_\mathrm{d}$ with $G \ne G_\mathrm{d}$ and $G,
G_\mathrm{d} \in L (\gprop)$.  We label these types by indicating
their respective types under both SIS and TIS.  See Table~\ref{tab:group-types-def}
for the explicit definitions.  In the following, we generally use short
labels when referring to one of the new types, but an instructive full
label indicating their types under SIS, TIS and CIS is also given in
Table~\ref{tab:group-types-def}.

Within each class, the groups of the types \ref{gcat:J},
\ref{gcat:JJJ}, \ref{gcat:JJJp}, \ref{gcat:JJJpp}, \ref{gcat:i-III},
\ref{gcat:iii-I}, \ref{gcat:iiip-III}, and \ref{gcat:iii-IIIp} are
all isomorphic.  Furthermore, the groups of the types \ref{gcat:JJ},
\ref{gcat:JJp}, \ref{gcat:i-II}, \ref{gcat:ii-I}, \ref{gcat:ii-III},
and \ref{gcat:iii-II} are all isomorphic.  A group of type
\ref{gcat:ii-II} is not isomorphic to any other group in the same
class but contains all of these as its subgroups.  See
Fig.~\ref{fig:hierarchy} for a description of general isomorphisms
and subgroup relations between magnetic-point-group types in a
given class.  Isomorphisms formed the basis of a prior
classification scheme of the 122 magnetic crystallographic
groups~\cite{kop79, kop06}.

Table~\ref{tab:group-types-def} provides comprehensive information
about all magnetic-point-group types, including associated categories of
multipole order and chirality.  These latter properties have been elucidated
in Secs.~\ref{sec:multipoles} and \ref{sec:chirality}.  As is already apparent
from Table~\ref{tab:group-types-def}, the 15 indicated group types map
onto particular combinations of a category of polarization with a category
of chirality.  For 11 types, this mapping is unique, i.e., for each of these,
the group type is synonymous with a distinct combination of polarization
with chirality.  The exception is the multipolar-parachiral combination that
subsumes four (rare sub)types \ref{gcat:JJJpp}, \ref{gcat:JJJp},
\ref{gcat:iiip-III} and \ref{gcat:iii-IIIp}.  Thus, Table~\ref{tab:group-types-def}
and Fig.~\ref{fig:hierarchy} constitute an alternative representation of
the standard model illustrated by Fig.~\ref{fig:standardmodel}.

The present classification treats $\theta$ as a black-white symmetry
like $i$ so that $i^2 = \theta^2 = e$.  This classification applies
to states with integer spin angular momentum (transforming according
to the single groups where $\theta^2 = e$), and it is relevant for
all macroscopic properties of a crystal structure such as those
characterized by Neumann's principle~\cite{voi10, nye85,
misc:doubleG2}.  The single-group description is also sufficient for
obtaining the invariant expansion of the electronic band structure
near $k=0$~\cite{bir74} as in Fig.~\ref{fig:polar}.  This is
illustrated by the fact that all building blocks in Eq.\
(\ref{eq:base-dipole}) for forming invariants in a spinful theory
transform according to integer IRs of $\Rit$ with $\ell=1$.  This
includes the operator of spin angular momentum $\vek{\sigma}$, Eq.\
(\ref{eq:spin-dipole}).  There are other properties of systems with
half-integer spin degrees of freedom that can only be properly
discussed using double groups where $\theta^2 = -e$~\cite{lud96}.
For example, double groups are needed to understand certain
degeneracies in the electronic band structure that may arise at the
Brillouin zone boundary~\cite{bir74}.  The typology and isomorphisms
discussed here do not hold for the respective double groups.

\begin{table*}[tbp]
  \caption{\label{tab:subclasses} Subclasses $L (G_0^\gamma, \gamma)$
  and groups in a class $L (\gprop)$.  Here $\gtprop$ denotes
  an index\=/2 subgroup of the class root $\gprop$.  The groups in $L
  (\gprop)$ are arranged in the upper part of the table
  according to the scheme shown in the center part of the table,
  where each group's type is indicated.   For
  clarity, we omit groups of multipolar-parachiral type
  (i.e., \ref{gcat:JJJpp}, \ref{gcat:JJJp}, \ref{gcat:iiip-III}, and
  \ref{gcat:iii-IIIp}).  The subclass $L (G_0^\theta, \theta)$ for each
  group is specified in the header of its column.  For groups of
  type \ref{gcat:JJ} or \ref{gcat:JJp}, their subclass $L (G_0^i, i)$
  is also given in the column header.  Groups of all other types
  have their subclass $L (G_0^i, i)$ specified in the header of their
  respective row.  Small $2\times
  2$ matrices mark congruences between tensors with signature $ss'$
  as discussed in detail in Appendix~\ref{app:tensors}.  The bottom part of
  the table lists the groups in the subclasses $L (G_0^{i\theta},
  i\theta) \equiv L (G_{0\mathrm{d}}^{i\theta}, i\theta)$.}
  \centering
  \newcommand{\mb}[2]{\makebox[0pt]{\makebox[\widthof{ii-III}]{#1}%
    \hspace{0.5em}\makebox[\widthof{magnetopolar}][l]{#2}}}%
  \newcommand{\nb}[2]{\makebox[0pt]{\makebox[\widthof{ii-III}]{#1}%
    \hspace{0.5em}\makebox[\widthof{antimagnetochiral}][l]{#2}}}%
  \newcommand{\mkb}[1]{\makebox[16em][l]{$#1$}}%
  \renewcommand{\arraystretch}{1.2}%
  \newcolumntype{q}{@{\hspace*{2.0em}}}%
  \begin{tabular*}{\linewidth}{nL*{3}{ELS{1.0em}C}q|S{1.5em}LS{1.0em}Cn}
    \hline \hline \rule{0pt}{3.8ex}
    \makebox[0pt][l]{$L (G_0^i, i)\, \backslash\, L (G_0^\theta, \theta)$}
    & \multicolumn{2}{C}{L (\gprop, \theta)}
    & \multicolumn{2}{C}{L (\gprop [\gtprop], \theta)}
    & \multicolumn{2}{Cq|}{L (\gprop \times \gsis, \theta)}
    & \multicolumn{2}{cn}{\tbox[1.2ex]{nLn}{%
      L (\gprop \times \gsis, \theta), \\
      L (\gprop \times \gtis, i)}}
    \\ \hline \rule{0pt}{2.8ex}
    L (\gprop, i) &
    \gprop & \mygroup{1}{1}{1}{1} &
    \gprop [\gtprop] & \mygroup{1}{1}{3}{3} &
    \gprop \times \gsis & \mygroup{1}{1}{0}{0} &
    \gprop \times \gstis & \mygroup{1}{0}{0}{1} \\
    L (\gprop (\gtprop), i) &
    \gprop (\gtprop) & \mygroup{1}{3}{1}{3} &
    \gprop \{ \gtprop \} & \mygroup{1}{3}{3}{1} &
    \gprop (\gtprop) \times \gsis & \mygroup{1}{3}{0}{0} &
    \gprop (\gtprop) \times \gstis & \mygroup{1}{0}{0}{3} \\
    L (\gprop \times \gtis, i) &
    \gprop \times \gtis & \mygroup{1}{0}{1}{0} &
    \gprop [\gtprop] \times \gtis & \mygroup{1}{0}{3}{0} &
    \gprop \times \gsistis & \mygroup{1}{0}{0}{0} &
    & \\
    \hline
    \rule{0pt}{2.8ex} &
    \multicolumn{2}{c}{\mb{\ref{gcat:J}}{\tbox{nln}{multipolar \\ multichiral}}} &
    \multicolumn{2}{c}{\nb{\ref{gcat:iii-I}}{\tbox{nln}{multipolar \\ magnetochiral}}} &
    \multicolumn{2}{cq|}{\mb{\ref{gcat:ii-I}}{\tbox{nln}{magnetopolar \\ magnetochiral}}} &
    \multicolumn{2}{Ern}{\makebox[0pt][r]{\makebox[1.6em][l]{\ref{gcat:JJ}}%
    \tbox{nln}{antimagnetopolar \\ antimagnetochiral}}} \\
    \multicolumn{1}{c}{} &
    \multicolumn{2}{c}{\mb{\ref{gcat:i-III}}{\tbox{nln}{multipolar \\ electrochiral}}} &
    \multicolumn{2}{c}{\nb{\ref{gcat:JJJ}}{\tbox{nln}{multipolar \\ antimagnetochiral}}} &
    \multicolumn{2}{cq|}{\mb{\ref{gcat:ii-III}}{\tbox{nln}{magnetopolar \\ parachiral}}} &
    \multicolumn{2}{Ern}{\makebox[0pt][r]{\makebox[1.6em][l]{\ref{gcat:JJp}}%
    \tbox{nln}{antimagnetopolar \\ parachiral}\hspace{0.2em}}} \\
    \multicolumn{1}{c}{} &
    \multicolumn{2}{c}{\mb{\ref{gcat:i-II}}{\tbox{nln}{electropolar \\ electrochiral}}} &
    \multicolumn{2}{c}{\nb{\ref{gcat:iii-II}}{\tbox{nln}{electropolar \\ parachiral}}} &
    \multicolumn{2}{cq|}{\mb{\ref{gcat:ii-II}}{\tbox{nln}{parapolar \\ parachiral}}} &
    \multicolumn{2}{c}{} \\
    \hline \hline
    \multicolumn{9}{nLn}{%
    \rule{0pt}{3.0ex}%
    \mkb{L (\gprop, i\theta)}
    \gprop, \;
    \gprop \{ \gtprop \}, \;
    \gprop \times \gstis} \\
    \multicolumn{9}{nLn}{%
    \mkb{L (\gprop (\gtprop), i\theta) \,\, \equiv
    L (\gprop [\gtprop], i\theta)}
    \gprop (\gtprop), \;
    \gprop [\gtprop], \;
    \gprop (\gtprop) \times \gstis \equiv
    \gprop [\gtprop] \times \gstis \hspace*{-12em}} \\
    \multicolumn{9}{nLn}{%
    \mkb{L (\gprop \times \gtis, i\theta) \equiv
    L (\gprop \times \gsis, i\theta)}
    \gprop \times \gtis, \;
    \gprop \times \gsis, \;
    \gprop \times \gsistis, \;
    \gprop [\gtprop] \times \gtis, \;
    \gprop (\gtprop) \times \gsis \hspace*{-12em}}
    \\ \hline \hline
  \end{tabular*}
\end{table*}

\subsection{Triadic relationships}
\label{app:triads}

The SIS-TIS duality (\ref{eq:dual-def}) appears naturally when
magnetic point groups are constructed in the usual way from proper
point groups by adding the black-white symmetries space inversion
$i$ and time inversion $\theta$ \cite{wey52, tav56} as reviewed at
the beginning of Appendix~\ref{app:StructClass}.  The duality
(\ref{eq:dual-def}) turns out to be useful to discuss physical
phenomena involving electric and magnetic fields, and we make
use of it throughout this work.  Here we briefly discuss the deeper
mathematical structure from which this duality emerges.

The full inversion group $\gsistis$ has order 4.  It is isomorphic to the
Klein four-group
\begin{equation}
  K_4 = \{e, \gamma', \gamma'', \tilde{\gamma} = \gamma' \gamma''
  : \gamma^{\prime\, 2} = \gamma^{\prime\prime\, 2} = \tilde{\gamma}^2 = e \}
\end{equation}
in abstract group theory \cite{lud96}.  This group treats the
elements $\gamma', \gamma''$, and $\tilde{\gamma} = \gamma' \gamma''$ fully
symmetrically.  Indeed, one can express the full inversion group
$\gsistis$ in three equivalent forms
\begin{equation}
  \gsistis = \gsis \times \gtis
  \equiv \gtis \times \gstis
  \equiv \gstis \times \gsis \,\, .
\end{equation}
Starting from a proper point group $\gprop$, it is thus
possible to construct all magnetic point groups in class
$L (\gprop)$ using any pair of inversion symmetries $\gamma',
\gamma'' \in \{ i, \theta, i\theta \}$ by first adding the
black-white symmetry $\gamma'$ and then adding $\gamma''$.
Therefore, the concept of duality exists for any pair of inversion
symmetries $\gamma', \gamma'' \in \{i, \theta, i\theta\}$.  The
resulting dualities are represented by the diagram
\begin{equation}
  \label{eq:triadDef}
  \unitlength 0.24ex
  \newcommand{\myarrow}{\begin{picture}(20,3)
    \put(0,0){\vector(1,0){53}}
    \put(10,0){\vector(-1,0){10}}
  \end{picture}}%
  \raisebox{-10ex}{\begin{picture}(90,80)
    \put(0,7){$G_\theta$}
    \put(80,7){$G_i$}
    \put(39,73){$G_{i\theta}$}
    \put(20,11){\myarrow}
    \put(30,0){$\theta \wlra i$}
    \put(10,22){\rotatebox{60}{\myarrow}}
    \put(66,22){\rotatebox{120}{\myarrow}}
    \put(61,59){\rotatebox{300}{$i\theta \wlra i$}}
    \put(8,32){\rotatebox{60}{$\theta \wlra i\theta$}}
  \end{picture}}
\end{equation}
with \emph{triadic partners} $G_i$, $G_\theta$, and $G_{i\theta}$
that are pairwise connected by duality relations similar to
Eq.\ (\ref{eq:dual-def}).

Triadic partners $G_i$, $G_\theta$, and $G_{i\theta}$ are isomorphic.
A set of triadic partners $G_i$, $G_\theta$, and $G_{i\theta}$
constitutes a \emph{triad} of magnetic groups.  Possibilities for
forming triads are shown in Table~\ref{tab:group-types-def}.  (See also
Fig.~\ref{fig:standardmodel} for an illustration.)
\emph{Self-triadic} groups are invariant under all dualities in the
diagram (\ref{eq:triadDef}).  They belong to the types \ref{gcat:J},
\ref{gcat:ii-II}, or \ref{gcat:JJJpp}.  The remaining groups form
triads with distinct partners $G_i$, $G_\theta$, and $G_{i\theta}$
as indicated in Table~\ref{tab:group-types-def}.  The index $\tilde{\gamma}
= \gamma' \gamma''$ of a group $G_{\tilde{\gamma}}$ in the diagram
(\ref{eq:triadDef}) indicates self-duality under the exchange
$\gamma' \leftrightarrow \gamma''$.  The equivalent expressions
provided in Table~\ref{tab:group-types-def} for groups \ref{gcat:JJJpp},
\ref{gcat:iii-II}, \ref{gcat:ii-III}, \ref{gcat:JJp},
\ref{gcat:iiip-III}, \ref{gcat:iii-IIIp}, and \ref{gcat:JJJp},
emphasize the respective self-dualities of these groups.

\subsection{Subclasses of magnetic point groups}
\label{app:subclasses}

To classify the tensors permitted by different groups $G$ in a class
$L (\gprop)$, the concept of subclasses of magnetic point groups
turns out to be useful.  Here we discuss the construction of subclasses,
details of their composition, and associated properties.

For $\gamma = i$, $\theta$, and $i \theta$, the \emph{subclasses}
$L (G_0^\gamma, \gamma)$ of a class $L (\gprop)$ are the subsets
of groups $G$
\begin{equation}
  \label{eq:subclass-def}
  L (G_0^\gamma, \gamma) = \{ G :
  G \mapsto G_0^\gamma \; \mbox{when} \; \gamma \mapsto e \} \; .
\end{equation}
For given $\gamma$, the subclasses $L (G_0^\gamma, \gamma)$ of
$L (\gprop)$ are disjoint, i.e., each group $G \in L (\gprop)$
belongs to one subclass $L (G_0^i, i)$, one subclass $L (G_0^\theta,
\theta)$, and one subclass $L (G_0^{i\theta}, i\theta)$.  The subclasses
derived for $\gamma = i$, $\theta$, and $i \theta$ hence represent
alternative schemes for partitioning $L (\gprop)$.  Conversely, for
a given class $L (\gprop)$ and fixed $\gamma = i$, $\theta$, or
$i \theta$, the subclasses $L (G_0^\gamma, \gamma)$ spawn the class
$L (\gprop)$, i.e.,
\begin{equation}
  \label{eq:jointSC}
  L (\gprop) = \bigcup_{G_0^\gamma} L (G_0^\gamma, \gamma) \quad
  \mbox{for fixed $\gamma = i$, $\theta$ or $i \theta$.}
\end{equation}

The \emph{subclass root} $G_0^\gamma$ of the subclass
$L (G_0^\gamma, \gamma)$ is $\gamma$\=/lacking by construction.  If
$G_0^\gamma = \gprop$ the subclass root $G_0^\gamma$ is actually
lacking all three inversions.  The associated subclasses $L (\gprop,
\gamma)$ for $\gamma = i$, $\theta$, and $i \theta$ comprise
$\gprop$, $\gprop \times \ggen$ and all $\gamma$\=/minor
groups $\gprop\gmin{\gtprop}_\gamma$.

If the subclass root $G_0^\gamma$ still contains another black-white
symmetry, it has to be different from $\gamma$.  For the cases $\gamma
= i$ or $\theta$, the black-white symmetry present in $G_0^\gamma$
will be
\begin{equation}
  \label{eq:gprDef}
  \gamma_\mathrm{d} = \left\{ \begin{array}{cl}
      i & \mbox{if $\gamma=\theta$} \\[0.1cm]
      \theta & \mbox{if $\gamma=i$} \; .
    \end{array} \right.
\end{equation}
The case $\gamma = i\theta$ requires a more elaborate discussion that is
relegated to Appendix~\ref{app:subclass:i-theta}.  Here we continue with a
brief consideration of the cases $\gamma = i$ or $\theta$.

If $G_0^\gamma$ is a $\gamma_\mathrm{d}$\=/major group, then
$G_0^\gamma = \gprop \times \ggend$.  For $\gamma = i$ or
$\theta$, the subclass $L (\gprop \times \ggend, \gamma)$ contains
all the $\gamma_\mathrm{d}$\=/major groups in class
$L (\gprop)$, as well as the type\=/\ref{gcat:JJ} and
type\=/\ref{gcat:JJp} groups in $L (\gprop)$.

If $G_0^\gamma$ is a $\gamma_\mathrm{d}$\=/minor group, then
$G_0^\gamma = \gprop \gmin{\gtprop}_{\gamma_\mathrm{d}}$.  For
$\gamma = i$ or $\theta$,
$L (\gprop \gmin{\gtprop}_{\gamma_\mathrm{d}}, \gamma)$ contains
$\gprop \gmin{\gtprop}_{\gamma_\mathrm{d}}$,
$\gprop \gmin{\gtprop}_{\gamma_\mathrm{d}} \times \ggen$ and
$\gprop \{ \gtprop \}$.  The
subclass $L (\gprop [\gtprop^{(1)}], \theta)$
furthermore contains $G_\mathrm{\ref{gcat:JJJp}} = \gprop
[\gtprop^{(1)}] \, (\gtprop^{(1')}
[\gttprop])$ (or the type-\ref{gcat:JJJpp} group of the
same form) and $G_\mathrm{\ref{gcat:iii-IIIp}} = \gprop
[\gtprop^{(1)}] \, (\gtprop^{(2)}
[\gttprop])$.  Similarly, $L (\gprop
(\gtprop^{(1)}), i)$ contains $G_\mathrm{\ref{gcat:JJJp}}$
(or the type-\ref{gcat:JJJpp} group of the same form) and
$G_\mathrm{\ref{gcat:iiip-III}} = \gprop (\gtprop^{(1)})
\, [\gtprop^{(2)} (\gttprop)]$.

The relation between group types and subclasses is summarized in
Table~\ref{tab:subclasses}.  Groups in a given row in the upper part
of this table belong to the same subclass $L (G_0^i, i)$, with $G_0^i$
being the group in the first column.  Similarly, groups in a given column
belong to the same subclass $L (G_0^\theta, \theta)$, with $G_0^\theta$
being the group in the first row.  The groups of type \ref{gcat:JJ} and
\ref{gcat:JJp} belong to the same subclasses $L (G_0^i, i)$ and
$L (G_0^\theta, \theta)$ as the group of type \ref{gcat:ii-II}.  Similarly,
the groups of type \ref{gcat:JJJp} or \ref{gcat:JJJpp} belong to the same
subclasses $L (G_0^i, i)$ and $L (G_0^\theta, \theta)$ as the group of
type \ref{gcat:JJJ}.  The composition of subclasses $L (G_0^{i\theta},
i\theta)$ is discussed in Appendix~\ref{app:subclass:i-theta}
and summarized in the lower part of Table~\ref{tab:subclasses}.

Application of the duality transformation $i \leftrightarrow \theta$
to the groups in a subclass $L (G_0^\gamma, \gamma)$ with $\gamma =
i$ or $\theta$ generates the subclass $L (G_{0\mathrm{d}}^{\gamma},
\gamma_\mathrm{d})$, with the subclass roots $G_0^{\gamma} \wlra
G_{0\mathrm{d}}^{\gamma}$ being dual partners.  This corresponds to
mapping rows onto columns and \textit{vice versa} in the upper part
of Table~\ref{tab:subclasses}.  In contrast, every subclass $L
(G_0^{i\theta}, i\theta) \equiv L (G_{0\mathrm{d}}^{i\theta},
i\theta)$ is mapped onto itself when the duality transformation is
applied to the groups it comprises.  Therefore, the subclass
classifications $L (G_0^\gamma, \gamma)$ do not reflect a triadic
relationship (\ref{eq:triadDef}).  More generally, for each pair
$\gamma', \gamma''$ of inversion symmetries, one can establish a
pair of $\gamma' \leftrightarrow \gamma''$-dual subclasses $L
(G_0^{\gamma'}, \gamma')$ and $L (G_0^{\gamma''}, \gamma'')$, as
well as the $\gamma' \leftrightarrow \gamma''$-self-dual subclass $L
(G_0^{\tilde{\gamma}}, \tilde{\gamma})$ with $\tilde{\gamma} =
\gamma' \gamma''$.  In this work, we adhere to the nomenclature
whereby the term duality without a specifier refers to SIS-TIS
duality associated with the exchange $i \leftrightarrow \theta$.

\subsection{Form of subclasses $\boldsymbol{L (G_0^{i\theta},
i\theta)}$}
\label{app:subclass:i-theta}

The concept of subclasses of magnetic point groups is introduced in
Appendix~\ref{app:subclasses}.  While the form of subclasses $L (G_0^i,
i)$ and $L (G_0^\theta, \theta)$ is straightforward, the form of
subclasses $L (G_0^{i\theta}, i\theta)$ discussed in the following
is more subtle.

Every magnetic point group $G\in L (\gprop)$ has the general
form~\cite{kop76}
\begin{equation}
  \label{eq:Gstruct}
  G = \gtprop + i \{ g_\nu^{(i)} \} + \theta \{
  g_\nu^{(\theta)} \} + i\theta \{ g_\nu^{(i\theta)} \} \,\, .
\end{equation}
Here $\gtprop \subseteq \gprop$ is a proper point
group, and $\{ g_\nu^{(\gamma)} \}$ denotes the set of operations
in the class root $\gprop$ that appears in $G$ in
combination with the black-white symmetry element $\gamma\in \{ i,
\theta, i\theta \}$.  In the following, we assume $G \supsetneq
\gprop$, i.e., at least one set $\{ g_\nu^{(\gamma)}
\}$ is nonempty.  The relation $\gamma^2 = e$ and general group
properties of $G$ imply the relations
\begin{subequations}
  \label{eq:Gelem}
  \begin{align}
    g_\nu^{(\gamma)} g_\mu^{(\gamma)} &\in \gtprop
    && \mbox{for any fixed $\gamma$} \,\, , \\ \label{eq:Gb}
    g_\nu^{(\gamma)} g_\mu^{(\gamma')} &\in \{ g_{\nu'}^{(\gamma\gamma')}
    \} && \mbox{for any pair $\gamma\ne\gamma'$} \,\, .
  \end{align}
\end{subequations}
It follows from Eq.\ (\ref{eq:Gb}) that either only one set $\{
g_\nu^{(\gamma)} \}$ is nonempty, or all three sets are nonempty.

To understand how the mapping $i\theta \mapsto e$ transforms the group
$G$ given in Eq.~(\ref{eq:Gstruct}), we first note
\begin{subequations}
  \begin{equation}
    \label{eq:eqiv1}
    \theta = i^2\, \theta = i\, i\theta
    \; \stackrel{i\theta \mapsto e}{\longmapsto} \; i
  \end{equation}
  to find
  \begin{equation}
    \label{eq:root1}
    G \; \stackrel{i\theta \mapsto e}{\longmapsto} \;
    G^{i\theta}_0 \equiv \gtprop + \{ g_\nu^{(i\theta)} \}
    + i\big( \{ g_\nu^{(i)} \} + \{ g_\nu^{(\theta)} \} \big) \, ,
  \end{equation}
\end{subequations}
where clearly $G^{i\theta}_0$ is again a group.
Alternatively, using
\begin{subequations}
  \begin{equation}
    \label{eq:eqiv2}
    i = i\, \theta^2 = i\theta\, \theta
    \; \stackrel{i\theta \mapsto e}{\longmapsto} \; \theta
  \end{equation}
  yields
  \begin{equation}
    \label{eq:root2}
    G \; \stackrel{i\theta \mapsto e}{\longmapsto} \; G_{0\mathrm{d}}^{i\theta}
    \equiv \gtprop + \{ g_\nu^{(i\theta)} \}
    + \theta \big( \{ g_\nu^{(i)} \} + \{ g_\nu^{(\theta)} \} \big) \,\, .
  \end{equation}
\end{subequations}
Thus we can identify two equivalent subclass roots $G_0^{i\theta} \leftrightarrow
G_{0\mathrm{d}}^{i\theta}$ that are related via the duality transformation
$i \leftrightarrow \theta$, and
\begin{equation}
  L (G_0^{i\theta}, i\theta) \equiv L (G_{0\mathrm{d}}^{i\theta}, i\theta) \; .
\end{equation}

Applying the duality transformation $i \leftrightarrow \theta$ to $G$
in Eq.\ (\ref{eq:Gstruct}) yields the dual partner
\begin{equation}
  \label{eq:GdualStruct}
  G \wlra G_\mathrm{d} = \gtprop + \theta \{ g_\nu^{(i)} \} +
  i \{ g_\nu^{(\theta)} \} + i\theta \{ g_\nu^{(i\theta)} \} \,\, .
\end{equation}
Equations (\ref{eq:eqiv1}) and (\ref{eq:eqiv2}) show that
$G_\mathrm{d}$ maps to the same subclass roots $G^{i\theta}_0
\leftrightarrow G_{0\mathrm{d}}^{i\theta}$ as $G$, i.e., dual
partners $G$ and $G_\mathrm{d}$ belong to the same subclass
$L (G_0^{i\theta}, i\theta) \equiv L (G_{0\mathrm{d}}^{i\theta},
i\theta)$.

Specializing the composition of a subclass $L (G_0^\gamma, \gamma)$
discussed generally in Appendix~\ref{app:subclasses} to the case
$\gamma = i\theta$, we find that $L (\gprop, i\theta)$ contains
not only $\gprop$ but also $G_\mathrm{\ref{gcat:JJ}} = \gprop \times
\gstis$ and all the type\=/\ref{gcat:JJJ} groups $\gprop \{\gtprop\}
\in L (\gprop)$.  It follows from Eqs.\ (\ref{eq:eqiv1}) and
(\ref{eq:eqiv2}) that the subclass $L (\gprop \times \gsis, i\theta)
\equiv L (\gprop \times \gtis, i\theta)$ contains the groups of type
\ref{gcat:ii-II}, \ref{gcat:ii-I} and \ref{gcat:i-II} as well as all
type-\ref{gcat:ii-III} and type-\ref{gcat:iii-II} groups in $L
(\gprop)$.  Groups in a subclass $L (\gprop [\gtprop], i\theta)
\equiv L (\gprop (\gtprop), i\theta)$ contain not only $\gprop
[\gtprop]$ and $\gprop (\gtprop)$, but also the type\=/\ref{gcat:JJp}
group $\gprop (\gtprop) \times \gstis$.  Furthermore,
$G_\mathrm{\ref{gcat:JJJp}} = \gprop [\gtprop^{(1)}] \,
(\gtprop^{(1')} [\gttprop])$ is in $L (\gprop [\gtprop^{(2)}],
i\theta) \equiv L (\gprop (\gtprop^{(2)}), i\theta)$, while
$G_\mathrm{\ref{gcat:iii-IIIp}} = \gprop [\gtprop^{(1)}] \,
(\gtprop^{(2)} [\gttprop])$ and $G_\mathrm{\ref{gcat:iiip-III}}=
\gprop [\gtprop^{(2)}] \, (\gtprop^{(1)} [\gttprop])$ both belong to
$L (\gprop [\gtprop^{(1)}], i\theta) \equiv L (\gprop
(\gtprop^{(1)}), i\theta)$.

The combination of $L (\gprop, i\theta)$ with $L (\gprop
 \times \gsis, i\theta) \equiv L (\gprop \times \gtis, i\theta)$
and all subclasses $L (\gprop [\gtprop], i\theta)
\equiv L (\gprop (\gtprop), i\theta)$ yields class $L (\gprop)$ [Eq.\ (\ref{eq:jointSC})].

\section{Classification of magnetic point groups}
\label{app:class-groups}

Having developed the classification of magnetic point groups within
a given class, we now apply this classification to the known
continuous and discrete point groups.

\begin{table}[b]
  \caption{\label{tab:spher:trafo} Spherical magnetic point groups
  $G_\mathrm{sp}$.  Symmetry elements present (absent) in different
  groups $G_\mathrm{sp}$ are labeled ``$\bullet$'' (``$\circ$'').
  Here $C$ represents an arbitrary proper rotation.  Labels in the
  first column refer to the classification of magnetic point groups
  introduced in Table~\ref{tab:group-types-def}.  The last column
  gives invariant tensors for these groups expressed in terms of
  irreducible tensors $T_0^{ss'} = \{ T_{00}^{ss'} \}$ of $\Rit$
  (see Sec.~\ref{sec:spherTens}) \cite{misc:char}.  A superscript
  ``$\any$'' indicates that invariant tensors may have any
  signature~$ss'$.}
  \centering
  \newcommand{\mb}[1]{\makebox[1.2em]{$#1$}}%
  \newcommand{\mc}[1]{\multicolumn{2}{l}{#1}}%
  \let\blt\bullet
  \let\tms\circ
  \renewcommand{\arraystretch}{1.2}%
  \begin{tabular*}{\linewidth}{ncs{0.6em}Ls{0.2em}LE*{7}{C}s{1em}Ln}
    \hline \hline
    \rule{0pt}{3.0ex} & \multicolumn{2}{C}{G_\mathrm{sp}}
    & \mb{C} & \mb{i} & \mb{i C}
    & \mb{\theta} & \mb{\theta C} & \mb{i\theta} & \mb{i\theta C}
    & T_0^{ss'} \\
    \hline \rule{0pt}{2.7ex}
    \ref{gcat:ii-II} & \Rit & = R \times \gsistis
    & \blt & \blt & \blt & \blt & \blt  & \blt & \blt & T_0^{++} \\
    \hline \rule{0pt}{2.7ex}
    \ref{gcat:i-II} & R_\theta & = R \times \gtis
    & \blt & \tms & \tms & \blt & \blt  & \tms & \tms & T_0^{-+} \\
    \ref{gcat:ii-I} & R_i & = R \times \gsis
    & \blt & \blt & \blt & \tms & \tms  & \tms & \tms & T_0^{+-} \\
    \ref{gcat:JJ} & R_{i \theta} & = R \times \gstis
    & \blt & \tms & \tms & \tms & \tms  & \blt & \blt & T_0^{--} \\
    \hline \rule{0pt}{2.7ex}
    \ref{gcat:J} & R & = R \times C_1 \equiv R
    & \blt & \tms & \tms & \tms & \tms  & \tms & \tms & T_0^\any \\
    \hline  \hline
  \end{tabular*}
\end{table}

\begin{table}[b]
  \caption{\label{tab:spher:point-group} Class $L(R)$ of spherical
  magnetic point groups arranged according to the scheme in
  Table~\ref{tab:subclasses} and indicated at the bottom part of the
  table.  See Table~\ref{tab:group-types-def} for the definitions of group
  types.  The groups of type \ref{gcat:J}, \ref{gcat:ii-I},
  \ref{gcat:i-II} and \ref{gcat:ii-II} delineate the main block of
  groups in class $L(R)$, where the duality transformation
  $i\leftrightarrow \theta$ maps rows onto columns and \textit{vice
  versa}.  The type-\ref{gcat:JJ} group outside the main block is
  self-dual, and so are the groups positioned on the main-block
  diagonal.  Small $2\times
  2$ matrices mark congruences between tensors with signature $ss'$
  as discussed in detail in Appendix~\ref{app:tensors}.}
  \centering
  \newcommand{\mb}[1]{\makebox[2.3em][l]{#1}}%
  \renewcommand{\arraystretch}{1.2}%
  \newcolumntype{Y}{LS{0.8em}C}%
  \begin{tabular*}{\linewidth}{nYEYs{2.0em}|S{2.0em}Yn}
    \hline \hline
    R & \mygroup{1}{1}{1}{1} &
    R \times \gsis & \mygroup{1}{1}{0}{0} &
    R \times \gstis & \mygroup{1}{0}{0}{1} \\
    R \times \gtis & \mygroup{1}{0}{1}{0} &
    R \times \gsistis & \mygroup{1}{0}{0}{0} \\
    \hline \hline
      \multicolumn{1}{c}{\mb{\ref{gcat:J}}} &
    & \multicolumn{1}{c}{\mb{\ref{gcat:ii-I}}} &
    & \multicolumn{1}{c}{\mb{\ref{gcat:JJ}}} \\
      \multicolumn{1}{c}{\mb{\ref{gcat:i-II}}} &
    & \multicolumn{1}{c}{\mb{\ref{gcat:ii-II}}} & \\
    \hline \hline
   \end{tabular*}
\end{table}

\subsection{Continuous spherical groups}
\label{app:spher-groups}

We have one spherical point group $R \equiv SO(3)$ of proper
rotations $C$.  By adding $\gamma$\=/\emph{improper} rotations
$\gamma C$ with $\gamma = i$, $\theta$ or $i\theta$, we obtain the
spherical magnetic groups $G_\mathrm{sp}$ listed in
Table~\ref{tab:spher:trafo}.  These form class $L(R)$.  As $R$
lacks an index\=/2 subgroup, $L(R)$ contains only groups of type
\ref{gcat:J}, \ref{gcat:ii-II}, \ref{gcat:JJ}, \ref{gcat:ii-I} and
\ref{gcat:i-II}.  See Table~\ref{tab:spher:point-group}.
We call $\Rit \equiv R \times \gsistis$ the full rotation group.
All continuous and discrete groups discussed in this work are
subgroups of $\Rit$.

\begin{table*}[tbp]
  \caption{\label{tab:c-cyclic:trafo} Continuous cyclic magnetic
  groups $G_{c\infty}$.  Symmetry elements present (absent) in
  different groups $G_{c\infty}$ are labeled ``$\bullet$''
  (``$\circ$'').  Here $C_z$ represents an arbitrary rotation about
  the fixed $z$ axis.  Labels in the first column refer to the
  classification of magnetic point groups introduced in
  Table~\ref{tab:group-types-def}.  The last column gives invariant
  tensors for these groups expressed in terms of irreducible tensor
  components $T_{\ell 0}^{ss'}$ of $\Rit$ (see
  Sec.~\ref{sec:spherTens} and Appendix~\ref{app:axial-groups})
  \cite{misc:char}.}
  \newcommand{\mb}[1]{\makebox[1.8em]{$#1$}}%
  \newcommand{\mc}[1]{\multicolumn{2}{l}{#1}}%
  \let\blt\bullet \let\tms\circ
  \renewcommand{\arraystretch}{1.2}%
  \begin{tabular*}{\linewidth}{ncs{1em}Ls{0.2em}Ls{1em}E*{7}{C}s{1em}Ln}
    \hline \hline
    \rule{0pt}{3.0ex} & \multicolumn{2}{C}{G_{c\infty}}
    & \mb{C_z} & \mb{i} & \mb{i C_z}
    & \mb{\theta} & \mb{\theta C_z} & \mb{i\theta} & \mb{i\theta C_z}
    & T_{\ell 0}^{ss'} \\
    \hline \rule{0pt}{2.7ex}
    \ref{gcat:ii-II}
    & C_\infty \times \gsistis & = C_{\infty h} \times \gtis
    & \blt & \blt & \blt & \blt & \blt & \blt & \blt
    & T_{\ell 0}^{++} \\
    \hline \rule{0pt}{2.7ex}
    \ref{gcat:i-II} & C_\infty \times \gtis &
    & \blt & \tms & \tms & \blt & \blt & \tms & \tms
    & T_{\ell 0}^{-+} \\
    \ref{gcat:ii-I} & C_\infty \times \gsis & = C_{\infty h}
    & \blt & \blt & \blt & \tms & \tms & \tms & \tms
    & T_{\ell 0}^{+-} \\
    \ref{gcat:JJ} & C_\infty \times \gstis & = C_{\infty h} (C_\infty)
    & \blt & \tms & \tms & \tms & \tms & \blt & \blt
    & T_{\ell 0}^{--} \\
    \hline \rule{0pt}{2.7ex}
    \ref{gcat:J} & C_\infty &
    & \blt & \tms & \tms & \tms & \tms & \tms & \tms
    & T_{\ell 0}^\any \\ \hline \hline
  \end{tabular*}
\end{table*}

\begin{table*}[tbp]
  \caption{\label{tab:c-dihedral:trafo} Continuous dihedral magnetic
  groups $G_{d\infty}$.  Symmetry elements present (absent) in
  different groups $G_{d\infty}$ are labeled ``$\bullet$''
  (``$\circ$'').  Here $C_z$ represents an arbitrary rotation about
  the fixed $z$ axis, and $C_2$ is a $\pi$ rotation about an axis
  perpendicular to the $z$ axis.  Labels in the first column refer
  to the classification of magnetic point groups introduced in
  Table~\ref{tab:group-types-def}.  The last two columns give
  invariant tensors for these groups expressed in terms of
  irreducible tensor components $T_{\ell 0}^{ss'}$ of $\Rit$ (see
  Sec.~\ref{sec:spherTens} and Appendix~\ref{app:axial-groups})
  \cite{misc:char}.  Here $T_{g0}^{ss'}$ stands for any $T_{\ell
  0}^{ss'}$ with $\ell$ even, while $T_{u0}^{ss'}$ can be any
  $T_{\ell 0}^{ss'}$ with $\ell$ odd.  The component $T_{u0}^{++}$ is
  also included for completeness, even though it is not an invariant
  quantity for any continuous dihedral group~$G_{d\infty}$.}
  \newcommand{\mb}[1]{\makebox[1.8em]{$#1$}}%
  \newcommand{\mc}[1]{\multicolumn{2}{l}{#1}}%
  \let\blt\bullet \let\tms\circ
  \renewcommand{\arraystretch}{1.2}%
  \begin{tabular*}{\linewidth}{ncs{1em}Ls{0.2em}Ls{1.1em}E*{11}{C}s{1em}LLn}
    \hline \hline
    \rule{0pt}{3.0ex} & \multicolumn{2}{C}{G_{d\infty}}
    & \mb{C_z} & \mb{i} & \mb{i C_z}
    & \mb{C_2} & \mb{i C_2}
    & \mb{\theta} & \mb{\theta C_z} & \mb{i\theta} & \mb{i\theta C_z}
    & \mb{\theta C_2} & \mb{i\theta C_2}
    & \multicolumn{2}{C}{T_{\ell 0}^{ss'}} \\
    \hline \rule{0pt}{2.7ex}
    \ref{gcat:ii-II} & D_\infty \times \gsistis & = D_{\infty h} \times \gtis
    & \blt & \blt & \blt & \blt & \blt & \blt & \blt & \blt & \blt & \blt & \blt
    & T_{g0}^{++} & \\
    \hline \rule{0pt}{2.7ex}
    \ref{gcat:i-II} & D_\infty \times \gtis &
    & \blt & \tms & \tms & \blt & \tms & \blt & \blt & \tms & \tms & \blt & \tms
    & T_{g0}^{-+} & \\
    \ref{gcat:ii-I} & D_\infty \times \gsis & = D_{\infty h}
    & \blt & \blt & \blt & \blt & \blt & \tms & \tms & \tms & \tms & \tms & \tms
    & T_{g0}^{+-} & \\
    \ref{gcat:JJ} & D_\infty \times \gstis & = D_{\infty h} (D_\infty)
    & \blt & \tms & \tms & \blt & \tms & \tms & \tms & \blt & \blt & \tms & \blt
    & T_{g0}^{--} & \\
    & &
    & \blt & \blt & \blt & \tms & \tms & \blt & \blt & \blt & \blt & \tms & \tms
    & & T_{u0}^{++} \\
    \ref{gcat:iii-II} & D_\infty [C_{\infty}] \times \gtis
    & = C_{\infty v} \times \gtis
    & \blt & \tms & \tms & \tms & \blt & \blt & \blt & \tms & \tms & \tms & \blt
    & & T_{u0}^{-+} \\
    \ref{gcat:ii-III} & D_\infty (C_\infty) \times \gsis
    & = D_{\infty h} (C_{\infty h})
    & \blt & \blt & \blt & \tms & \tms & \tms & \tms & \tms & \tms & \blt & \blt
    & & T_{u0}^{+-} \\
    \ref{gcat:JJp} & D_\infty (C_\infty) \times \gstis
    & = D_{\infty h} (C_{\infty v})
    & \blt & \tms & \tms & \tms & \blt & \tms & \tms & \blt & \blt & \blt & \tms
    & & T_{u0}^{--} \\ \hline \rule{0pt}{2.7ex}
    \ref{gcat:iii-I} & D_\infty [C_{\infty}] & = C_{\infty v}
    & \blt & \tms & \tms & \tms & \blt & \tms & \tms & \tms & \tms & \tms & \tms
    & T_{g0}^{+-} & T_{u0}^{-\pm} \\
    \ref{gcat:i-III} & D_\infty (C_\infty) &
    & \blt & \tms & \tms & \tms & \tms & \tms & \tms & \tms & \tms & \blt & \tms
    & T_{g0}^{-+} & T_{u0}^{\pm-} \\
    \ref{gcat:JJJ} & D_\infty \{ C_\infty\} & = C_{\infty v} (C_\infty)
    & \blt & \tms & \tms & \tms & \tms & \tms & \tms & \tms & \tms & \tms & \blt
    & T_{g0}^{--} & T_{u0}^{\mp\pm} \\
    \hline \rule{0pt}{2.7ex}
    \ref{gcat:J} & D_\infty &
    & \blt & \tms & \tms & \blt & \tms & \tms & \tms & \tms & \tms & \tms & \tms
    & T_{g0}^\any & \\ \hline \hline
  \end{tabular*}
\end{table*}

\begin{table*}[tbp]
  \caption{\label{tab:ax:point-group} Continuous axial magnetic
  point groups organized into classes.  Within each class, the
  arrangement of groups follows the scheme in
  Table~\ref{tab:subclasses} and indicated in the bottom part of the
  table.  See Table~\ref{tab:group-types-def} for the definitions of group
  types.  The groups of type \ref{gcat:J}, \ref{gcat:ii-I},
  \ref{gcat:i-II} and \ref{gcat:ii-II} delineate the main block of
  groups in a given class, where the duality transformation
  $i\leftrightarrow \theta$ maps rows onto columns and \textit{vice
  versa}.  The type-\ref{gcat:JJ} and type-\ref{gcat:JJp} groups
  outside the main block are self-dual, and so are the groups
  positioned on a main-block diagonal.  Small $2\times
  2$ matrices mark congruences between tensors with signature $ss'$
  as discussed in detail in Appendix~\ref{app:tensors}.  A group with exactly
  one filled symbol, pertaining to signature $ss'$ represents the
  maximal group of the vector component $T_{10}^{ss'}$ (see
  Sec.~\ref{sec:spherTens}).}
  \centering
  \newcommand{\mb}[1]{\makebox[2.3em][l]{#1}}%
  \renewcommand{\arraystretch}{1.2}%
  \begin{tabular*}{\linewidth}{nLS{0.5em}C*{2}{ELS{0.2em}LS{0.5em}C}<{\hspace{1em}}|ELS{0.2em}LS{0.5em}C}
    \hline \hline
    C_\infty & \mygroup{2}{2}{2}{2} & & & &
    C_{\infty h} & = C_\infty \times \gsis & \mygroup{2}{2}{0}{0} &
    C_{\infty h} (C_\infty) & = C_\infty \times \gstis
    & \mygroup{2}{0}{0}{2} \\
    C_\infty \times \gtis & \mygroup{2}{0}{2}{0} & & & &
    C_{\infty h} \times \gtis & = C_\infty \times \gsistis
    & \mygroup{2}{0}{0}{0} \\
    \hline
    D_\infty & \mygroup{1}{1}{1}{1} &
    C_{\infty v} & = D_\infty [C_{\infty}] & \mygroup{1}{1}{4}{4} &
    D_{\infty h} & = D_\infty \times \gsis & \mygroup{1}{1}{0}{0} &
    D_{\infty h} (D_\infty) & = D_\infty \times \gstis
    & \mygroup{1}{0}{0}{1} \\
    D_\infty (C_\infty) & \mygroup{1}{4}{1}{4} &
    C_{\infty v} (C_\infty) & = D_\infty \{C_\infty\} & \mygroup{1}{4}{4}{1} &
    D_{\infty h} (C_{\infty h}) & = D_\infty (C_\infty) \times \gsis
    & \mygroup{1}{4}{0}{0} &
    D_{\infty h} (C_{\infty v}) & = D_\infty (C_\infty) \times \gstis
    & \mygroup{1}{0}{0}{4} \\
    D_\infty \times \gtis & \mygroup{1}{0}{1}{0} &
    C_{\infty v} \times \gtis & = D_\infty [C_{\infty}] \times \gtis
    & \mygroup{1}{0}{4}{0} &
    D_{\infty h} \times \gtis & = D_\infty \times \gsistis
    & \mygroup{1}{0}{0}{0} \\
    \hline \hline
      \multicolumn{2}{c}{\mb{\ref{gcat:J}}}
    & \multicolumn{3}{c}{\mb{\ref{gcat:iii-I}}}
    & \multicolumn{2}{c}{\mb{\ref{gcat:ii-I}}} &
    & \multicolumn{2}{Ec}{\mb{\ref{gcat:JJ}}} \\
      \multicolumn{2}{c}{\mb{\ref{gcat:i-III}}}
    & \multicolumn{3}{c}{\mb{\ref{gcat:JJJ}}}
    & \multicolumn{2}{c}{\mb{\ref{gcat:ii-III}}} &
    & \multicolumn{2}{Ec}{\mb{\ref{gcat:JJp}}} \\
      \multicolumn{2}{c}{\mb{\ref{gcat:i-II}}}
    & \multicolumn{3}{c}{\mb{\ref{gcat:iii-II}}}
    & \multicolumn{2}{c}{\mb{\ref{gcat:ii-II}}} & \\
    \hline \hline
  \end{tabular*}
\end{table*}

\subsection{Continuous axial groups}
\label{app:axial-groups}

There exist two proper continuous axial point groups, the cyclic
group $C_\infty \equiv SO(2)$ and the dihedral group $D_\infty$.  By
adding SIS and TIS to these groups, one obtains the cyclic magnetic
groups $G_{c\infty}$ listed in Table~\ref{tab:c-cyclic:trafo} [class
$L (C_\infty)$] and the dihedral magnetic groups $G_{d\infty}$
listed in Table~\ref{tab:c-dihedral:trafo} [class $L (D_\infty)$].
We call $D_\infty \times \gsistis$ the full axial rotation group.
See also Refs.~\cite{tav61, hli14, lit15, jan17, che21} for related
work on axial point groups.

Like $R$, the group $C_\infty$ lacks an index\=/2 subgroup, so
the structure of class $L(C_\infty)$ mirrors that of $L(R)$.
See the upper part of Table~\ref{tab:ax:point-group}.  In contrast,
group $D_\infty$ has the index\=/2 subgroup $C_\infty$.
Class $L(D_\infty)$ thus has a richer composition that includes
groups of type \ref{gcat:i-III}, \ref{gcat:JJJ}, \ref{gcat:ii-III}
and \ref{gcat:JJp}, i.e., the types indicated in the middle row of
the scheme shown in the bottom part of
Table~\ref{tab:ax:point-group}.

The continuous axial groups are the maximal groups of many important
physical quantities, see Table~\ref{tab:spher:phys-example}.
Therefore, in the following, we review in more detail the IRs of the
continuous axial groups that we refer to repeatedly in this work.
The cyclic groups $G_{c\infty}$ are Abelian so that all their IRs
are one-dimensional.  Under these groups, the transformation
behavior of tensor components $T_{\ell m}^{ss'}$ of $\Rit$ depends
only on $m$ but not on $\ell$; for fixed $m$ but different values of
$\ell$ (with $|m| \le \ell$), the components $T_{\ell m}^{ss'}$
transform the same way under groups $G_{c\infty}$.  The invariant
tensors of groups $G_{c\infty}$ are then the tensor components
$T_{\ell 0}^{ss'}$ (Table~\ref{tab:c-cyclic:trafo}).  Similar to
$T_{00}^{++}$ under the spherical groups, tensor components $T_{\ell
0}^{++}$ are invariant under all cyclic groups $G_{c\infty}$.

The dihedral groups $G_{d\infty}$ are non-Abelian.  Under groups
$G_{d\infty}$, the tensor components $T_{\ell 0}^{ss'}$ of $\Rit$
transform according to one-dimensional IRs, whereas tensor
components $T_{\ell, \pm m}^{ss'}$ with $m \ne 0$ pairwise transform
according to two-dimensional IRs.  More specifically, the
transformation properties depend on the parity of $\ell$.  For even
$\ell$ (``$g$''), tensor components $T_{g 0}^{ss'}$ are even under
$\pi$ rotations $C_2$ about an axis perpendicular to the main axis
of the dihedral groups (chosen as the $z$ axis in
Table~\ref{tab:c-dihedral:trafo}), whereas for odd $\ell$ (``$u$''),
tensor components $T_{u 0}^{ss'}$ are odd under rotations $C_2$
\cite{misc:char}.  The invariant tensor components $T_{\ell
0}^{ss'}$ under the dihedral groups $G_{d\infty}$ thus distinguish
between $\ell$ even or odd, see Table~\ref{tab:c-dihedral:trafo}.
But similar to tensor components $T_{\ell 0}^{ss'}$ under the cyclic
groups, tensor components with different even or different odd
values of $\ell$ behave the same.

The cyclic groups in Table~\ref{tab:c-cyclic:trafo} are subgroups of
the dihedral groups in Table~\ref{tab:c-dihedral:trafo}.  The
maximal groups of tensor components $T_{\ell 0}^{ss'}$ are thus
generally the dihedral groups that provide a more fine-grained
classification of the symmetries of $T_{\ell 0}^{ss'}$ based on the
parity of $\ell$, see Table~\ref{tab:c-dihedral:trafo}.  As noted
above, tensor components $T_{u 0}^{ss'}$ change sign under $\pi$
rotations $C_2$ about an axis perpendicular to the $z$ axis
\cite{misc:char}.  The transformation behavior of these tensor
components under the full axial group $D_\infty \times \gsistis$ as
listed in Table~\ref{tab:c-dihedral:trafo} thus defines \emph{polar
directions} with signature $ss'$.  For the three polar directions
$T_{u 0}^{-+}$, $T_{u 0}^{+-}$, and $T_{u 0}^{--}$ the respective
maximal groups are listed in Table~\ref{tab:c-dihedral:trafo}.

Interestingly, inspection of Table~\ref{tab:c-dihedral:trafo} shows
that an electrotoroidal tensor component $T_{u0}^{++}$ is not
invariant under any of the continuous dihedral groups.  It is only
invariant under continuous cyclic groups $G_{c\infty}$, including
$C_\infty \times \gsistis$, that do not distinguish between tensor
components $T_{u0}^{++}$ and electric tensor components
$T_{g0}^{++}$.  This fact has the remarkable consequence that, very
generally, one cannot define a setting that would allow one to
distinguish between electrotoroidal tensor components $T_{u0}^{++}$
and electric tensor components $T_{g0}^{++}$.  Beyond the fact that
tensors $T_u^{++}$ have no suitable building blocks in
Table~\ref{tab:tensor-pow-spher}, this is the reason why
electrotoroidal tensors $T_u^{++}$ are not studied further in this
work.

\begin{sidewaystable*}
  \vspace*{12ex}
  \caption{\label{tab:c-infinity-objects} Axially symmetric finite
  objects illustrating the categories of chirality for classes
  $L(C_\infty)$ and $L(D_\infty)$, see
  Figs.~\ref{fig:chiral:c-infty}, \ref{fig:chiral:d-infty}
  and~\ref{fig:chiral:dc-infty}.}
\renewcommand{\marr}[2][@{}c@{}]{\makebox[0pt]{\renewcommand{\arraystretch}{0.90}%
  \tabcolsep 0pt\begin{tabular}{#1} #2 \end{tabular}}}%
  \newcommand{\mb}[1]{\makebox[7em][c]{#1}}%
  \renewcommand{\arraystretch}{1.2}%
  \setlength{\extrarowheight}{0.4ex}%
\begin{tabular*}{\linewidth}{ncs{1em}E*{8}{C}n}
\hline \hline
\rule{0pt}{9.0ex}%
\raisebox{2ex}{object} &
\includegraphics[height=8.0ex]{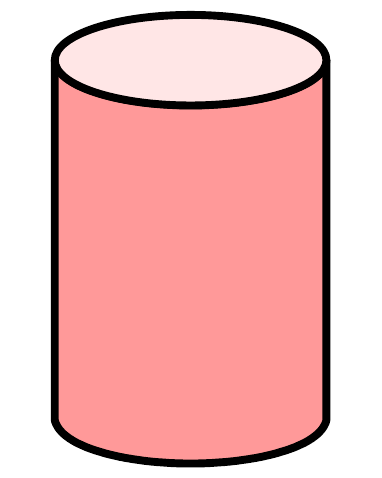} &
\includegraphics[height=8.0ex]{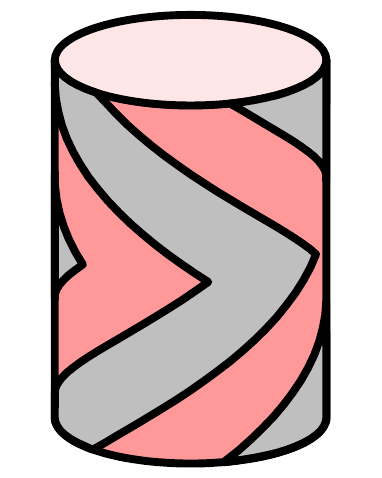} &
\includegraphics[height=8.0ex]{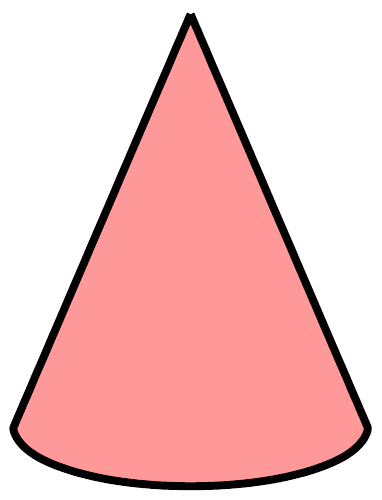} &
\includegraphics[height=8.0ex]{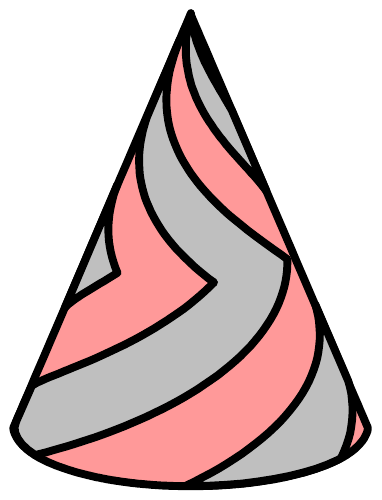} &
\includegraphics[height=8.0ex]{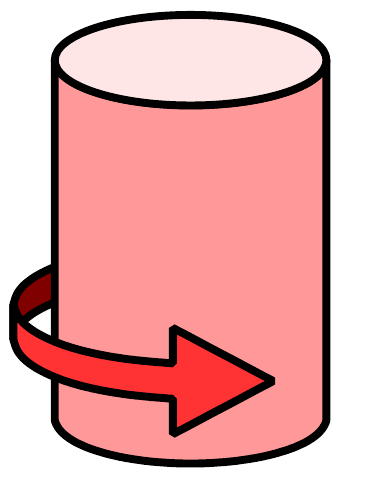} &
\includegraphics[height=8.0ex]{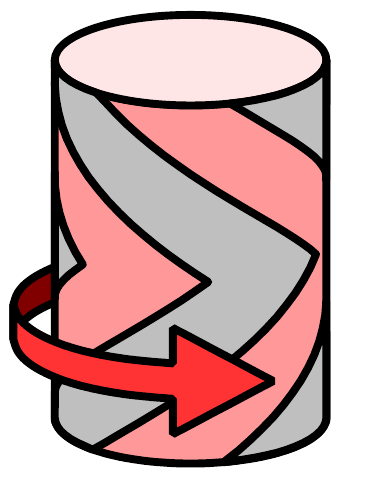} &
\includegraphics[height=8.0ex]{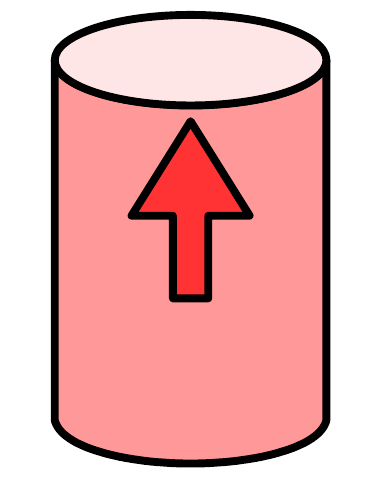} &
\includegraphics[height=8.0ex]{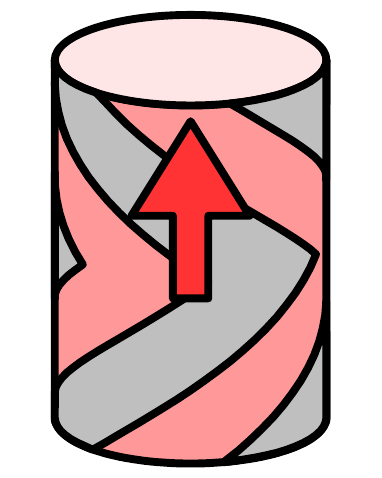} \\[-0.8ex] &
\mb{\marr{plain cylinder}} & \mb{\marr{marbled cylinder}} &
\mb{\marr{plain cone}} & \mb{\marr{marbled cone}} &
\mb{\marr{rotating cylinder}} & \mb{\marr{rotating marbled\\ cylinder}} & \mb{\marr{moving cylinder}} &
\mb{\marr{moving marbled\\ cylinder}} \\ \hline
point group & D_\infty \times\gsistis & C_\infty \times \gsistis &
D_\infty [C_\infty] \times \gtis  & C_\infty \times \gtis &
D_\infty (C_\infty) \times \gsis  & C_\infty \times \gsis &
D_\infty (C_\infty) \times \gstis & C_\infty \times \gstis \\
\marr{inversion\\[-0.3ex] symmetries} &
i, \theta, i\theta & i, \theta, i\theta & \theta & \theta & i & i &
i\theta & i\theta \\
& i C_2, \theta C_2, i\theta C_2, & & i C_2, i\theta C_2, & &
\theta C_2, i\theta C_2, & & i C_2, \theta C_2, & \\[-0.3ex]
\raisebox{2.1ex}[0pt]{\marr{improper\\ rotations}} &
i C_z, \theta C_z, i\theta C_z & i C_z, \theta C_z, i\theta C_z &
\theta C_z & \theta C_z  & i C_z & i C_z & i\theta C_z & i\theta C_z \\
\marr{category of\\[-0.04ex] chirality} &
\marr{parachiral} & \marr{parachiral} &
\marr{parachiral} & \marr{electrochiral} &
\marr{parachiral} & \marr{magnetochiral} &
\marr{parachiral} & \marr{antimagneto-\\ chiral} \\ \hline
\rule{0pt}{9.0ex}%
\raisebox{2ex}{object} &
\includegraphics[height=8.0ex]{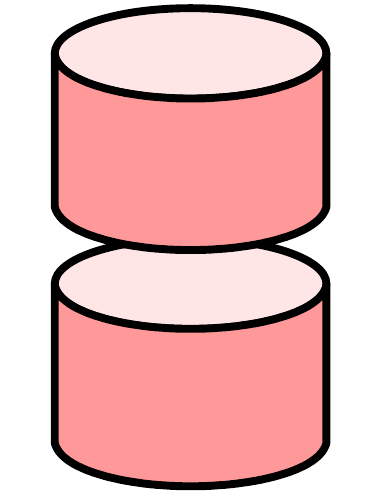} &
\includegraphics[height=8.0ex]{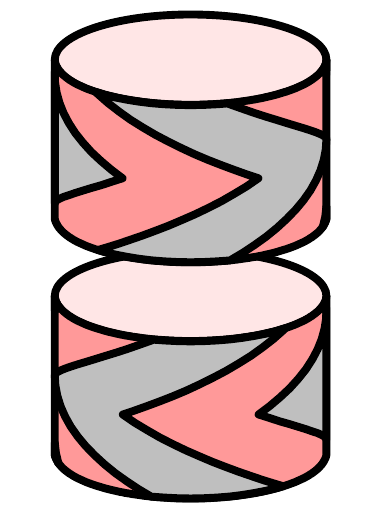} &
\includegraphics[height=8.0ex]{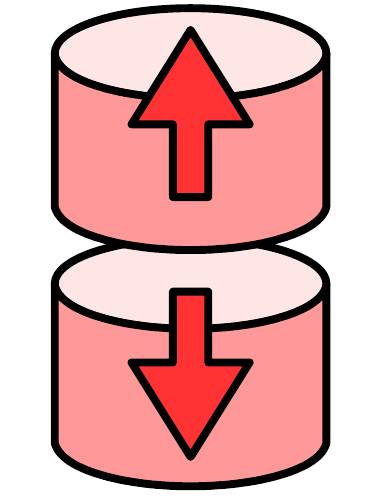} &
\includegraphics[height=8.0ex]{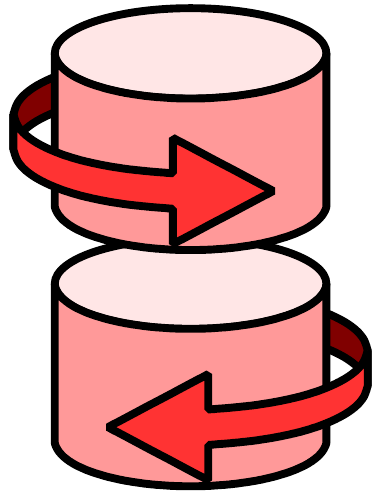} & &
\includegraphics[height=8.0ex]{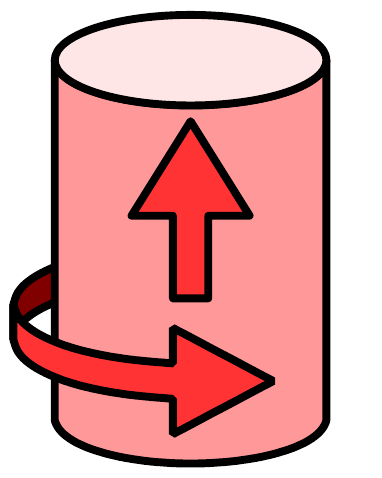} &
\includegraphics[height=8.0ex]{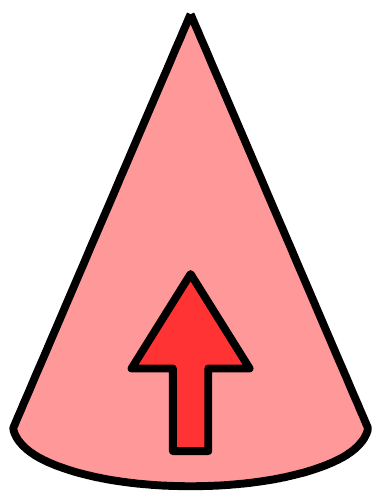} &
\includegraphics[height=8.0ex]{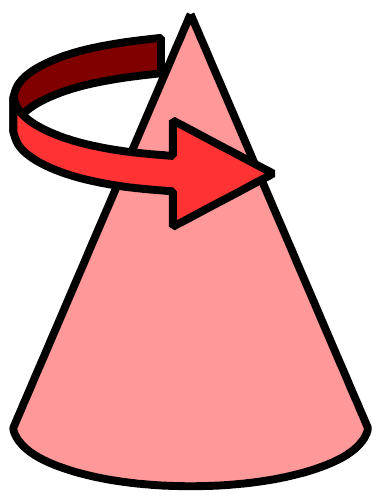} \\[-0.5ex] &
\mb{\marr{plain divided\\ cylinders}} & \mb{\marr{marbled divided\\ cylinders}}&
\mb{\marr{countermoving\\ cylinders}} & \mb{\marr{counterrotating\\ cylinders}}&
& \mb{\marr{moving rotating\\ cylinder}}
& \mb{\marr{moving cone}}
& \mb{\marr{rotating cone}} \\ \hline
point group & D_\infty \times \gsistis & D_\infty \times \gtis &
D_\infty \times \gsis & D_\infty \times \gstis & &
D_\infty (C_\infty) & D_\infty [C_\infty] & D_\infty \{C_\infty\} \\
\marr{inversion\\[-0.3ex] symmetries} &
i, \theta, i\theta & \theta & i & i\theta \\
& i C_2, \theta C_2, i\theta C_2, & \theta C_2, & i C_2, & i\theta C_2, & &
\theta C_2 & i C_2 & i\theta C_2 \\[-0.3ex]
\raisebox{2.1ex}[0pt]{\marr{improper\\ rotations}} &
i C_z, \theta C_z, i\theta C_z & \theta C_z & i C_z & i\theta C_z \\
\marr{category of\\[-0.04ex] chirality} &
\marr{parachiral} &
\marr{electrochiral} &
\marr{magnetochiral} &
\marr{antimagneto-\\ chiral} & &
\marr{electrochiral} &
\marr{magnetochiral} &
\marr{antimagneto-\\ chiral} \\ \hline \hline
\end{tabular*}
\end{sidewaystable*}

According to the general definition of group types given in
Table~\ref{tab:group-types-def}, the continuous dihedral groups have
the group types listed in the first column of
Table~\ref{tab:c-dihedral:trafo}.  A comparison of
Tables~\ref{tab:categories} and~\ref{tab:c-dihedral:trafo} shows
that the range of multipole densities $T_\ell^{ss'}$ permitted by
these continuous dihedral groups falls short of the range of
densities generally permitted by the categories of polarization
associated with these group types.  For example, the group $D_\infty
\times \gtis$ has the type~\ref{gcat:i-II} that belongs to the
electropolar category.  However, the group $D_\infty \times \gtis$ does
not allow odd\=/$\ell$ electric multipoles ($-+$).  This result can
be traced back to the fact that the groups in class
$L(D_\infty)$ can also be viewed as limiting groups $n \rightarrow
\infty$ for the groups in class $L (D_n)$.  The lowest degree
$\ell_{\min}^{(e,u)}$ of odd\=/$\ell$ electric multipoles permitted
by groups $D_n \times \gtis$ grows monotonically with increasing
$n$, with $\ell_{\min}^{(e,u)} \rightarrow \infty$ for $n
\rightarrow \infty$ (see Table~\ref{tab:tensor-nonmag}).  This does
not affect our classification of categories of polarized matter as
crystals are characterized by finite crystallographic point groups.

Continuous axial symmetries can be useful to represent the limiting
symmetries of physical systems \cite{tav61}.  Illustrating these by
simple geometrical shapes has a long history \cite{tav61, hli14,
jan17}, and such simple objects can also be employed as depictions
of finite chiral systems \cite{bar86}.  Inspired by these earlier
approaches, we use axially symmetric objects (such as those shown in
Table~\ref{tab:c-infinity-objects}) to explore the five categories
of chirality, see Figs.~\ref{fig:chiral:c-infty},
\ref{fig:chiral:d-infty} and~\ref{fig:chiral:dc-infty}.  To ease
cross-referencing between previous works~\cite{hli14, che21}
discussing axial quantities and ours, we provide
Table~\ref{tab:ax-previous} relating respective notations for axial
quantities.

\begin{table}[tb]
  \caption{\label{tab:ax-previous} Notation for axial quantities in
  the literature.  Reference~\cite{hli14} uses the symbols from
  Ref.~\cite{dub90} for tensor components $T_{u0}^{ss'}$.}
  \centering
  \renewcommand{\arraystretch}{1.2}%
  \begin{tabular*}{\linewidth}{nl*{2}{s{0.6em}*{4}{EC}}n}
    \hline \hline \rule{0pt}{2.7ex}This work &
    T_{g0}^{++} & T_{g0}^{-+} & T_{g0}^{+-} & T_{g0}^{--} &
    T_{u0}^{++} & T_{u0}^{-+} & T_{u0}^{+-} & T_{u0}^{--} \\[0.2ex]
    Ref.~\cite{hli14} &
    \vek{N} & \vek{C} & \vek{L} & \vek{F} &
    \vek{G} & \vek{P} & \vek{M} & \vek{T} \\ Ref.~\cite{che21} &
    \mathcal{D} & \mathcal{C} & \mathcal{D}' & \mathcal{C}' &
    \scalebox{1.2}{\mbox{\ensuremath{\displaystyle
    \mathcal{a}}}} & \mathcal{P} &
    \scalebox{1.2}{\mbox{\ensuremath{\displaystyle
    \mathcal{a}}}}' & \mathcal{P}' \\ \hline \hline
  \end{tabular*}
\end{table}

\begin{sidewaystable*}[tbp]
  \caption{\label{tab:mag-point-group} The 122 crystallographic
  magnetic point groups organized into (Laue) classes.  Within each
  class, the arrangement of groups follows the scheme in
  Table~\ref{tab:subclasses} and indicated at the bottom part of the
  table.  See Table~\ref{tab:group-types-def} for the definitions of group
  types.  The groups of type \ref{gcat:J}, \ref{gcat:ii-I},
  \ref{gcat:i-II} and \ref{gcat:ii-II} delineate the main block of
  groups in a given class, where the duality transformation
  $i\leftrightarrow \theta$ maps rows onto columns and \textit{vice
  versa}.  The type-\ref{gcat:JJ}, type-\ref{gcat:JJp},
  type-\ref{gcat:JJJp}, and type-\ref{gcat:JJJpp} groups outside the
  main block are self-dual, and so are the groups positioned on a
  main-block diagonal.  Small $2\times
  2$ matrices mark congruences between tensors with signature $ss'$
  as discussed in detail in Appendix~\ref{app:tensors}.}
  \newcommand{\mb}[1]{\makebox[2.3em]{#1}}%
  \extrarowheight 0.6ex
  \renewcommand{\arraystretch}{0.8}%
\begin{tabular*}{1.00\linewidth}{LS{0.5em}C*{3}{ELS{0.2em}LS{0.5em}C}<{\hspace{1em}}|ELS{0.2em}LS{0.5em}C}
    \hline \hline
    \multicolumn{14}{c}{triclinic (5 groups)} \\ \hline
    C_1 & \mygroup{2}{2}{2}{2} & & & & & & &
    C_i & = C_1 \times \gsis & \mygroup{2}{2}{0}{0} &
    C_i (C_1) & = C_1 \times \gstis & \mygroup{2}{0}{0}{2} \\
    C_1 \times \gtis & \mygroup{2}{0}{2}{0} & & & & & & &
    C_i \times \gtis & = C_1 \times \gsistis & \mygroup{2}{0}{0}{0} \\
    \hline \multicolumn{14}{c}{monoclinic (11 groups)} \\ \hline
    C_2 & \mygroup{2}{2}{2}{2} &
    C_s & = C_2 [C_1] & \mygroup{2}{2}{4}{4} & & & &
    C_{2h} & = C_2 \times \gsis & \mygroup{2}{2}{0}{0} &
    C_{2h} (C_2) & = C_2 \times \gstis  & \mygroup{2}{0}{0}{2} \\
    C_2 (C_1) & \mygroup{2}{4}{2}{4} &
    C_s (C_1) & = C_2 \{C_1\} & \mygroup{2}{4}{4}{2} & & & &
    C_{2h} (C_i) & = C_2 (C_1) \times \gsis & \mygroup{2}{4}{0}{0} &
    C_{2h} (C_s) & = C_2 (C_1) \times \gstis & \mygroup{2}{0}{0}{4} \\
    C_2 \times \gtis & \mygroup{2}{0}{2}{0} &
    C_s \times \gtis & = C_2 [C_1] \times \gtis & \mygroup{2}{0}{4}{0} & & & &
    C_{2h} \times \gtis & = C_2 \times \gsistis
    & \mygroup{2}{0}{0}{0} \\
    \hline \multicolumn{14}{c}{(ortho-) rhombic (12 groups)} \\ \hline
    D_2 & \mygroup{1}{1}{1}{1} &
    C_{2v} & = D_2 [C_2] & \mygroup{1}{1}{4}{4} & & & &
    D_{2h} & = D_2 \times \gsis & \mygroup{1}{1}{0}{0} &
    D_{2h} (D_2) & = D_2 \times \gstis & \mygroup{1}{0}{0}{1} \\
    D_2 (C_2) & \mygroup{1}{4}{1}{4} &
    C_{2v} (C_2) & = D_2 \{C_2\} & \mygroup{1}{4}{4}{1}
    & & & &
    D_{2h} (C_{2h}) & = D_2 (C_2) \times \gsis & \mygroup{1}{4}{0}{0} &
    D_{2h} (C_{2v}) & = D_2 (C_2) \times \gstis & \mygroup{1}{0}{0}{4} \\
    D_2 \times \gtis & \mygroup{1}{0}{1}{0} &
    C_{2v} \times \gtis  & = D_2 [C_2] \times \gtis & \mygroup{1}{0}{4}{0}
    & & & &
    D_{2h} \times \gtis & = D_2 \times \gsistis
    & \mygroup{1}{0}{0}{0} \\ \cline{1-11}
    & &
    C_{2v} (C_s) & = D_2 [C_2] \, (C_2' [C_1]) & \mygroup{1}{4}{4}{4} \\
    \hline \multicolumn{14}{c}{tetragonal ($11+20$ groups)} \\ \hline
    C_4 & \mygroup{2}{2}{2}{2} &
    S_4 & = C_4 [C_2] & \mygroup{2}{2}{3}{3} & & & &
    C_{4h} & = C_4 \times \gsis & \mygroup{2}{2}{0}{0} &
    C_{4h} (C_4) & = C_4 \times \gstis & \mygroup{2}{0}{0}{2} \\
    C_4 (C_2) & \mygroup{2}{3}{2}{3} &
    S_4 (C_2) & = C_4 \{C_2\} & \mygroup{2}{3}{3}{2}
    & & & &
    C_{4h} (C_{2h}) & = C_4 (C_2) \times \gsis & \mygroup{2}{3}{0}{0} &
    C_{4h} (S_4) & = C_4 (C_2) \times \gstis & \mygroup{2}{0}{0}{3} \\
    C_4 \times \gtis & \mygroup{2}{0}{2}{0} &
    S_4 \times \gtis & = C_4 [C_2] \times \gtis & \mygroup{2}{0}{3}{0}
    & & & &
    C_{4h} \times \gtis & = C_4 \times \gsistis
    & \mygroup{2}{0}{0}{0} \\
    \hline
    D_4 & \mygroup{1}{1}{1}{1} &
    C_{4v} & = D_4 [C_4] & \mygroup{1}{1}{4}{4} &
    D_{2d} & = D_4 [D_2] & \mygroup{1}{1}{5}{5} &
    D_{4h} & = D_4 \times \gsis & \mygroup{1}{1}{0}{0} &
    D_{4h} (D_4) & = D_4 \times \gstis & \mygroup{1}{0}{0}{1} \\
    D_4 (C_4) & \mygroup{1}{4}{1}{4} &
    C_{4v} (C_4) & = D_4 \{ C_4 \} & \mygroup{1}{4}{4}{1} &
    D_{2d} (S_4) & = D_4 [D_2] \, (C_4 [C_2]) & \mygroup{1}{4}{5}{5} &
    D_{4h} (C_{4h}) & = D_4 (C_4) \times \gsis & \mygroup{1}{4}{0}{0} &
    D_{4h} (C_{4v}) & = D_4 (C_4) \times \gstis & \mygroup{1}{0}{0}{4} \\
    D_4 (D_2) & \mygroup{1}{5}{1}{5} &
    C_{4v} (C_{2v}) & = D_4 [C_4] \, (D_2 [C_2]) & \mygroup{1}{5}{4}{5} &
    D_{2d} (D_2) & = D_4 \{ D_2\} & \mygroup{1}{5}{5}{1} &
    D_{4h} (D_{2h}) & = D_4 (D_2) \times \gsis & \mygroup{1}{5}{0}{0} &
    D_{4h} (D_{2d}) & = D_4 (D_2) \times \gstis & \mygroup{1}{0}{0}{5} \\
    D_4 \times \gtis & \mygroup{1}{0}{1}{0} &
    C_{4v} \times \gtis & = D_4 [C_4] \times \gtis & \mygroup{1}{0}{4}{0} &
    D_{2d} \times \gtis & = D_4 [D_2] \times \gtis & \mygroup{1}{0}{5}{0} &
    D_{4h} \times \gtis & = D_4 \times \gsistis
    & \mygroup{1}{0}{0}{0} \\ \cline{1-11}
    & & & & &
    D_{2d} (C_{2v}) & = D_4 [D_2] \, (D_2' [C_2]) & \mygroup{1}{5}{5}{4} \\
    \hline \multicolumn{14}{c}{trigonal ($5+11$ groups)} \\ \hline
    C_3 & \mygroup{2}{2}{2}{2}
    & & & & & & &
    C_{3i} & = C_3 \times \gsis & \mygroup{2}{2}{0}{0} &
    C_{3i} (C_3) & = C_3 \times \gstis & \mygroup{2}{0}{0}{2} \\
    C_3 \times \gtis & \mygroup{2}{0}{2}{0}
    & & & & & & &
    C_{3i} \times \gtis & = C_3 \times \gsistis
    & \mygroup{2}{0}{0}{0} \\
    \hline
    D_3 & \mygroup{1}{1}{1}{1} &
    C_{3v} & = D_3 [C_3] & \mygroup{1}{1}{4}{4}
    & & & &
    D_{3d} & = D_3 \times \gsis & \mygroup{1}{1}{0}{0} &
    D_{3d} (D_3) & = D_3 \times \gstis & \mygroup{1}{0}{0}{1} \\
    D_3 (C_3) & \mygroup{1}{4}{1}{4} &
    C_{3v} (C_3) & = D_3 \{C_3\} & \mygroup{1}{4}{4}{1} & & & &
    D_{3d} (C_{3i}) & = D_3 (C_3) \times \gsis & \mygroup{1}{4}{0}{0} &
    D_{3d} (C_{3v}) & = D_3 (C_3) \times \gstis & \mygroup{1}{0}{0}{4} \\
    D_3 \times \gtis & \mygroup{1}{0}{1}{0} &
    C_{3v} \times \gtis & = D_3 [C_3] \times \gtis & \mygroup{1}{0}{4}{0}
    & & & &
    D_{3d} \times \gtis & = D_3 \times \gsistis
    & \mygroup{1}{0}{0}{0} \\
    \hline \multicolumn{14}{c}{hexagonal ($11+20$ groups)} \\ \hline
    C_6 & \mygroup{2}{2}{2}{2} &
    C_{3h} & = C_6 [C_3] & \mygroup{2}{2}{3}{3}
    & & & &
    C_{6h} & = C_6 \times \gsis & \mygroup{2}{2}{0}{0} &
    C_{6h} (C_6) & = C_6 \times \gstis & \mygroup{2}{0}{0}{2} \\
    C_6 (C_3) & \mygroup{2}{3}{2}{3} &
    C_{3h} (C_3) & = C_6 \{C_3\} & \mygroup{2}{3}{3}{2}
    & & & &
    C_{6h} (C_{3i}) & = C_6 (C_3) \times \gsis & \mygroup{2}{3}{0}{0} &
    C_{6h} (C_{3h}) & = C_6 (C_3) \times \gstis & \mygroup{2}{0}{0}{3} \\
    C_6 \times \gtis & \mygroup{2}{0}{2}{0} &
    C_{3h} \times \gtis & = C_6 [C_3] \times \gtis & \mygroup{2}{0}{3}{0}
    & & & &
    C_{6h} \times \gtis & = C_6 \times \gsistis
    & \mygroup{2}{0}{0}{0} \\
    \hline
    D_6 & \mygroup{1}{1}{1}{1} &
    C_{6v} & = D_6 [C_6] & \mygroup{1}{1}{4}{4} &
    D_{3h} & = D_6 [D_3] & \mygroup{1}{1}{5}{5} &
    D_{6h} & = D_6 \times \gsis & \mygroup{1}{1}{0}{0} &
    D_{6h} (D_6) & = D_6 \times \gstis & \mygroup{1}{0}{0}{1} \\
    D_6 (C_6) & \mygroup{1}{4}{1}{4} &
    C_{6v} (C_6) & = D_6 \{ C_6 \} & \mygroup{1}{4}{4}{1} &
    D_{3h} (C_{3h}) & = D_6 [D_3] \, (C_6 [C_3]) & \mygroup{1}{4}{5}{5} &
    D_{6h} (C_{6h}) & = D_6 (C_6) \times \gsis & \mygroup{1}{4}{0}{0} &
    D_{6h} (C_{6v}) & = D_6 (C_6) \times \gstis & \mygroup{1}{0}{0}{4} \\
    D_6 (D_3) & \mygroup{1}{5}{1}{5} &
    C_{6v} (C_{3v}) & = D_6 [C_6] \, (D_3 [C_3]) & \mygroup{1}{5}{4}{5} &
    D_{3h} (D_3) & = D_6 \{ D_3\} & \mygroup{1}{5}{5}{1} &
    D_{6h} (D_{3d}) & = D_6 (D_3) \times \gsis & \mygroup{1}{5}{0}{0} &
    D_{6h} (D_{3h}) & = D_6 (D_3) \times \gstis & \mygroup{1}{0}{0}{5} \\
    D_6 \times \gtis & \mygroup{1}{0}{1}{0} &
    C_{6v} \times \gtis & = D_6 [C_6] \times \gtis & \mygroup{1}{0}{4}{0} &
    D_{3h} \times \gtis & = D_6 [D_3] \times \gtis & \mygroup{1}{0}{5}{0} &
    D_{6h} \times \gtis & = D_6 \times \gsistis
    & \mygroup{1}{0}{0}{0} \\ \cline{1-11}
    & & & & &
    D_{3h} (C_{3v}) & = D_6 [D_3] \, (D_3' [C_3]) & \mygroup{1}{5}{5}{4} \\
    \hline \multicolumn{14}{c}{cubic ($5+11$ groups)} \\ \hline
    T & \mygroup{1}{1}{1}{1}
    & & & & & & &
    T_h & = T \times \gsis & \mygroup{1}{1}{0}{0} &
    T_h (T) & = T \times \gstis & \mygroup{1}{0}{0}{1} \\
    T \times \gtis & \mygroup{1}{0}{1}{0}
    & & & & & & &
    T_h \times \gtis & = T \times \gsistis & \mygroup{1}{0}{0}{0} \\
    \hline
    O & \mygroup{1}{1}{1}{1} &
    T_d & = O [T] & \mygroup{1}{1}{3}{3}
    & & & &
    O_h & = O \times \gsis & \mygroup{1}{1}{0}{0} &
    O_h(O) & = O \times \gstis & \mygroup{1}{0}{0}{1} \\
    O(T) & \mygroup{1}{3}{1}{3} &
    T_d(T) & = O \{T\} & \mygroup{1}{3}{3}{1}
    & & & &
    O_h (T_h) & = O (T) \times \gsis & \mygroup{1}{3}{0}{0} &
    O_h (T_d) & = O (T) \times \gstis & \mygroup{1}{0}{0}{3} \\
    O \times \gtis & \mygroup{1}{0}{1}{0} &
    T_d \times \gtis & = O [T] \times \gtis & \mygroup{1}{0}{3}{0}
    & & & &
    O_h \times \gtis & = O \times \gsistis & \mygroup{1}{0}{0}{0} \\
    \hline \hline
      \multicolumn{2}{c}{\mb{\ref{gcat:J}}}
    & \multicolumn{6}{c}{\mb{\ref{gcat:iii-I}}}
    & \multicolumn{2}{c}{\mb{\ref{gcat:ii-I}}} &
    & \multicolumn{2}{Ec}{\mb{\ref{gcat:JJ}}} \\
      \multicolumn{2}{c}{\mb{\ref{gcat:i-III}}}
    & \multicolumn{6}{c}{\mb{\ref{gcat:JJJ}, \ref{gcat:iiip-III},
     and \ref{gcat:iii-IIIp}}}
    & \multicolumn{2}{c}{\mb{\ref{gcat:ii-III}}} &
    & \multicolumn{2}{Ec}{\mb{\ref{gcat:JJp}}} \\
      \multicolumn{2}{c}{\mb{\ref{gcat:i-II}}}
    & \multicolumn{6}{c}{\mb{\ref{gcat:iii-II}}}
    & \multicolumn{2}{c}{\mb{\ref{gcat:ii-II}}} & \\ \cline{1-11}
    & & \multicolumn{6}{c}{\mb{\ref{gcat:JJJp} and \ref{gcat:JJJpp}}} \\
    \hline \hline
  \end{tabular*}
  \vspace*{-0.37\textheight}
\end{sidewaystable*}

\begin{table*}
  \caption{\label{tab:groups-multipoles} Lowest rank $\ell_{\min}$
  of nontrivial multipole densities permitted by the 122
  crystallographic magnetic point groups.  In the parapolar case, we
  consider even\=/$\ell$ electric multipole densities (signature
  $++$).  In the remaining cases we consider odd\=/$\ell$ electric
  ($-+$), odd\=/$\ell$ magnetic ($+-$), and odd\=/$\ell$
  magnetotoroidal ($--$) multipole densities.  Numbers in
  parentheses indicate that multiple multipolar groups $\gamma = i$,
  $\theta$, and $i\theta$ permit the same rank\=/$\ell_{\min}$
  multipole density.}
  \renewcommand{\arraystretch}{1.1}%
  \extrarowheight 0.6ex
  \newcolumntype{M}{S{2em}}%
  \begin{tabular*}{\linewidth}{nLs{2em}LERMRELMLMCn}
    \hline\hline \rule{0pt}{2.7ex} \ell_{\min} &
    \multicolumn{1}{c}{parapolar} &
    \multicolumn{2}{c}{unipolar} &
    \multicolumn{3}{c}{multipolar}
    \\ \hline
    1 & & C_2 \gmin{C_1}_{\bar{\gamma}} \times \ggen & C_1 \times \ggen & & & C_1 \\
      & & D_2 \gmin{C_2}_{\bar{\gamma}} \times \ggen & C_2 \times \ggen & D_2 \gmin{C_2}_\gamma \;(2) & C_2 \gmin{C_1}_\gamma \;(3) & C_2 \\
      & & D_4 \gmin{C_4}_{\bar{\gamma}} \times \ggen & C_4 \times \ggen & D_4 \gmin{C_4}_\gamma \;(2) & C_4 \gmin{C_2}_\gamma & C_4 \\
      & & D_3 \gmin{C_3}_{\bar{\gamma}} \times \ggen & C_3 \times \ggen & D_3 \gmin{C_3}_\gamma \;(2) & & C_3 \\
      & & D_6 \gmin{C_6}_{\bar{\gamma}} \times \ggen & C_6 \times \ggen & D_6 \gmin{C_6}_\gamma \;(2) & C_6 \gmin{C_3}_\gamma & C_6 \\
      & & & & \multicolumn{3}{C}{D_2 [C_2] (C_2' [C_1])} \\
      & & & & \multicolumn{3}{C}{D_4 \gmin{D_2}_\gamma \gmin{D_2' \gmin{C_2}_\gamma}_{\bar{\gamma}}} \\
      & & & & \multicolumn{3}{C}{D_6 \gmin{D_3}_\gamma \gmin{D_3' \gmin{C_3}_\gamma}_{\bar{\gamma}}} \\ \hline
    2 & C_1 \times \gsistis, C_2 \times \gsistis, D_2 \times \gsistis, \\
      & C_4 \times \gsistis, D_4 \times \gsistis, C_3 \times \gsistis, \\
      & D_3 \times \gsistis, C_6 \times \gsistis, D_6 \times \gsistis \\
    3 & & C_4 \gmin{C_2}_{\bar{\gamma}} \times \ggen \\
      & & D_4 \gmin{D_2}_{\bar{\gamma}} \times \ggen & D_2 \times \ggen & D_4 \gmin{D_2}_\gamma \;(2) & & D_2  \\
      & & C_6 \gmin{C_3}_{\bar{\gamma}} \times \ggen \\
      & & D_6 \gmin{D_3}_{\bar{\gamma}} \times \ggen & D_3 \times \ggen & D_6 \gmin{D_3}_\gamma \;(2) & & D_3          \\
      & & O \gmin{T}_{\bar{\gamma}} \times \ggen  & T \times \ggen  & O \gmin{T}_\gamma \hspace{1.35em} (2) & & T           \\
    4 & T \times \gsistis, O \times \gsistis \\
    5 & & & D_4 \times \ggen & & & D_4 \\
    7 & & & D_6 \times \ggen & & & D_6 \\
    9 & & & O   \times \ggen & & & O \\
    \hline \hline
  \end{tabular*}
\end{table*}

\subsection{Discrete groups}
\label{app:discr-groups}

There are 11 proper crystallographic point groups: $C_1$, $C_2$,
$D_2$, $C_4$, $D_4$, $C_3$, $D_3$, $C_6$, $D_6$, $T$, and $O$,
representing the roots for 11 classes.  By adding SIS, one obtains
21 improper (with respect to SIS) point groups~\cite{lud96}; 11 of
these are $i$\=/major groups and 10 are $i$\=/minor groups.  Jointly,
these groups represent the familiar 32 nonmagnetic crystallographic
point groups (ignoring TIS).  By also adding TIS, the 122 magnetic
crystallographic point groups are obtained~\cite{tav56, cor58a,
bra72, lan8e}.  The 32 type\=/II (i.e., $\theta$\=/major) groups
among these are the crystallographic point groups representing
paramagnetic solids.  The same groups also characterize
antiferromagnetic crystals with nonsymmorphic space groups
containing symmetry elements that combine nonprimitive translations
with time inversion \cite{cor58a, cam24}, see Sec~\ref{sec:msg}.
All other magnetic point groups describe crystals with magnetic
order where TIS is broken.

The full set of 122 magnetic crystallographic point groups can be
organized into 11 classes~\cite{kop79, kop06, gri91, gri94}.  These
classes are generalizations of the \emph{Laue} classes that are
usually defined in terms of the improper groups $\gprop \times
\gsis$ \cite{bur13}.  In a similar spirit, the Laue classes of
magnetic crystallographic point groups have previously been defined
in terms of the paramagnetic groups $\gprop \times \gsistis$
\cite{gri91}.  Here we find it more advantageous to organize these
Laue classes in terms of their respective proper point groups
$\gprop$~\cite{kop06}.  See Table~\ref{tab:mag-point-group}.

The composition of the 11 (Laue) classes of magnetic point groups
derived from the proper crystallographic point groups partly
mirrors, but also extends that of the continuous groups.  In
particular, the structure of $L(C_1)$, $L(C_3)$ and $L(T)$ is
entirely analogous to that of $L(R)$ and $L(C_\infty)$ (compare
Tables~\ref{tab:ax:point-group} and \ref{tab:mag-point-group}),
which is due to the fact that, like $R$ and $C_\infty$, the class
roots $C_1$, $C_3$ and $T$ have no index\=/2 subgroup.  Similarly,
the structure of $L(C_2)$, $L(C_4)$, $L(D_3)$, $L(C_6)$ and $L(O)$
mirrors that of $L(D_\infty)$, because the respective class roots all
have one index\=/2 subgroup.  Class $L(D_2)$ exhibits the basic
$L(D_\infty)$ structure, but it contains, in addition, the group
$C_{2v} (C_s) = D_2 [C_2] \, (C_2 [C_1])$ which is the only
crystallographic point group of type \ref{gcat:JJJpp}.
Classes $L(D_4)$ and $L(D_6)$ exhibit greater complexity because
they have two distinct index\=/2 subgroups, which doubles the
possibilities for constructing groups of type \ref{gcat:iii-I},
\ref{gcat:i-III}, \ref{gcat:JJJ}, \ref{gcat:iii-II},
\ref{gcat:ii-III}, \ref{gcat:JJp} and also yields a triad of groups
with respective types \ref{gcat:JJJp}, and \ref{gcat:iiip-III}, and
\ref{gcat:iii-IIIp}.

Tables~\ref{tab:spher:point-group}, \ref{tab:ax:point-group} and
\ref{tab:mag-point-group} systematize classes of magnetic point
groups in terms of the SIS-TIS-symmetric group types summarized in
Table~\ref{tab:group-types-def}.  The placement of individual groups
within rows and columns as per the schemes given at the bottom of
each table illustrates SIS-TIS duality, see
Appendix~\ref{app:StructClass}.  In particular, for each class $L
(\gprop)$, the \emph{main block} has the groups of type
\ref{gcat:J}, \ref{gcat:ii-I}, \ref{gcat:i-II} and \ref{gcat:ii-II}
at its four corners.  Within this block, the duality transformation
based on the exchange $i \leftrightarrow \theta$ maps rows on
columns and vice versa.  Groups placed on the diagonal of the main
block are self-dual, and so are the type \ref{gcat:JJ},
\ref{gcat:JJp}, \ref{gcat:JJJp}, and \ref{gcat:JJJpp} groups
positioned outside of the main block.  Using SIS-TIS duality of
magnetic point groups within a class as the organizing principle
differentiates the arrangement of Table~\ref{tab:mag-point-group}
from the previously developed periodic arrangement of the 122
crystallographic magnetic point groups shown, e.g., in Table~2 of
Ref.~\cite{gri91}.

The organizing principle of Tables~\ref{tab:spher:point-group},
\ref{tab:ax:point-group} and \ref{tab:mag-point-group} follows that
shown for a general class in the upper part of Table~\ref{tab:subclasses}.
Thus, in all these tables, groups within the main block of a class
$L (\gprop)$ \cite{misc:mainBlock} are placed such that groups
in a given row belong to the same subclass $L (G_0^i, i)$, where $G_0^i$
is the group in the first column.  Groups in a given column of the main
block belong to the same subclass $L (G_0^\theta, \theta)$, $G_0^\theta$
being the group in the first row.  See Appendices~\ref{app:subclasses}
and~\ref{app:subclass:i-theta} for a detailed discussion of subclass
structures.

Subgroup relations and isomorphisms between groups in a given class
$L (\gprop)$ follow Fig.~\ref{fig:hierarchy}.  See also explicit
tabulations of subgroup relations for the ordinary crystallographic
point groups~\cite{kos63, jan75} and for the magnetic
groups~\cite{asc65, dam81}.  Table~\ref{tab:group-types-def} lists
the number of point groups of each type among the 122
crystallographic magnetic point groups.  The triadic structure
underlies the appearance of certain \emph{magic numbers}
\cite{ope74, kop76, sch08} when groups are collated based on
particular physical characteristics described by tensor quantities.
See Appendix~\ref{app:tensors} for a detailed discussion of this
point.

Table~\ref{tab:groups-multipoles} lists the lowest rank
$\ell_{\min}$ of nontrivial multipole densities permitted by
all 122 crystallographic magnetic point groups.

\section{Compound multipoles}
\label{app:compound-multipoles}

\begin{table*}
  \caption{\label{tab:ir-prod} Multiplication table for electric and
  magnetic IRs $D_\ell^{ss'}$ of the full rotation group $\Rit$ with
  $\ell \le 3$.  We have $D_{1,2,3}^{ss'} \equiv D_1^{ss'} +
  D_2^{ss'} + D_3^{ss'}$ etc.  Bold subscripts $\ell$ indicate that
  an IR contained in the product representation represents a
  toroidal multipole.}
  \centering
  \renewcommand{\arraystretch}{1.4}%
  \newcommand{\q}[1]{\bm{#1}}%
  \begin{tabular*}{\linewidth}{*{4}{Les{2em}LE}s{3em}|C}
    \hline \hline \rule{0pt}{2.8ex}
    D_{0}^{++} & D_{0}^{--} & D_{1}^{-+} & D_{1}^{+-} &
    D_{2}^{++} & D_{2}^{--} & D_{3}^{-+} & D_{3}^{+-} & D_\ell^{ss'} \\
    \hline \rule{0pt}{2.6ex}
    D_{0}^{++} & D_{0}^{--} & D_{1}^{-+} & D_{1}^{+-} &
    D_{2}^{++} & D_{2}^{--} & D_{3}^{-+} & D_{3}^{+-} & D_{0}^{++} \\
               & D_{0}^{++} & D_{1}^{+-} & D_{1}^{-+} &
    D_{2}^{--} & D_{2}^{++} & D_{3}^{+-} & D_{3}^{-+} & D_{0}^{--} \\
               &            & D_{0,1,2}^{++} & D_{0,\q{1},2}^{--} &
    D_{1,\q{2},3}^{-+} & D_{1,\q{2},3}^{+-} & D_{2,\q{3},4}^{++} & D_{2,\q{3,}4}^{--} & D_{1}^{-+} \\
               &            &              & D_{0,1,2}^{++} &
    D_{1,\q{2},3}^{+-} & D_{1,\q{2},3}^{-+} & D_{2,\q{3},4}^{--} & D_{2,\q{3},4}^{++} & D_{1}^{+-} \\
               &            &              &              &
    D_{0,1,2,3,4}^{++} & D_{0,\q{1},2,\q{3},4}^{--} & D_{1,\q{2},3,\q{4},5}^{-+} & D_{1,\q{2},3,\q{4},5}^{+-} & D_{2}^{++} \\
               &            &              &              &
    & D_{0,1,2,3,4}^{++} & D_{1,\q{2},3,\q{4},5}^{+-} & D_{1,\q{2},3,\q{4},5}^{-+} & D_{2}^{--} \\
               &            &              &              &
    & & D_{0,1,2,3,4,5,6}^{++} & D_{0,\q{1},2,\q{3},4,\q{5},6}^{--} & D_{3}^{-+} \\
               &            &              &              &
    & & & D_{0,1,2,3,4,5,6}^{++} & D_{3}^{+-} \\ \hline \hline
  \end{tabular*}
\end{table*}

\subsection{General theory}
\label{app:compound-multipoles-general}

To illustrate how compound tensors can be formed from electric and
magnetic tensors, Table~\ref{tab:ir-prod} gives the multiplication
table for electric and magnetic IRs $D_\ell^{ss'}$ of the full
rotation group $\Rit$ with $\ell \le 3$.  The range of ranks $\ell$
contained in a product representation is given by Eq.\
(\ref{eq:l:ang-tot}).

For example, Table~\ref{tab:ir-prod} shows that a magnetic monopole
$T_0^{--}$ can arise as a product of an electric and a magnetic
multipole with the same $\ell > 0$.  These multipoles with $\ell >
0$ break spherical symmetry so that the symmetry group of a system
realizing such a synthetic magnetic monopole cannot be the spherical
magnetic group $R_{i \theta}$ (that would be the symmetry group of
an elementary magnetic monopole if it existed).  However, a product of
electric and magnetic multipoles with large $\ell$ can be
arbitrarily close to spherical symmetry, which indicates an avenue
for realizing ``realistic'' synthetic magnetic
monopoles~\cite{her04, cas08}.

Products (\ref{eq:tensor-prod}) of electric and magnetic tensors
yield not only electric and magnetic compound tensors but also
electrotoroidal and magnetotoroidal compound tensors (marked
with bold subscripts $\ell$ in Table~\ref{tab:ir-prod}).  For
example, an electric quadrupole $T_2^{++}$ and an electric dipole
$T_1^{-+}$ jointly give rise to an electrotoroidal quadrupole
$T_2^{-+}$.

Note that an electrotoroidal compound tensor $T_1^{++}$ does not
appear in Table~\ref{tab:ir-prod}.  Electrotoroidal compound tensors
$T_\ell^{++}$ with odd $\ell$ are realized by antisymmetric
(noncommuting) products $(T_{\ell_1}^{ss'} \times
T_{\ell_1}^{ss'})_\mathrm{a}$ of the same tensor $T_{\ell_1}^{ss'}$
or by products $T_{\ell_1}^{ss'} \times \tilde{T}_{\ell_2}^{ss'}$ of
two distinct tensors $T_{\ell_1}^{ss'}$ and
$\tilde{T}_{\ell_2}^{ss'}$ with the same signature $ss'$
\cite{edm60}.  However, odd\=/$\ell$ IRs $D_\ell^{++}$ appearing
off-diagonally in the multiplication table have $\ell \ge 3$.  The
IR $D_1^{++}$ arises only on the diagonal.  Assuming that for each
$\ell$ a system has only one electric and one magnetic multipole
$T_\ell^{ss'}$, second-order products of these multipoles thus
cannot realize an electrotoroidal compound tensor $T_1^{++}$.  This
is in line with the difficulties encountered in
Table~\ref{tab:tensor-pow-spher} and Appendix~\ref{app:axial-groups}
to realize and interpret such tensors.

Toroidal scalars $T_0^{\mp\pm}$ likewise do not arise as
second-order products of electric and magnetic tensors, see
Table~\ref{tab:ir-prod}, but they do arise in third-order products
of electric and magnetic tensors, for example:
\begin{equation}
  \label{eq:tens-prod:low}
  \newcommand{\tritens}[5]{\left.
    \marr[nLn]{%
    \raisebox{0pt}[0pt]{$\left. \marr[nLn]{#1 \\[1ex] #2} \right\}$} \; #3 \\[4ex]
    #4} \right\} \; #5}%
  \tritens{T_2^{++}}{T_4^{++}}{T_{3}^{++}}{T_3^{\mp\pm}}{T_0^{\mp\pm}}.
\end{equation}
Similar to magnetic scalars $T_0^{--}$, the toroidal scalars
$T_0^{-+}$ and $T_0^{+-}$ can only arise in electromagnetic media
that break spherical symmetry.

Inspection of Table~\ref{tab:ir-prod} and Eq.\
(\ref{eq:tens-prod:low}) shows that all nontrivial compound tensors
(except $T_1^{++}$ \cite{misc:T1++}) can be obtained as products
combining one nontrivial and one or two trivial tensors.  By
definition, trivial tensors have the signature $++$.  Therefore,
toroidal compound tensors inherit the signature of the nontrivial
electromagnetic tensor they are composed of, while the index $\ell'$
of toroidal compound tensors has the opposite parity as the index
$\ell$ of the nontrivial tensor they are composed of.  Consistent
with Tables~\ref{tab:tensor-pow-spher}
and~\ref{tab:tensor-pow-cryst}, we thus interpret toroidal
compound tensors as manifestations of the nontrivial electromagnetic
tensor they are composed of.

The above group-theoretical analysis of compound tensors is
independent of specific realizations of the tensors $T_\ell^{ss'}$
as discussed in Table~\ref{tab:tensor-pow-spher}.  It is equally
applicable to finite and infinite periodic systems.

Consistent with our definition of categories of multipole order
based on the signature $ss'$, the terms electropolarization
(signature $-+$), magnetopolarization ($+-$), and
antimagnetopolarization ($--$) refer jointly to the respective
odd\=/$\ell$ electric, odd\=/$\ell$ magnetic, and even\=/$\ell$
magnetic multipoles, as well as the corresponding toroidal compound
moments with the same signature, unless we want to emphasize the
composition of toroidal compound tensors.

\subsection{Examples of compound moments}
\label{app:examples-compound}

Reference~\cite{win23} contains a detailed study of how electric and
magnetic multipoles manifest themselves via characteristic terms in
the electronic band structure of a crystal, using variations of
lonsdaleite and diamond as examples.  A careful inspection shows
that a number of examples discussed in that work rely on compound
moments.

For instance, we can have three qualitatively distinct homogeneous
polynomials in $k,\sigma$ of degree~2 in a crystal, two of which are
compound moments.  (i) Crystal structures with point groups $T$ and
$O$ permit the crystallographic tensor $\vek{\sigma} \cdot \kk$.  This
term transforms like an electrotoroidal compound scalar $T_0^{-+}$
that arises as a product (\ref{eq:tens-prod:low}) of a nontrivial
odd\=/$\ell$ electric multipole and two trivial even\=/$\ell$
electric multipoles.  See also Sec.~\ref{sec:chirality}.  (ii)
Electropolar crystal structures such as wurtzite (point group
$C_{6v}$, ignoring TIS) permit an electric dipole density that
manifests itself via the Rashba term $k_x \sigma_y - k_y \sigma_x$
\cite{ras59a}.  Under $\Rit$, this term transforms like the
component $T_{10}^{-+}$ of an electric dipole density.  (iii) The
cubic zincblende structure (point group $T_d$) permits an electric
octupole density ($\ell = 3$) that manifests itself via the
Dresselhaus term $\sigma_x k_x (k_y^2-k_z^2) + \cp$, where $\cp$
denotes cyclic permutation \cite{dre55}.  Under $\Rit$, this term
transforms like the component $T_{3,-2}^{-+} + T_{3,2}^{-+}$ of an
electric octupole density ($\ell = 3$).  When uniaxial strain is
applied to a zincblende structure (in [001] direction, thus reducing
the crystal symmetry from $T_d$ to $D_{2d}$), we obtain the crystallographic
tensor $k_x \sigma_x - k_y \sigma_y$.  Under $\Rit$, this term
transforms like the component $T_{2,-2}^{-+} + T_{2,2}^{-+}$ of an
electrotoroidal quadrupole density that arises as a product
(\ref{eq:tensor-prod}) of the nontrivial electric octupole density
($\ell=3$) that exists already in unstrained zincblende and a
trivial strain-induced electric quadrupole density ($\ell = 2$).
The electrotoroidal quadrupole density represents the lowest
even-rank electrotoroidal moment density permitted in a structure
with point group $D_{2d}$ (Appendix~\ref{app:spherTensTab},
Table~\ref{tab:tensor-nonmag}).

Pristine lonsdaleite has four identical atoms per unit cell; its
crystallographic point group is $D_{6h} = D_6 \times C_i$ (ignoring
TIS).  If the four sublattices are occupied by two types of atoms as
in Fig.~2(b) of Ref.~\cite{win23}, this results in an electric
octupolarization ($\ell = 3$), and the point-group symmetry is
reduced to $D_{3h}$.  In addition, two new invariants arise in
the band structure (to lowest order in~$\kk$) \cite{win23}
\begin{subequations}
  \label{eq:lons:e3:H}
  \begin{align}
    H_1 & = c_1 \, \sigma_z \, k_y \left( 3 k_x^2 - k_y^2 \right) \\*
    H_2 & = c_2 \, k_z
    \left[ \sigma_x k_x k_y + \frack{1}{2} \sigma_y (k_x^2 - k_y^2) \right] \,,
  \end{align}
\end{subequations}
where $c_1$ and $c_2$ are material-specific prefactors.  These terms can be
rearranged as
\begin{subequations}
  \label{eq:lons:e3:H-alt}
  \begin{align}
    H^{(3)} & = \frack{5}{8} (H_1 - 2 H_2) \\*
    H^{(4)} & = \frack{3}{8} (H_1 + 6 H_2) \, .
  \end{align}
\end{subequations}
Here, the first term transforms like the component $T_{3,-3}^{-+} +
T_{3,3}^{-+}$ of an electric octupole density ($\ell = 3$), while
the second term transforms like the component $T_{4,-3}^{-+} +
T_{4,3}^{-+}$ of an electrotoroidal hexadecapole density ($\ell =
4$).  The latter represents the lowest even-rank electrotoroidal
moment density permitted in a structure with point group
$D_{3h}$ (Appendix~\ref{app:spherTensTab},
Table~\ref{tab:tensor-nonmag}).  It is realized as a compound
moment density that combines the nontrivial electric
octupole density ($\ell = 3$) with the trivial electric quadrupole
density ($\ell = 2$) present already in pristine lonsdaleite.

An important example of compound moments arises in the context of
antimagnetopolarizations.  Very generally, unlike
electropolarizations and magnetopolarizations, even\=/$\ell$
antimagnetopolarizations $T_\ell^{--}$ have no indicators in the
band structure associated with them, see
Table~\ref{tab:tensor-pow-spher}.
However, antimagnetopolarizations combined with trivial
even\=/$\ell$ parapolarizations give rise to odd\=/$\ell$
magnetotoroidal densities $T_\ell^{--}$ that manifest themselves in
the band structure via terms proportional to odd powers in $k$
(Table~\ref{tab:tensor-pow-spher}).  Therefore, we associate
antimagnetopolarizations with such terms $k^{2n+1}$.  While
odd\=/$\ell$ electropolarizations and odd\=/$\ell$
magnetopolarizations permit a simple correspondence between the rank
$\ell$ and a minimal-degree polynomial representation of the
associated tensors (Table~\ref{tab:tensor-pow-spher}), no such
correspondence exists for even\=/$\ell$ antimagnetopolarizations.

For example, in a diamond crystal structure, a magnetic quadrupole
density ($\ell = 2$) gives rise to a term $k_z (k_x^2 -
k_y^2)$~\cite{win20, win23}.  Under $\Rit$, this term transforms
like the component $T_{3,-2}^{--} + T_{3,2}^{--}$ of a
magnetotoroidal octupole density ($\ell = 3$).  This term arises as
a product (\ref{eq:tensor-prod}) of a nontrivial magnetic quadrupole
density ($\ell=2$) and a trivial electric hexadecapole density
($\ell = 4$) present already in pristine diamond.

A complete review of the examples in Ref.~\cite{win23} is beyond the
scope of the present work.  Tables \ref{tab:tensor-nonmag} and
\ref{tab:tensor-mag} in Appendix~\ref{app:spherTensTab} give the
lowest ranks of electromagnetic multipoles and toroidal compound
multipoles in crystal structures with different crystallographic
point groups $G$.

\section{Spherical tensors and multipole order permitted by the 122
crystallographic magnetic point groups}
\label{app:spherTensTab}

This appendix provides explicit details about which spherical
tensors $T_\ell^{ss'}$ are permitted by the 122 crystallographic
magnetic point groups.  The information given here is directly
relevant for discussing electric, magnetic and toroidal multipole
order in crystals, see Sec.~\ref{sec:categories-multipole}.  As
discussed below, the information is furthermore relevant for
materials and property tensors that are commonly represented by
Cartesian rank\=/$N$ tensors.

\begin{table*}[tbp]
  \caption{\label{tab:tensor-nonmag} Number of independent
  components of $\theta$\=/even spherical tensors $T_\ell^{s+}$
  permitted by type\=/II magnetic point groups (considering ranks
  $\ell \le 3$).  Tensors $T_\ell^{++}$ represent parapolarizations
  and $T_\ell^{-+}$ are electropolarizations.  A ``0'' indicates
  that the tensor is forbidden by symmetry.  The electric
  monopolarization $T_0^{++}$ has been omitted, as it is trivially
  always allowed.  The quantities
  $\ell_\mathrm{min}^{(\mathrm{e/et,g/u})}$ indicate, respectively,
  for each group, the lowest symmetry allowed even (``$g$'') and odd
  (``$u$'') orders $\ell$ of electric (``e'') and electrotoroidal
  (``et'') multipole densities $\ell > 0$, i.e., again ignoring
  the electric and electrotoroidal monopolarizations $T_0^{++}$ and
  $T_0^{-+}$.  Both the Sch\"onflies and Hermann-Mauguin
  notations~\cite{bur06} are given for each group, as well as the
  group's type according to the classification in
  Table~\ref{tab:group-types-def}.}
 \vspace{2pt}
\let\mc\multicolumn
\newcommand* {\mb}[2][1.3em]{\makebox[#1]{$#2$}}%
\newcommand{\mtab}[2][@{}l@{}]{\mbox{\renewcommand{\arraystretch}{1.0}%
  \tabcolsep 0pt\begin{tabular}{#1} #2 \end{tabular}}}%
\tabcolsep 0.05em
\newcolumntype {T}[1]{CCs{0.8em}*{#1}{C}}%
\begin{tabular*}{\linewidth}{nlLs{0.15em}LECcT{3}s{1.5em}T{4}n} \hline \hline
\rule{0pt}{2.5ex} & & & & & \mc{5}{c}{parapolar} & \mc{6}{c}{electropolar} \\[-0.5ex]
Crystal system & \mc{2}{c}{Sch\"onflies} & \mtab{Hermann-\\ Mauguin} & type &
\mb{\ell_{\min}^{(e,g)}}  &
\mb{\ell_{\min}^{(et,u)}} &
\mb{T_1^{++}} & \mb{T_2^{++}} & \mb{T_3^{++}} &
\mb{\ell_{\min}^{(et,g)}} &
\mb{\ell_{\min}^{(e,u)}}  &
\mb{T_0^{-+}} & \mb{T_1^{-+}} & \mb{T_2^{-+}} & \mb{T_3^{-+}} \\ \hline \rule{0pt}{2.5ex}%
Triclinic & C_1 \times \gtis & = C_1 \times \gtis & 1' & i-II & 2 & 1 & 3 & 5 & 7 & 2 & 1 & 1 & 3 & 5 & 7 \\
 & C_1 \times \gsistis & = \gsis \times \gtis & \bar{1}1' & ii-II & 2 & 1 & 3 & 5 & 7 & \infty & \infty & 0 & 0 & 0 & 0 \\ \hline \rule{0pt}{2.5ex}%
Monoclinic & C_2 \times \gtis & = C_2 \times \gtis & 21' & i-II & 2 & 1 & 1 & 3 & 3 & 2 & 1 & 1 & 1 & 3 & 3 \\
 & C_2 [C_1] \times \gtis & = C_s \times \gtis & m1' & iii-II & 2 & 1 & 1 & 3 & 3 & 2 & 1 & 0 & 2 & 2 & 4 \\
 & C_2 \times \gsistis & = C_{2h} \times \gtis & 2/m1' & ii-II & 2 & 1 & 1 & 3 & 3 & \infty & \infty & 0 & 0 & 0 & 0 \\ \hline \rule{0pt}{2.5ex}%
Orthorhombic & D_2 \times \gtis & = D_2 \times \gtis & 2221' & i-II & 2 & 3 & 0 & 2 & 1 & 2 & 3 & 1 & 0 & 2 & 1 \\
 & D_2 [C_2] \times \gtis & = C_{2v} \times \gtis & mm21' & iii-II & 2 & 3 & 0 & 2 & 1 & 2 & 1 & 0 & 1 & 1 & 2 \\
 & D_2 \times \gsistis & = D_{2h} \times \gtis & mmm1' & ii-II & 2 & 3 & 0 & 2 & 1 & \infty & \infty & 0 & 0 & 0 & 0 \\ \hline \rule{0pt}{2.5ex}%
Tetragonal & C_4 \times \gtis & = C_4 \times \gtis & 41' & i-II & 2 & 1 & 1 & 1 & 1 & 2 & 1 & 1 & 1 & 1 & 1 \\
 & C_4 [C_2] \times \gtis & = S_4 \times \gtis & \bar{4}1' & iii-II & 2 & 1 & 1 & 1 & 1 & 2 & 3 & 0 & 0 & 2 & 2 \\
 & C_4 \times \gsistis & = C_{4h} \times \gtis & 4/m1' & ii-II & 2 & 1 & 1 & 1 & 1 & \infty & \infty & 0 & 0 & 0 & 0 \\ \cline{2-16} \rule{0pt}{2.5ex}%
 & D_4 \times \gtis & = D_4 \times \gtis & 4221' & i-II & 2 & 5 & 0 & 1 & 0 & 2 & 5 & 1 & 0 & 1 & 0 \\
 & D_4 [C_4] \times \gtis & = C_{4v} \times \gtis & 4mm1' & iii-II & 2 & 5 & 0 & 1 & 0 & 4 & 1 & 0 & 1 & 0 & 1 \\
 & D_4 [D_2] \times \gtis & = D_{2d} \times \gtis & \bar{4}2m1' & iii-II & 2 & 5 & 0 & 1 & 0 & 2 & 3 & 0 & 0 & 1 & 1 \\
 & D_4 \times \gsistis & = D_{4h} \times \gtis & 4/mmm1' & ii-II & 2 & 5 & 0 & 1 & 0 & \infty & \infty & 0 & 0 & 0 & 0 \\ \hline \rule{0pt}{2.5ex}%
Trigonal & C_3 \times \gtis & = C_3 \times \gtis & 31' & i-II & 2 & 1 & 1 & 1 & 3 & 2 & 1 & 1 & 1 & 1 & 3 \\
 & C_3 \times \gsistis & = C_{3i} \times \gtis & \bar{3}1' & ii-II & 2 & 1 & 1 & 1 & 3 & \infty & \infty & 0 & 0 & 0 & 0 \\ \cline{2-16} \rule{0pt}{2.5ex}%
 & D_3 \times \gtis & = D_3 \times \gtis & 321' & i-II & 2 & 3 & 0 & 1 & 1 & 2 & 3 & 1 & 0 & 1 & 1 \\
 & D_3 [C_3] \times \gtis & = C_{3v} \times \gtis & 3m1' & iii-II & 2 & 3 & 0 & 1 & 1 & 4 & 1 & 0 & 1 & 0 & 2 \\
 & D_3 \times \gsistis & = D_{3d} \times \gtis & \bar{3}m1' & ii-II & 2 & 3 & 0 & 1 & 1 & \infty & \infty & 0 & 0 & 0 & 0 \\ \hline \rule{0pt}{2.5ex}%
Hexagonal & C_6 \times \gtis & = C_6 \times \gtis & 61' & i-II & 2 & 1 & 1 & 1 & 1 & 2 & 1 & 1 & 1 & 1 & 1 \\
 & C_6 [C_3] \times \gtis & = C_{3h} \times \gtis & \bar{6}1' & iii-II & 2 & 1 & 1 & 1 & 1 & 4 & 3 & 0 & 0 & 0 & 2 \\
 & C_6 \times \gsistis & = C_{6h} \times \gtis & 6/m1' & ii-II & 2 & 1 & 1 & 1 & 1 & \infty & \infty & 0 & 0 & 0 & 0 \\ \cline{2-16} \rule{0pt}{2.5ex}%
 & D_6 \times \gtis & = D_6 \times \gtis & 6221' & i-II & 2 & 7 & 0 & 1 & 0 & 2 & 7 & 1 & 0 & 1 & 0 \\
 & D_6 [C_6] \times \gtis & = C_{6v} \times \gtis & 6mm1' & iii-II & 2 & 7 & 0 & 1 & 0 & 6 & 1 & 0 & 1 & 0 & 1 \\
 & D_6 [D_3] \times \gtis & = D_{3h} \times \gtis & \bar{6}m21' & iii-II & 2 & 7 & 0 & 1 & 0 & 4 & 3 & 0 & 0 & 0 & 1 \\
 & D_6 \times \gsistis & = D_{6h} \times \gtis & 6/mmm1' & ii-II & 2 & 7 & 0 & 1 & 0 & \infty & \infty & 0 & 0 & 0 & 0 \\ \hline \rule{0pt}{2.5ex}%
Cubic & T \times \gtis & = T \times \gtis & 231' & i-II & 4 & 3 & 0 & 0 & 1 & 4 & 3 & 1 & 0 & 0 & 1 \\
 & T \times \gsistis & = T_h \times \gtis & m\bar{3}1' & ii-II & 4 & 3 & 0 & 0 & 1 & \infty & \infty & 0 & 0 & 0 & 0 \\
  \cline{2-16} \rule{0pt}{2.5ex}%
 & O \times \gtis & = O \times \gtis & 4321' & i-II & 4 & 9 & 0 & 0 & 0 & 4 & 9 & 1 & 0 & 0 & 0 \\
 & O [T] \times \gtis & = T_d \times \gtis & \bar{4}3m1' & iii-II & 4 & 9 & 0 & 0 & 0 & 6 & 3 & 0 & 0 & 0 & 1 \\
 & O \times \gsistis & = O_h \times \gtis & m\bar{3}m1' & ii-II & 4 & 9 & 0 & 0 & 0 & \infty & \infty & 0 & 0 & 0 & 0 \\ \hline
\mc{7}{nl}{Number of crystallographic groups allowing tensor \rule{0pt}{2.5ex}} & 13 & 27 & 21 & & & 11 & 10 & 13 & 18 \\ \hline \rule{0pt}{2.5ex}%
Axial & C_\infty \times \gtis & = C_\infty \times \gtis & \infty 1' & i-II & 2 & 1 & 1 & 1 & 1 & 2 & 1 & 1 & 1 & 1 & 1 \\
 & C_\infty \times \gsistis & = C_{\infty h} \times \gtis & \infty /m1' & ii-II & 2 & 1 & 1 & 1 & 1 & \infty & \infty & 0 & 0 & 0 & 0 \\ \cline{2-16} \rule{0pt}{2.5ex}%
 & D_\infty \times \gtis & = D_\infty \times \gtis & \infty 21' & i-II & 2 & \infty & 0 & 1 & 0 & 2 & \infty & 1 & 0 & 1 & 0 \\
 & D_\infty [C_\infty] \times \gtis & = C_{\infty v} \times \gtis & \infty m1' & iii-II & 2 & \infty & 0 & 1 & 0 & \infty & 1 & 0 & 1 & 0 & 1 \\
 & D_\infty \times \gsistis & = D_{\infty h} \times \gtis & \infty /mm1' & ii-II & 2 & \infty & 0 & 1 & 0 & \infty & \infty & 0 & 0 & 0 & 0 \\ \hline \rule{0pt}{2.5ex}%
Spherical & R \times \gtis & = R_\theta & \infty \infty 1' & i-II & \infty & \infty & 0 & 0 & 0 & \infty & \infty & 1 & 0 & 0 & 0 \\
 & R \times \gsistis & = \Rit & \infty \infty m1' & ii-II & \infty & \infty & 0 & 0 & 0 & \infty & \infty & 0 & 0 & 0 & 0 \\ \hline \hline
\end{tabular*}
\end{table*}

Table~\ref{tab:tensor-nonmag} is devoted to $\theta$\=/even tensors
$T_\ell^{s+}$.  It lists the number of independent components such
tensors with rank $\ell \le 3$ have under the nonmagnetic
(type\=/II) groups belonging to the 11 Laue classes $L (\gprop)$.
For comparison, the same information is also provided for type\=/II
groups in the three classes of continuous cyclic, dihedral and
spherical point groups.  To streamline the presentation, the
universally allowed electric monopolarization $T_0^{++}$ has been
omitted.  Basic physical properties can be read off from this table.
For example, groups allowing $T_0^{-+}$ represent electrochiral
systems, see Sec.~\ref{sec:electro-chiral}, and groups allowing
$T_1^{-+}$ represent pyroelectric and ferroelectric
systems~\cite{nye85}.

Table~\ref{tab:tensor-nonmag} together with
Table~\ref{tab:mag-point-group} can be used to infer the properties
of tensors $T_\ell^{s+}$ in all 122 magnetic point groups.
Specifically, Table~\ref{tab:tensor-nonmag} gives the properties of
tensors for the crystallographic point groups $G_0^\theta \times
\gtis$ that appear in Table~\ref{tab:mag-point-group} in the last
row of the main block~\cite{misc:mainBlock} for a class $L
(\gprop)$.  The structure of tensors $T_\ell^{s+}$ allowed
under $G_0^\theta \times \gtis$ is repeated for all members of the
subclass $L (G_0^\theta, \theta) \subset L (\gprop)$, which
are the groups positioned in the column headed by $G_0^\theta$
within the same class's main block.  The formal basis for these
congruences is elaborated in detail in Appendix~\ref{app:tensors}.

Table~\ref{tab:tensor-nonmag} also tabulates the lowest even
(``$g$'') and odd (``$u$'') orders $\ell$ of electric (``e'') and
electrotoroidal (``et'') multipole densities permitted in each
crystal class.  The electric monopolarization $T_0^{++}$ is
trivially allowed under all point groups and, therefore, is not
considered for determining $\ell_\mathrm{min}^{(\mathrm{e,g})}$.
Instead, to convey information that is specific to each group, we
tabulate the lowest \emph{nonzero} even order of
parapolarizations allowed under each group.

\begin{table*}[tbp]
  \caption{\label{tab:tensor-mag} Number of independent components
  of $\theta$\=/odd spherical tensors $T_\ell^{s-}$ permitted by
  type\=/III magnetic point groups (considering ranks $\ell \le 3$).
  Tensors $T_\ell^{+-}$ represent magnetopolarizations and
  $T_\ell^{--}$ are antimagnetopolarizations.  A ``0'' indicates
  that the corresponding tensor is forbidden by symmetry.  The
  magnetic monopolarization $T_0^{+-}$ has been omitted, as it is
  universally forbidden under type\=/III groups.  The quantities
  $\ell_\mathrm{min}^{(\mathrm{m/mt,g/u})}$ indicate, respectively,
  for each group, the lowest symmetry-allowed even (``$g$'') and odd
  (``$u$'') orders $\ell$ of magnetic (``m'') and magnetotoroidal
  (``mt'') multipole densities $\ell > 0$, i.e., ignoring
  monopolarizations.  Both the Sch\"onflies and Hermann-Mauguin
  notations~\cite{bur06} are given for each group, as well as the
  group's type according to the classification in
  Table~\ref{tab:group-types-def}.}
 \vspace{2pt}
\let\mc\multicolumn
\newcommand* {\mb}[2][1.3em]{\makebox[#1]{$#2$}}%
\newcommand{\mtab}[2][@{}l@{}]{\mbox{\renewcommand{\arraystretch}{1.0}%
  \tabcolsep 0pt\begin{tabular}{#1} #2 \end{tabular}}}%
\tabcolsep 0.05em
\newcolumntype {T}[1]{CCs{0.8em}*{#1}{C}}%
\begin{tabular*}{\linewidth}{nlLs{0.15em}LECcT{3}s{1.0em}T{4}n} \hline \hline
\rule{0pt}{2.5ex} & & & & & \mc{5}{c}{magnetopolar} & \mc{6}{c}{antimagnetopolar} \\[-0.5ex]
Crystal system & \mc{2}{c}{Sch\"onflies} & \mtab{Hermann-\\ Mauguin} & type &
\mb[1.7em]{\ell_{\min}^{(mt,g)}} &
\mb[1.7em]{\ell_{\min}^{(m,u)}}  &
\mb{T_1^{+-}} & \mb{T_2^{+-}} & \mb{T_3^{+-}} &
\mb[1.7em]{\ell_{\min}^{(m,g)}} &
\mb[1.7em]{\ell_{\min}^{(mt,u)}}  &
\mb{T_0^{--}} & \mb{T_1^{--}} & \mb{T_2^{--}} & \mb{T_3^{--}} \\ \hline \rule{0pt}{2.5ex}%
Triclinic & C_1 \times \gstis & = \gsis (C_1) & \bar{1}' & JJ & \infty & \infty & 0 & 0 & 0 & 2 & 1 & 1 & 3 & 5 & 7 \\ \hline \rule{0pt}{2.5ex}%
Monoclinic & C_2 (C_1) & = C_2 (C_1) & 2' & i-III & 2 & 1 & 2 & 2 & 4 & 2 & 1 & 0 & 2 & 2 & 4 \\
 & C_2  \{ C_1 \} & = C_s (C_1) & m' & JJJ & 2 & 1 & 2 & 2 & 4 & 2 & 1 & 1 & 1 & 3 & 3 \\
 & C_2 (C_1) \times \gsis & = C_{2h} (\gsis) & 2'/m' & ii-III & 2 & 1 & 2 & 2 & 4 & \infty & \infty & 0 & 0 & 0 & 0 \\
 & C_2 \times \gstis & = C_{2h} (C_2) & 2/m' & JJ & \infty & \infty & 0 & 0 & 0 & 2 & 1 & 1 & 1 & 3 & 3 \\
 & C_2 (C_1) \times \gstis & = C_{2h} (C_s) & 2'/m & JJ' & \infty & \infty & 0 & 0 & 0 & 2 & 1 & 0 & 2 & 2 & 4 \\ \hline \rule{0pt}{2.5ex}%
Orthorhombic & D_2 (C_2) & = D_2 (C_2) & 2'2'2 & i-III & 2 & 1 & 1 & 1 & 2 & 2 & 1 & 0 & 1 & 1 & 2 \\
 & D_2  \{ C_2 \} & = C_{2v} (C_2) & m'm'2 & JJJ & 2 & 1 & 1 & 1 & 2 & 2 & 3 & 1 & 0 & 2 & 1 \\
 & D_2[C_2] (C_2'[C_1]) & = C_{2v} (C_s) & 2'm'm & JJJ'' & 2 & 1 & 1 & 1 & 2 & 2 & 1 & 0 & 1 & 1 & 2 \\
 & D_2 (C_2) \times \gsis & = D_{2h} (C_{2h}) & mm'm' & ii-III & 2 & 1 & 1 & 1 & 2 & \infty & \infty & 0 & 0 & 0 & 0 \\
 & D_2 \times \gstis & = D_{2h} (D_2) & m'm'm' & JJ & \infty & \infty & 0 & 0 & 0 & 2 & 3 & 1 & 0 & 2 & 1 \\
 & D_2 (C_2) \times \gstis & = D_{2h} (C_{2v}) & mmm' & JJ' & \infty & \infty & 0 & 0 & 0 & 2 & 1 & 0 & 1 & 1 & 2 \\ \hline \rule{0pt}{2.5ex}%
Tetragonal & C_4 (C_2) & = C_4 (C_2) & 4' & i-III & 2 & 3 & 0 & 2 & 2 & 2 & 3 & 0 & 0 & 2 & 2 \\
 & C_4  \{  C_2 \} & = S_4 (C_2) & \bar{4}' & JJJ & 2 & 3 & 0 & 2 & 2 & 2 & 1 & 1 & 1 & 1 & 1 \\
 & C_4 (C_2) \times \gsis & = C_{4h} (C_{2h}) & 4'/m & ii-III & 2 & 3 & 0 & 2 & 2 & \infty & \infty & 0 & 0 & 0 & 0 \\
 & C_4 \times \gstis & = C_{4h} (C_4) & 4/m' & JJ & \infty & \infty & 0 & 0 & 0 & 2 & 1 & 1 & 1 & 1 & 1 \\
 & C_4 (C_2) \times \gstis & = C_{4h} (S_4) & 4'/m' & JJ' & \infty & \infty & 0 & 0 & 0 & 2 & 3 & 0 & 0 & 2 & 2 \\ \cline{2-16} \rule{0pt}{2.5ex}%
 & D_4 (D_2) & = D_4 (D_2) & 4'22' & i-III & 2 & 3 & 0 & 1 & 1 & 2 & 3 & 0 & 0 & 1 & 1 \\
 & D_4 (C_4) & = D_4 (C_4) & 42'2' & i-III & 4 & 1 & 1 & 0 & 1 & 4 & 1 & 0 & 1 & 0 & 1 \\
 & D_4[C_4] (D_2[C_2]) & = C_{4v} (C_{2v}) & 4'mm' & iii'-III & 2 & 3 & 0 & 1 & 1 & 2 & 3 & 0 & 0 & 1 & 1 \\
 & D_4  \{ C_4 \} & = C_{4v} (C_4) & 4m'm' & JJJ & 4 & 1 & 1 & 0 & 1 & 2 & 5 & 1 & 0 & 1 & 0 \\
 & D_4  \{ D_2 \} & = D_{2d} (D_2) & \bar{4}'2m' & JJJ & 2 & 3 & 0 & 1 & 1 & 2 & 5 & 1 & 0 & 1 & 0 \\
 & D_4[D_2] (D_2'[C_2]) & = D_{2d} (C_{2v}) & \bar{4}'m2' & JJJ' & 2 & 3 & 0 & 1 & 1 & 4 & 1 & 0 & 1 & 0 & 1 \\
 & D_4[D_2] (C_4[C_2]) & = D_{2d} (S_4) & \bar{4}2'm' & iii-III' & 4 & 1 & 1 & 0 & 1 & 2 & 3 & 0 & 0 & 1 & 1 \\
 & D_4 (D_2) \times \gsis & = D_{4h} (D_{2h}) & 4'/mmm' & ii-III & 2 & 3 & 0 & 1 & 1 & \infty & \infty & 0 & 0 & 0 & 0 \\
 & D_4 (C_4) \times \gsis & = D_{4h} (C_{4h}) & 4/mm'm' & ii-III & 4 & 1 & 1 & 0 & 1 & \infty & \infty & 0 & 0 & 0 & 0 \\
 & D_4 \times \gstis & = D_{4h} (D_4) & 4/m'm'm' & JJ & \infty & \infty & 0 & 0 & 0 & 2 & 5 & 1 & 0 & 1 & 0 \\
 & D_4 (C_4) \times \gstis & = D_{4h} (C_{4v}) & 4/m'mm & JJ' & \infty & \infty & 0 & 0 & 0 & 4 & 1 & 0 & 1 & 0 & 1 \\
 & D_4 (D_2) \times \gstis & = D_{4h} (D_{2d}) & 4'/m'm'm & JJ' & \infty & \infty & 0 & 0 & 0 & 2 & 3 & 0 & 0 & 1 & 1 \\ \hline \rule{0pt}{2.5ex}%
Trigonal & C_3 \times \gstis & = C_{3i} (C_3) & \bar{3}' & JJ & \infty & \infty & 0 & 0 & 0 & 2 & 1 & 1 & 1 & 1 & 3 \\ \cline{2-16} \rule{0pt}{2.5ex}%
 & D_3 (C_3) & = D_3 (C_3) & 32' & i-III & 4 & 1 & 1 & 0 & 2 & 4 & 1 & 0 & 1 & 0 & 2 \\
 & D_3  \{ C_3 \} & = C_{3v} (C_3) & 3m' & JJJ & 4 & 1 & 1 & 0 & 2 & 2 & 3 & 1 & 0 & 1 & 1 \\
 & D_3 (C_3) \times \gsis & = D_{3d} (C_{3i}) & \bar{3}m' & ii-III & 4 & 1 & 1 & 0 & 2 & \infty & \infty & 0 & 0 & 0 & 0 \\
 & D_3 \times \gstis & = D_{3d} (D_3) & \bar{3}'m' & JJ & \infty & \infty & 0 & 0 & 0 & 2 & 3 & 1 & 0 & 1 & 1 \\
 & D_3 (C_3) \times \gstis & = D_{3d} (C_{3v}) & \bar{3}'m & JJ' & \infty & \infty & 0 & 0 & 0 & 4 & 1 & 0 & 1 & 0 & 2 \\ \hline \rule{0pt}{2.5ex}%
Hexagonal & C_6 (C_3) & = C_6 (C_3) & 6' & i-III & 4 & 3 & 0 & 0 & 2 & 4 & 3 & 0 & 0 & 0 & 2 \\
 & C_6  \{ C_3 \} & = C_{3h} (C_3) & \bar{6}' & JJJ & 4 & 3 & 0 & 0 & 2 & 2 & 1 & 1 & 1 & 1 & 1 \\
 & C_6 (C_3) \times \gsis & = C_{6h} (C_{3i}) & 6'/m' & ii-III & 4 & 3 & 0 & 0 & 2 & \infty & \infty & 0 & 0 & 0 & 0 \\
 & C_6 \times \gstis & = C_{6h} (C_6) & 6/m' & JJ & \infty & \infty & 0 & 0 & 0 & 2 & 1 & 1 & 1 & 1 & 1 \\
 & C_6 (C_3) \times \gstis & = C_{6h} (C_{3h}) & 6'/m & JJ' & \infty & \infty & 0 & 0 & 0 & 4 & 3 & 0 & 0 & 0 & 2 \\ \cline{2-16} \rule{0pt}{2.5ex}%
 & D_6 (D_3) & = D_6 (D_3) & 6'22' & i-III & 4 & 3 & 0 & 0 & 1 & 4 & 3 & 0 & 0 & 0 & 1 \\
 & D_6 (C_6) & = D_6 (C_6) & 62'2' & i-III & 6 & 1 & 1 & 0 & 1 & 6 & 1 & 0 & 1 & 0 & 1 \\
 & D_6[C_6] (D_3[C_3]) & = C_{6v} (C_{3v}) & 6'mm' & iii'-III & 4 & 3 & 0 & 0 & 1 & 4 & 3 & 0 & 0 & 0 & 1 \\
 & D_6  \{ C_6 \} & = C_{6v} (C_6) & 6m'm' & JJJ & 6 & 1 & 1 & 0 & 1 & 2 & 7 & 1 & 0 & 1 & 0 \\
 & D_6  \{ D_3 \} & = D_{3h} (D_3) & \bar{6}'2m' & JJJ & 4 & 3 & 0 & 0 & 1 & 2 & 7 & 1 & 0 & 1 & 0 \\
 & D_6[D_3] (D_3'[C_3]) & = D_{3h} (C_{3v}) & \bar{6}'m2' & JJJ' & 4 & 3 & 0 & 0 & 1 & 6 & 1 & 0 & 1 & 0 & 1 \\
 & D_6[D_3] (C_6[C_3]) & = D_{3h} (C_{3h}) & \bar{6}m'2' & iii-III' & 6 & 1 & 1 & 0 & 1 & 4 & 3 & 0 & 0 & 0 & 1 \\
 & D_6 (D_3) \times \gsis & = D_{6h} (D_{3d}) & 6'/m'mm' & ii-III & 4 & 3 & 0 & 0 & 1 & \infty & \infty & 0 & 0 & 0 & 0 \\
 & D_6 (C_6) \times \gsis & = D_{6h} (C_{6h}) & 6/mm'm' & ii-III & 6 & 1 & 1 & 0 & 1 & \infty & \infty & 0 & 0 & 0 & 0 \\
 & D_6 \times \gstis & = D_{6h} (D_6) & 6/m'm'm' & JJ & \infty & \infty & 0 & 0 & 0 & 2 & 7 & 1 & 0 & 1 & 0 \\
 & D_6 (C_6) \times \gstis & = D_{6h} (C_{6v}) & 6/m'mm & JJ' & \infty & \infty & 0 & 0 & 0 & 6 & 1 & 0 & 1 & 0 & 1 \\
 & D_6 (D_3) \times \gstis & = D_{6h} (D_{3h}) & 6'/mmm' & JJ' & \infty & \infty & 0 & 0 & 0 & 4 & 3 & 0 & 0 & 0 & 1 \\ \hline \hline
\end{tabular*}
\end{table*}

\begin{table*}[tbp]
  \centerline{\tablename~\ref{tab:tensor-mag}.  (\emph{Continued.})}
 \vspace{2pt}
\let\mc\multicolumn
\newcommand* {\mb}[2][1.3em]{\makebox[#1]{$#2$}}%
\newcommand{\mtab}[2][@{}l@{}]{\mbox{\renewcommand{\arraystretch}{1.0}%
  \tabcolsep 0pt\begin{tabular}{#1} #2 \end{tabular}}}%
\tabcolsep 0.05em
\newcolumntype {T}[1]{CCs{0.8em}*{#1}{C}}%
\begin{tabular*}{\linewidth}{nlLs{0.15em}LECcT{3}s{1.0em}T{4}n} \hline \hline
\rule{0pt}{2.5ex} & & & & & \mc{5}{c}{magnetopolar} & \mc{6}{c}{antimagnetopolar} \\[-0.5ex]
Crystal system & \mc{2}{c}{Sch\"onflies} & \mtab{Hermann-\\ Mauguin} & type &
\mb[1.7em]{\ell_{\min}^{(mt,g)}} &
\mb[1.7em]{\ell_{\min}^{(m,u)}}  &
\mb{T_1^{+-}} & \mb{T_2^{+-}} & \mb{T_3^{+-}} &
\mb[1.7em]{\ell_{\min}^{(m,g)}} &
\mb[1.7em]{\ell_{\min}^{(mt,u)}}  &
\mb{T_0^{--}} & \mb{T_1^{--}} & \mb{T_2^{--}} & \mb{T_3^{--}} \\ \hline \rule{0pt}{2.5ex}%
Cubic & T \times \gstis & = T_h (T) & m'\bar{3}'
 & JJ & \infty & \infty & 0 & 0 & 0 & 4 & 3 & 1 & 0 & 0 & 1 \\ \cline{2-16} \rule{0pt}{2.5ex}%
 & O (T) & = O (T) & 4'32' & i-III & 6 & 3 & 0 & 0 & 1 & 6 & 3 & 0 & 0 & 0 & 1 \\
 & O  \{ T \} & = T_d (T) & \bar{4}'3m' & JJJ & 6 & 3 & 0 & 0 & 1 & 4 & 9 & 1 & 0 & 0 & 0 \\
 & O (T) \times \gsis & = O_h (T_h) & m\bar{3}m'
 & ii-III & 6 & 3 & 0 & 0 & 1 & \infty & \infty & 0 & 0 & 0 & 0 \\
 & O \times \gstis & = O_h (O) & m'\bar{3}'m'
 & JJ & \infty & \infty & 0 & 0 & 0 & 4 & 9 & 1 & 0 & 0 & 0 \\
 & O (T) \times \gstis & = O_h (T_d) & m'\bar{3}'m
 & JJ' & \infty & \infty & 0 & 0 & 0 & 6 & 3 & 0 & 0 & 0 & 1 \\ \hline
\mc{7}{nl}{Number of crystallographic groups allowing tensor \rule{0pt}{2.5ex}} & 18 & 15 & 37 & & & 21 & 21 & 29 & 40 \\ \hline \rule{0pt}{2.5ex}%
Axial & C_\infty \times \gstis & = C_{\infty h} (C_\infty) & \infty /m' & JJ & \infty & \infty & 0 & 0 & 0 & 2 & 1 & 1 & 1 & 1 & 1 \\ \cline{2-16} \rule{0pt}{2.5ex}%
 & D_\infty (C_\infty) & = D_\infty (C_\infty) & \infty 2' & i-III & \infty & 1 & 1 & 0 & 1 & \infty & 1 & 0 & 1 & 0 & 1 \\
 & D_\infty  \{ C_\infty \} & = C_{\infty v} (C_\infty) & \infty m' & JJJ & \infty & 1 & 1 & 0 & 1 & 2 & \infty & 1 & 0 & 1 & 0 \\
 & D_\infty (C_\infty) \times \gsis & = D_{\infty h} (C_{\infty h}) & \infty /mm' & ii-III & \infty & 1 & 1 & 0 & 1 & \infty & \infty & 0 & 0 & 0 & 0 \\
 & D_\infty \times \gstis & = D_{\infty h} (D_\infty) & \infty /m'm' & JJ & \infty & \infty & 0 & 0 & 0 & 2 & \infty & 1 & 0 & 1 & 0 \\
 & D_\infty (C_\infty) \times \gstis & = D_{\infty h} (C_{\infty v}) & \infty /m'm & JJ' & \infty & \infty & 0 & 0 & 0 & \infty & 1 & 0 & 1 & 0 & 1 \\ \hline \rule{0pt}{2.5ex}%
Spherical & R \times \gstis & = R_{i\theta} & \infty \infty m' & JJ & \infty & \infty & 0 & 0 & 0 & \infty & \infty & 1 & 0 & 0 & 0 \\ \hline \hline
\end{tabular*}

\end{table*}

Table~\ref{tab:tensor-mag} pertains to $\theta$\=/odd tensors $T_\ell^{s-}$.
It lists the number of independent components for such tensors with
rank $\ell \le 3$ under the type\=/III magnetic point groups (58
crystallographic, 7 continuous).  According to the classification
introduced in Appendix~\ref{app:StructClass}, these comprise the groups
of type \ref{gcat:i-III}, \ref{gcat:ii-III}, \ref{gcat:iiip-III}, \ref{gcat:iii-IIIp},
\ref{gcat:JJJ}, \ref{gcat:JJJp}, \ref{gcat:JJJpp}, \ref{gcat:JJ} and
\ref{gcat:JJp}.  In the other types, $\theta$\=/odd tensors are either
equivalent to the $\theta$\=/even ones listed in Table~\ref{tab:tensor-nonmag}
(types \ref{gcat:J}, \ref{gcat:ii-I} and \ref{gcat:iii-I}), or they are forbidden
by symmetry (types \ref{gcat:i-II}, \ref{gcat:ii-II} and \ref{gcat:iii-II}).
More generally, the entries in Table~\ref{tab:tensor-mag} are related to
those given in Table~\ref{tab:tensor-nonmag} as per the
relationships illustrated by small $2 \times 2$ matrices in
Tables~\ref{tab:spher:point-group}, \ref{tab:ax:point-group}, and
\ref{tab:mag-point-group}, and further elucidated in
Appendix~\ref{app:tensors}.  As the magnetotoroidal monopolarization
$T_0^{+-}$ is forbidden by symmetry under all type\=/III groups, it
has been omitted to simplify the presentation.
Table~\ref{tab:tensor-mag} also lists the lowest even- and
odd\=/$\ell$ magnetic and magnetotoroidal multipole densities
with $\ell>0$ permitted under a given type\=/III magnetic point group.

A variety of magnetism-related physical properties can be inferred
from Table~\ref{tab:tensor-mag}.  For example, ferromagnetism occurs
for groups allowing the tensor $T_1^{+-}$~\cite{tav58, cor58a}.  In
groups where $T_1^{+-}$ is forbidden, the presence of $T_\ell^{+-}$
with odd $\ell \ge 3$ signals altermagnetism~\cite{bho24, win23}.
Axion electrodynamics can be associated with a finite
$T_0^{--}$~\cite{heh08a}.

\section{Rationalizing symmetry-allowed tensor components for
magnetic point groups}
\label{app:tensors}

This appendix discusses commonalities regarding the shape of tensors
describing systems belonging to the same Laue class $L (\gprop)$.
Combining rotations in a proper point group $\gprop$ with space
inversion $i$ and time inversion $\theta$ yields the magnetic point
groups of class $L (\gprop)$.  The relations between
symmetry transformations for these groups manifest in commonalities
regarding the form of symmetry-allowed tensors.  Similarities in the
form of crystal-property matrices are apparent from their
tabulations (see, e.g., Refs.~\cite{nye85, gri91, new05} and
Appendix~\ref{app:spherTensTab} above), and they also underpin the
observation of Opechowski's magic relations~\cite{kop06}.

We visualize the commonalities between tensor configurations for
groups within a class via small $2 \times 2$ matrices such as
$\mygroup{1}{4}{5}{5}$ shown in Tables~\ref{tab:group-types-def},
\ref{tab:subclasses}, \ref{tab:spher:point-group},
\ref{tab:ax:point-group}, and \ref{tab:mag-point-group}.  Symbols in
these matrices mark congruences between tensors with signature $++$
(upper left), $+-$ (upper right), $-+$ (lower left) and $--$ (lower
right) for a given group and between different groups within a given
class.  Filled symbols indicate that a rank\=/1 tensor is allowed
(compare Tables~\ref{tab:tensor-nonmag}
and~\ref{tab:tensor-mag}), i.e., filled symbols reflect the fact
that a group possesses a polar direction as discussed in
Sec.~\ref{sec:polar:bands}.  A missing symbol indicates that tensors
with the respective signature are forbidden for any $\ell$.  For
example, the matrix $\mygroup{1}{4}{5}{5}$ listed in
Table~\ref{tab:mag-point-group} for the group $D_{2d} (S_4) = D_4
[D_2] \, (C_4 [C_2])$ [class $L (D_4)$] indicates that this group
allows tensors $T_\ell^{ss'}$ for all four signatures $ss'$.  In
particular, this group allows rank\=/1 tensors with signature $+-$,
and rank\=/$\ell$ tensors with signature $-+$ have the same
structure as rank\=/$\ell$ tensors with signature $--$ (for $\ell
\ge \ell_\mathrm{min}$ as listed in Tables~\ref{tab:tensor-nonmag}
and~\ref{tab:tensor-mag}).  Finally, repeated appearances of the
symbol $\mcirc$ in the matrices listed for other groups in class
$L (D_4)$ indicate that the respective tensors $T_\ell^{ss'}$
allowed for these groups have the same structure as the tensors
$T_\ell^{++}$ allowed for the group $D_{2d} (S_4)$ [and similarly
for the other symbols in $\mygroup{1}{4}{5}{5}$].
These congruences of tensor configurations for groups within a given
class $L (\gprop)$ imply that knowing the structure of rank\=/$\ell$
tensors for nonmagnetic groups of type \ref{gcat:J} and
\ref{gcat:iii-I} (as discussed, e.g., by Nye~\cite{nye85}) is
sufficient to infer the structure of rank\=/$\ell$ tensors for all
groups in $L (\gprop)$.

In the following, we use the concepts of duality (\ref{eq:dual-def})
and triadic relationships (\ref{eq:triadDef}) between magnetic point
groups to formally derive, for each signature $ss'$, the congruences
of tensors $T_\ell^{ss'}$ among the groups in a given class $L
(\gprop)$.  In particular, we show how it is possible to relate the
structure of tensors $T_\ell^{s-}$ allowed for magnetic groups in $L
(\gprop)$ to the structure of the tensors $T_\ell^{s+}$ allowed for
nonmagnetic groups in $L (\gprop)$.  While our discussion focuses on
spherical tensors $T_\ell^{ss'}$ \cite{edm60, tun85} that are
relevant, e.g., for describing multipole order in solids (see
discussion in Sec.~\ref{sec:multipoles}), the congruences also apply
to Cartesian tensors representing materials properties~\cite{nye85,
bir74, kop76, kop79, gri91, lit94, gri94, kop06}, see
Appendix~\ref{app:spherTensTab}.

\subsection{Tensors $\boldsymbol{T_\ell^{++}}$: Parapolarizations}

Parapolarizations, having signature $++$, are agnostic with respect
to both SIS and TIS; hence, their structure is determined by symmetry
under proper rotations only.  Tensors $T_\ell^{++}$ thus have the
same structure for all groups in a given class $L (\gprop)$, and
this structure depends only on the symmetries present in the class
root $\gprop$.

\subsection{Tensors $\boldsymbol{T_\ell^{-+}}$: Electropolarizations}

Tensors that are odd under SIS and even under TIS are unaffected by
whether TIS may be broken.  Tensors $T_\ell^{-+}$ thus have
the same structure for all groups in a subclass $L (G_0^\theta,
\theta) \subset L (\gprop)$ , coinciding with the structure of
$T_\ell^{-+}$ for the subclass root $G_0^\theta$.  For the groups in
the subclass $L (\gprop, \theta)$, the structure of tensors
$T_\ell^{-+}$ is the same as the structure of tensors $T_\ell^{++}$
because these tensors have the same structure for the subclass root
$\gprop$.

In contrast, the structure of tensors $T_\ell^{-+}$ for the groups
in $L (\gprop [\gtprop], \theta)$ is, in
general, different from the structure of the tensors $T_\ell^{++}$.
No tensors of the form $T_\ell^{-+}$ are allowed for any groups in
a subclass $L (\gprop \times \gsis, \theta)$.
Table~\ref{tab:subclasses} illustrates the congruence of tensors
$T_\ell^{-+}$ between groups in a given subclass $L (G_0^\theta,
\theta)$ (corresponding to columns in the upper part of the table).

\subsection{Tensors $\boldsymbol{T_\ell^{+-}}$: Magnetopolarizations}

Tensors with signature $+-$ are unaffected by whether SIS may be
broken.  For all groups in a subclass $L (G_0^i, i) \subset L
(\gprop)$ (corresponding to rows in the upper part of
Table~\ref{tab:subclasses}) tensors $T_\ell^{+-}$ thus have the same
structure, coinciding with the structure of $T_\ell^{+-}$ for the
subclass root $G_0^i$.  In particular, for the groups in $L
(\gprop, i)$ tensors $T_\ell^{+-}$ have the same structure as
tensors $T_\ell^{++}$, while tensors of the form $T_\ell^{+-}$ are
forbidden by symmetry for all groups in $L (\gprop \times
\gtis, i)$.

For the groups in $L (\gprop (\gtprop), i)$, the
structure of tensors $T_\ell^{+-}$ is, in general, different from
that of the tensors $T_\ell^{++}$.  However, by virtue of the
SIS-TIS duality relation under exchange $i \leftrightarrow \theta$,
the structure of magnetopolarizations $T_\ell^{+-}$ permitted by the
groups in $L (\gprop (\gtprop), i)$ coincides
with the structure of electropolarizations $T_\ell^{-+}$ permitted
by the groups in $L (\gprop [\gtprop], \theta)$.

\subsection{Tensors $\boldsymbol{T_\ell^{--}}$: Antimagnetopolarizations}

While tensors $T_\ell^{--}$ are odd under $i$ and $\theta$ applied
individually, these tensors are even under the combination
$i\theta$.  For all groups in a Laue subclass $L (G_0^{i\theta},
i\theta) \subset L (\gprop)$, tensors $T_\ell^{--}$ thus have
the same structure, coinciding with the structure of $T_\ell^{--}$
for the subclass root $G_0^{i\theta}$.  For the groups in $L
(\gprop, i\theta)$, the structure of tensors $T_\ell^{--}$ is
the same as the structure of tensors $T_\ell^{++}$, and such tensors
are forbidden for all groups in $L (\gprop \times \gtis,
i\theta) \equiv L (\gprop \times \gsis, i\theta)$.

For the groups in $L (\gprop (\gtprop), i\theta)
\equiv L (\gprop [\gtprop], i\theta)$, the
structure of tensors $T_\ell^{--}$ is, in general, different from
the structure of the tensors $T_\ell^{++}$.  However, for the
type\=/\ref{gcat:i-III} group $\gprop (\gtprop)$,
tensors $T_\ell^{+-}$ and $T_\ell^{--}$ have the same structure.
This group is in the subclass $L (\gprop
(\gtprop) , i\theta)$.  We thus find that for the
groups in $L (\gprop (\gtprop), i\theta)$ the
structure of tensors $T_\ell^{--}$ is the same as the structure of
tensors $T_\ell^{+-}$ for the groups in $L (\gprop
(\gtprop), i)$ and the structure of
tensors $T_\ell^{-+}$ for the groups in $L (\gprop
[\gtprop], \theta)$.  See the bottom part of
Table~\ref{tab:subclasses} for the composition of subclasses $L
(G_0^{i\theta}, i\theta)$.

\subsection{Inferences facilitated by triadic relationships}

By itself, the concept of SIS-TIS duality is insufficient to infer
congruences of tensors $T_\ell^{--}$ for different groups in a class
$L (\gprop)$, as the signature $--$ is mapped onto itself
under the duality transformation $i \leftrightarrow \theta$.
Nevertheless, in the preceding subsection, we leveraged the identical
properties of $T_\ell^{+-}$ and $T_\ell^{--}$ tensors under
type\=/\ref{gcat:i-III} groups to predict the form of the tensors
for all groups in the subclass $L (\gprop
(\gtprop), i\theta)$.  Here we utilize the notion of
triadic relationships discussed in Appendix~\ref{app:triads} to reveal
further close links between the properties of tensors $T_\ell^{--}$
and tensors with mixed signatures $+-$ and $-+$.

A triad of magnetic point groups consists of a dual pair
$G_\theta\wlra G_i$ augmented by a self-dual group $G_{i\theta}$.
The diagram (\ref{eq:triadDef}) illustrates our notation, and
Table~\ref{tab:group-types-def} displays the general form of
magnetic point groups forming a triad.  For triadic partners, the
structure of tensors with signature $-+$, $+-$ and $--$ is
permuted.  As a simple nontrivial example, consider the triad
formed by groups of type \ref{gcat:JJp}, \ref{gcat:iii-II} and
\ref{gcat:ii-III} within a class, according to their general forms
given in Table~\ref{tab:group-types-def}.  Here the form of tensors
$T_\ell^{--}$ permitted by the type\=/\ref{gcat:JJp} member
$G_{i\theta}$ of this triad mirrors the form of the tensors
$T_\ell^{+-}$ allowed under the group $G_i$ of type
\ref{gcat:ii-III}, and the tensors $T_\ell^{-+}$ associated with
$G_i$'s dual partner $G_\theta$ of type \ref{gcat:iii-II}.  Thus,
the triads embody a close interrelation of tensors with signatures
$-+$, $+-$ and $--$ for magnetic point groups within a given class.

\section{Chiral materials candidates}
\label{app:chiral-materials}

This appendix tabulates examples of materials candidates
realizing the categories of chirality introduced in
Sec.~\ref{sec:chiralCat}.

\subsection{Electrochiral materials}

The physical property of electropolar electrochiral materials that
is currently attracting the most interest is the expected radial
spin texture $\vek{\sigma} \cdot \kk$ \cite{cha18} and the
associated phenomenon of a current-induced
magnetization~\cite{fur21, cal22}.  Typical illustrations of the
chirality-dependent spin texture look like
Fig.~\ref{fig:chiral-en-cub}(d).  The crucial importance of
crystallographic symmetries for shaping spin textures arising from
electrochirality has only recently been recognized~\cite{gos23, hua20}.
Insights from Ref.~\cite{gos23} and our more detailed discussion
above are useful for interpreting and guiding precision measurements
such as those performed recently for the paradigmatic chiral metal
Te~\cite{sak20, gat20}.  More electropolar electrochiral materials
candidates are shown in Table~\ref{tab:materialsElMulC}.  Selected
examples for symmetry-appropriate spin textures are shown in
Figs.~\ref{fig:chiral-en-cub} and \ref{fig:chiral-en-d4}.  See also
Ref.~\cite{gos23}.

As opposed to the nonmagnetic electropolar electrochiral materials,
multipolar electrochiral materials show either ferromagnetism or
antiferromagnetism.  The topological properties of the chiral
antiferromagnets have been discussed in Refs.~\cite{kno22, gao23}.
Materials candidates for this type are also listed in
Table~\ref{tab:materialsElMulC}.

\subsection{Magnetochiral materials}

The presence of a magnetotoroidal scalar $T_0^{+-}$ is a sufficient
condition for magnetochirality.  Some physical consequences of having
$T_0^{+-}$ are discussed in Ref.~\cite{hay23}, focusing on the
effect of terms coupling $T_0^{+-}$ to scalar products of dipoles
transforming like $T_0^{+-}$ (i.e., $T_1^{ss'}\cdot T_1^{s (-s')}$).

A more explicit electronic-structure implication of magnetochirality is
altermagnetism in materials candidates where ferromagnetism is
forbidden (``inadmissible'' magnetochiral antiferromagnets).  Candidates
for magnetochiral materials (both magnetopolar and multipolar types)
are listed in Table~\ref{tab:materialsMagnetoC}.  A recently investigated
magnetochiral materials example is Co$_2$SiO$_4$~\cite{hay25}.

\subsection{Antimagnetochiral materials}

Prominently discussed materials examples for this category are
antimagnetopolar LiMnPO$_4$~\cite{spa13} and multipolar
LuFeO$_3$~\cite{fog19}.  Interest in these compounds has been fueled
by their realization of a monopolization $T_0^{--}$ whose value has
been estimated using computational-chemistry
approaches~\cite{spa13, fog19}.  The main physical implication is the
isotropic linear magnetoelectric response (``axion
electrodynamics''~\cite{heh08a, ahn22}).  Candidates for
antimagnetochiral materials (both antimagnetopolar and multipolar
types) are listed in Table~\ref{tab:materialsAntimagnetoC}.

\subsection{Multichiral materials}

Candidates for multichiral materials are listed in
Table~\ref{tab:materialsMulC}.  These are necessarily multipolar
(see Table~\ref{tab:group-types-def}).  The existence of four
enantiomorphs is a distinctive property of such materials.
A recently investigated multichiral materials example is
BaCoSiO$_4$~\cite{xu22}.

\begin{subtables}
\begin{table*}[t]
  \caption{\label{tab:materialsElMulC} Materials candidates for the
  electrochiral category of chirality.  Examples for electropolar
  electrochiral materials are taken from Ref.~\cite{cha18}.  For the
  multipolar electrochiral category, candidate materials were
  extracted from the Bilbao magnetic-materials database~\cite{gal16,
  gal16a}.  Some of the latter have been shown previously in Table~I
  of Ref.~\cite{gao23}.  The electrochiral magnetic-point-group
  types exhibit true chirality according to the conventional
  classification~\cite{bar04}.}
  \renewcommand{\arraystretch}{1.2}%
  \begin{tabular*}{\linewidth}{nlE*{2}{les{1em}CE}n}
    \hline\hline
    crystal system &
    \multicolumn{2}{c}{electropolar electrochiral} &
    \multicolumn{2}{c}{multipolar electrochiral} \\ \hline
    cubic & Mg$_3$Ru$_2$ & O\times\gtis & BaCuTe$_2$O$_6$ & O(T) \\
    & $\beta$-RhSi & T\times\gtis \\ \hline
    hexagonal & Hf$_5$Ir$_3$ & D_6\times\gtis & EuIn$_2$As$_2$ & D_6(C_6) \\
    & & & $-$ & D_6(D_3) \\
    & $\alpha$-In$_2$Se$_3$ & C_6\times\gtis & YMnO$_3$ & C_6(C_3) \\ \hline
    trigonal & SrIr$_2$P$_2$ & D_3\times\gtis
    & BaCu$_3$V$_2$O$_8$(OD)$_2$ & D_3(C_3) \\
    & LaBSiO$_5$ & C_3\times\gtis \\ \hline
    tetragonal & MgAs$_4$ & D_4\times\gtis & CeMn$_2$Ge$_2$ & D_4(C_4) \\
    & & & Er$_2$Ge$_2$O$_7$  & D_4(D_2) \\
    & Sr$_2$As$_2$O$_7$ & C_4\times\gtis & $-$ & C_4(C_2) \\ \hline
    orthorhombic & $\alpha$-Ag$_2$Se & D_2\times\gtis
    & Ca$_2$CoSi$_2$O$_7$ & D_2(C_2) \\ \hline \hline
  \end{tabular*}
\end{table*}

\begin{table*}[t]
  \caption{\label{tab:materialsMagnetoC} Materials candidates for
  the magnetochiral category of chirality.  All materials examples
  have been extracted from the Bilbao magnetic-materials
  database~\cite{gal16, gal16a}.  See also Ref.~\cite{hay23}
  (especially Table~I in that work) for a related discussion.  The
  category of magnetochirality introduced in our work subsumes
  point-group types that have previously been considered as
  achiral~\cite{bar04}.  They are also different from preexisting
  notions of magnetic chirality~\cite{che22}.}
  \renewcommand{\arraystretch}{1.2}%
  \begin{tabular*}{\linewidth}{nlE*{2}{les{1em}RLE}n}
    \hline\hline
    crystal system &
    \multicolumn{3}{c}{magnetopolar magnetochiral} &
    \multicolumn{3}{c}{multipolar magnetochiral} \\ \hline
    cubic &  $-$ & O_h & = O\times\gsis & $-$ & T_d & = O [T] \\
    & NiS$_2$ & T_h & = T\times\gsis \\ \hline
    hexagonal & $-$ & D_{6h} & = D_6\times\gsis & HoMnO$_3$
    & C_{6v} & = D_6 [C_6] \\
    & & & & Ba$_3$CoSb$_2$O$_9$ & D_{3h} & = D_6 [D_3] \\
    & FeF$_3$ & C_{6h} & = C_6\times\gsis & $-$ & C_{3h}
    & = C_6 [C_3] \\  \hline
    trigonal & $\alpha$-Fe$_2$O$_3$ & D_{3d} & = D_3\times\gsis & PbNiO$_3$
    & C_{3v} & = D_3 [C_3] \\
    & CaFe$_3$Ti$_4$O$_{12}$ & C_{3i} & = C_3\times\gsis \\ \hline
    tetragonal & CdYb$_2$S$_4$ & D_{4h} & = D_4\times\gsis & $-$
    & C_{4v} & = D_4 [C_4] \\
    & & & & Ba$_2$MnSi$_2$O$_7$ & D_{2d} & = D_4 [D_2] \\
    & MnV$_2$O$_4$ & C_{4h} & = C_4\times\gsis & $-$ & S_4
    & = C_4 [C_2] \\ \hline
    orthorhombic & Co$_2$SiO$_4$ & D_{2h} & = D_2\times\gsis
    & HoCrWO$_6$ & C_{2v} & = D_2 [C_2] \\ \hline
  \end{tabular*}
\end{table*}

\begin{table*}[t]
  \caption{\label{tab:materialsAntimagnetoC} Materials candidates for
  the antimagnetochiral categories of chirality.  All examples have
  been extracted from the Bilbao magnetic-materials
  database~\cite{gal16, gal16a}.  The antimagnetochiral
  magnetic-point-group types coincide with those considered to
  be falsely chiral according to prior convention~\cite{bar04}.}
  \renewcommand{\arraystretch}{1.2}%
  \begin{tabular*}{\linewidth}{nlE*{2}{les{1em}RLE}n}
    \hline\hline
    crystal system &
    \multicolumn{3}{c}{antimagnetopolar antimagnetochiral} &
    \multicolumn{3}{c}{multipolar antimagnetochiral} \\ \hline
    cubic & $-$ & O_h (O)& = O\times\gstis & $-$ & T_d(T)& = O \{T\} \\
    & $-$ & T_h (T) & = T\times\gstis \\ \hline
    hexagonal & $-$ & D_{6h} (D_6) & = D_6\times\gstis & LuFeO$_3$
    & C_{6v} (C_6) & = D_6\{C_6\} \\
    & & & & TmAgGe & D_{3h} (D_3) & = D_6 \{D_3\} \\
    & PbMn$_2$Ni$_6$Te$_3$O$_{18}$ & C_{6h} (C_6)
    & = C_6\times\gstis & Cu$_{0.82}$Mn$_{1.18}$As & C_{3h} (C_3)
    & = C_6 \{C_3\}   \\ \hline
    trigonal & Cr$_2$O$_3$ & D_{3d} (D_3) & = D_3\times\gstis
    & CaBaCo$_2$Fe$_2$O$_7$ & C_{3v} (C_3) & = D_3 \{ C_3\} \\
    & MnTiO$_3$ & C_{3i} (C_3) & = C_3\times\gstis & \\ \hline
    tetragonal & GdB$_4$ & D_{4h} (D_4) & = D_4\times\gstis
    & CeCoGe$_3$ & C_{4v} (C_4) & = D_4 \{C_4\} \\
    & & & & Pb$_2$MnO$_4$ & D_{2d} (D_2) & = D_4 \{D_2\} \\
    & TlFe$_{1.6}$Se$_2$ & C_{4h} (C_4) & = C_4\times\gstis
    & CsCoF$_4$ & S_4 (C_2) & = C_4 \{ C_2\} \\ \hline
    orthorhombic & LiMnPO$_4$ & D_{2h} (D_2) & = D_2\times\gstis
    & SrMnSb$_2$ & C_{2v} (C_2) & = D_2 \{C_2\} \\ \hline
  \end{tabular*}
\end{table*}

\begin{table}[t]
  \caption{\label{tab:materialsMulC} Materials candidates for the
  multichiral category of chirality.  These were extracted from the Bilbao
  magnetic-materials database~\cite{gal16, gal16a}, and some
  of them have been shown previously in Table~I of Ref.~\cite{gao23}.
  The multichiral magnetic-point-group types exhibit true chirality
  according to the conventional classification~\cite{bar04}.}
  \renewcommand{\arraystretch}{1.2}%
  \begin{tabular*}{\linewidth}{nlE*{1}{les{2em}CE}n}
    \hline\hline
    crystal system &
    \multicolumn{2}{c}{multipolar multichiral} \\ \hline
    cubic & SrCuTe$_2$O$_6$ & O \\
    & Mn$_3$CoGe & T \\ \hline
    hexagonal & $-$ & D_6 \\
    & BaCoSiO$_4$ & C_6 \\ \hline
    trigonal & La$_{0.33}$Sr$_{0.67}$FeO$_3$ & D_3 \\
    & Cu$_2$OSeO$_3$ & C_3 \\ \hline
    tetragonal & Ho$_2$Ge$_2$O$_7$ & D_4 \\
    & Ce$_5$TeO$_8$ & C_4 \\ \hline
    orthorhombic & FePO$_4$ & D_2 \\ \hline \hline
  \end{tabular*}
\end{table}
\end{subtables}

\cleardoublepage


\begin{thebibliography}{205}
\bibitem [{misc:doubleG2()}]{misc:doubleG2}%
  \BibitemOpen
  \bibinfo {note} {For TIS to be a black-white symmetry requires $\theta^2 =
  e$, whereas $\theta^2=-1$ for half-integer spin states \cite{lud96}. The case
  $\theta^2 = e$ is relevant for a range of important quantities and phenomena.
  The present work focuses on macroscopically observable quantities such as
  electric and magnetic multipoles that must transform according to the integer
  irreducible representations (IRs) of the full rotation group $\Rit$ for which
  $i^2 = \theta^2 = e$. Moreover, we are interested in electronic band
  structures as in Fig.~\ref{fig:polar}. Double-group IRs where $\theta^2=-e$
  are relevant when analyzing eigenstates like Bloch functions in a spinful
  theory. However, the operator of spin angular momentum $\vek{\sigma}$ by
  itself transforms like a vector ($\ell=1$), see Eq.\ (\ref{eq:spin-dipole}).
  Therefore, single-group IRs are sufficient when the relevant effective
  Hamiltonians for a spinful band theory are derived, e.g., via the theory of
  invariants~\cite{bir74}. Hence, for the range of quantities that are of
  interest here, the group theory of the symmetries $i$ and $\theta$ is the
  same.}\BibitemShut {Stop}%
\bibitem [{misc:sigtens()}]{misc:sigtens}%
  \BibitemOpen
  \bibinfo {note} {Previous work does not offer a universally accepted notation
  for the four basic types of tensors. In Refs.~\cite{gri91, gri94}, tensors
  with signatures $++$, $+-$, $-+$, and $--$ are denoted as i-tensor, s-tensor,
  t-tensor, and u-tensor, respectively, while Refs.~\cite{kop76, kop79, kop06}
  use the symbols $1^+_e$, $1^+_m$, $1^-_e$, and $1^-_m$. See also Table~4 in
  Ref.~\cite{sch08}.}\BibitemShut {Stop}%
\bibitem [{misc:specinv()}]{misc:specinv}%
  \BibitemOpen
  \bibinfo {note} {If $G$ contains none of the transformations $i$, $\theta$,
  or $i\theta$ as individual elements, then $\tilde{G} = G$ and $C_\gin \equiv
  C_1$. If $G$ contains all three of these transformations as elements, then
  $C_\gin \equiv \gsistis$. The only remaining possibility is that $G$ contains
  exactly one inversion symmetry $\gamma$ as an individual group element, in
  which case $C_\gin \equiv C_\gamma$.}\BibitemShut {Stop}%
\bibitem [{misc:T1++()}]{misc:T1++}%
  \BibitemOpen
  \bibinfo {note} {The electrotoroidal tensor $T_1^{++}$ is an interesting
  exception among all spherical tensors. It cannot be expressed as a compound
  tensor combining one nontrivial and one or two trivial tensors. This tensor
  does not arise in Table~\ref{tab:tensor-pow-spher} either. Also, the tensor
  component $T_{10}^{++}$ is not an invariant tensor for any continuous
  dihedral group $G_{d \infty}$; it is only invariant under the continuous
  cyclic group $G_{c \infty} \times \gsistis$, so it cannot be distinguished
  from even\=/$\ell$ electric components $T_{g0}^{++}$, see
  Appendix~\ref{app:axial-groups}. Therefore $T_1^{++}$ is not considered
  further in this work.}\BibitemShut {Stop}%
\bibitem [{misc:theoinv()}]{misc:theoinv}%
  \BibitemOpen
  \bibinfo {note} {For conceptual clarity, the current analysis is restricted
  to bands without orbital degeneracies near the $\Gamma$ point $k=0$ of the
  Brillouin zone. All arguments can be extended to bands with orbital
  degeneracies and other expansion points in the Brillouin zone
  \cite{bir74}.}\BibitemShut {Stop}%
\bibitem [{misc:weak()}]{misc:weak}%
  \BibitemOpen
  \bibinfo {note} {Different symmetry classifications identifying point and
  space group symmetries permitting weak ferromagnetism have been developed by
  Tavger \cite{tav58} and Turov \cite{tur62}.}\BibitemShut {Stop}%
\bibitem [{misc:chiral()}]{misc:chiral}%
  \BibitemOpen
  \bibinfo {note} {See Ref.~\cite{wag07} for a general overview. The relation
  between modern uses for the term ``handedness'' and the precise meaning of
  ``chirality'' is discussed, e.g., in Ref.~\cite{fec22}.}\BibitemShut {Stop}%
\bibitem [{misc:allBackground()}]{misc:allBackground}%
  \BibitemOpen
  \bibinfo {note} {The definition, basic structure and general properties of
  magnetic-point-group classes are discussed in Appendix~\ref{app:StructClass}.
  The application to continuous and discrete point groups is discussed in
  Appendix~\ref{app:class-groups}.}\BibitemShut {Stop}%
\bibitem [{misc:noether()}]{misc:noether}%
  \BibitemOpen
  \bibinfo {note} {The chiral charges $T_0^{ss'}$ are different from the chiral
  charges encountered in relativistic field theories based on Noether's
  theorem, i.e., based on a conservation law reflecting a continuous chiral
  symmetry \cite{pes95}. The chiral charges $T_0^{ss'}$ reflect the discrete
  symmetries in a crystalline environment.}\BibitemShut {Stop}%
\bibitem [{misc:relateToBarron()}]{misc:relateToBarron}%
  \BibitemOpen
  \bibinfo {note} {Systems considered to be chiral according to the definitions
  of Ref.~\cite{bar04} could be either electrochiral or multichiral. Both the
  parachiral and magnetochiral categories would conventionally be considered as
  achiral, and antimagnetochirality is equivalent to false chirality. Magnetic
  chirality as discussed, e.g., in Ref.~\cite{che22} constitutes a subset of
  the electrochiral category.}\BibitemShut {Stop}%
\bibitem [{misc:L-text()}]{misc:L-text}%
  \BibitemOpen
  \bibinfo {note} {Recent work has studied chirality-induced
  orbital-angular-momentum textures~\cite{yan23a, kim23, bri24}.}\BibitemShut
  {Stop}%
\bibitem [{misc:char()}]{misc:char}%
  \BibitemOpen
  \bibinfo {note} {The rows of Tables~\ref{tab:spher:trafo},
  \ref{tab:c-cyclic:trafo}, and \ref{tab:c-dihedral:trafo}, listing only one
  invariant tensor on the right, can be interpreted as sections of the
  character tables for the $\ell=0$ IRs of the full rotation group $\Rit$
  (Table~\ref{tab:spher:trafo}) and the $m=0$ IRs of $C_\infty \times \gsistis$
  (Table~\ref{tab:c-cyclic:trafo}) and the full axial rotation group $D_\infty
  \times \gsistis$ (Table~\ref{tab:c-dihedral:trafo}), respectively. All of
  these IRs are one-dimensional and ``$\bullet$'' (``$\circ$'') indicates that
  the respective character is $+1$ ($-1$). The invariant tensors on the right
  then represent basis functions for these IRs.}\BibitemShut {Stop}%
\bibitem [{misc:mainBlock()}]{misc:mainBlock}%
  \BibitemOpen
  \bibinfo {note} {In Tables~\ref{tab:spher:point-group},
  \ref{tab:ax:point-group} and \ref{tab:mag-point-group}, the main block of
  groups from a given class has the groups of type \ref{gcat:J},
  \ref{gcat:ii-I}, \ref{gcat:i-II} and \ref{gcat:ii-II} from this class at its
  four corners. See table captions for the discussion of how placement of
  magnetic groups within a class's main block illustrates SIS-TIS
  duality.}\BibitemShut {Stop}%
\end{thebibliography}
\end{document}